\documentclass{ieeeaccess}
\usepackage{cite}
\usepackage{amsmath,amssymb,amsfonts}
\usepackage{algorithmic}
\usepackage{graphicx}
\usepackage{textcomp}
\usepackage{float}
\usepackage{soul}
\usepackage{subcaption}
\usepackage{multirow}
\usepackage{changepage}
\usepackage{hyperref}
\usepackage{ulem}
\usepackage{adjustbox}
\usepackage{changepage}
\makeatother
\usepackage{comment}
\usepackage{xcolor}
\usepackage{pict2e}
 
\newcommand\orcidicon[1]{\href{https://orcid.org/#1}{\usebox{\ORCIDlogo}}}
\newsavebox{\ORCIDlogo}
\savebox{\ORCIDlogo}{%
\setlength{\unitlength}{\dimexpr 1em/256\relax}%
\begin{picture}(256,256)%
  \color[HTML]{A6CE39}\put(128,128){\circle*{256}}%
  \color{white}%
  \put(78.6,199.2){\circle*{20}}%
  \moveto(70.9,176,9)\lineto(86.3,176,9)\lineto(86.3,69.8)\lineto(70.9,69.8)%
  \closepath\fillpath%
  \moveto(108.9,176.9)\lineto(150.5,176.9)%
  \curveto(190.1,176.9)(207.5,148.6)(207.5 ,123.3)%
  \curveto(207.5,95,8)(186,69.7)(150.7,69.7)%
  \lineto(108.9,69.7)%
  \closepath\fillpath%
  \color[HTML]{A6CE39}%
  \moveto(124.3,83.6)\lineto(148.8,83.6)%
  \curveto(183.7,83.6)(191.7,110.1)(191.7,123.3)%
  \curveto(191.7,144.8)(178,163)(148,163)%
  \lineto(124.3,163)%
  \closepath\fillpath%
\end{picture}%
}
\def\BibTeX{{\rm B\kern-.05em{\sc i\kern-.025em b}\kern-.08em
T\kern-.1667em\lower.7ex\hbox{E}\kern-.125emX}}
\usepackage{CJKutf8}
\def\BibTeX{{\rm B\kern-.05em{\sc i\kern-.025em b}\kern-.08em
T\kern-.1667em\lower.7ex\hbox{E}\kern-.125emX}}
\usepackage{xcolor}
\usepackage{pict2e}
\savebox{\ORCIDlogo}{%
\setlength{\unitlength}{\dimexpr 1em/256\relax}%
\begin{picture}(256,256)%
  \color[HTML]{A6CE39}\put(128,128){\circle*{256}}%
  \color{white}%
  \put(78.6,199.2){\circle*{20}}%
  \moveto(70.9,176,9)\lineto(86.3,176,9)\lineto(86.3,69.8)\lineto(70.9,69.8)%
  \closepath\fillpath%
  \moveto(108.9,176.9)\lineto(150.5,176.9)%
  \curveto(190.1,176.9)(207.5,148.6)(207.5 ,123.3)%
  \curveto(207.5,95,8)(186,69.7)(150.7,69.7)%
  \lineto(108.9,69.7)%
  \closepath\fillpath%
  \color[HTML]{A6CE39}%
  \moveto(124.3,83.6)\lineto(148.8,83.6)%
  \curveto(183.7,83.6)(191.7,110.1)(191.7,123.3)%
  \curveto(191.7,144.8)(178,163)(148,163)%
  \lineto(124.3,163)%
  \closepath\fillpath%
\end{picture}%
}
\usepackage{hyperref} 
\hypersetup{
  colorlinks=false,
  linkbordercolor=white,
 urlbordercolor=white,
pdfborder={0 0 0}
}
\usepackage{CJKutf8}

\begin{document}
\history{Date of publication xxxx 00, 0000, date of current version xxxx 00, 0000.}
\doi{10.1109/ACCESS.2017.DOI}
\title{A Methodological and Structural Review of Parkinson's Disease Detection Across Diverse Data Modalities}
\author{\uppercase{Abu Saleh Musa Miah\ \orcidicon{0000-0002-1238-0464} \authorrefmark{1}, (IEEE Member)},
\uppercase{TARO SUZUKI}\authorrefmark{2},
\uppercase{Jungpil Shin}\ \orcidicon{0000-0002-7476-2468}\authorrefmark{3}, \IEEEmembership{Senior Member, IEEE},
}
 \address[1,2,3]{School of Computer Science and Engineering, The University of Aizu, Aizuwakamatsu, Japan (musa@u-aizu.ac.jp, taro@u-aizu.ac.jp)}
\markboth
{....}
{This paper is currently under review for possible publication in IEEE Access.}

\corresp{Corresponding author:Jungpil Shin (jpshin@u-aizu.ac.jp).}
\begin{abstract}
Parkinson’s Disease (PD) is a progressive neurological disorder that primarily affects motor functions and can lead to mild cognitive impairment (MCI) and dementia in its advanced stages. With approximately 10 million people diagnosed globally—1 to 1.8 per 1,000 individuals, according to reports by the Japan Times and the Parkinson Foundation—early and accurate diagnosis of PD is crucial for improving patient outcomes. While numerous studies have utilized machine learning (ML) and deep learning (DL) techniques for PD recognition, existing surveys are limited in scope, often focusing on single data modalities and failing to capture the potential of multimodal approaches.
To address these gaps, this study presents a comprehensive review of PD recognition systems across diverse data modalities, including Magnetic Resonance Imaging (MRI), gait-based pose analysis, gait sensory data, handwriting analysis, speech test data, Electroencephalography (EEG), and multimodal fusion techniques. Based on over 347 articles from leading scientific databases, this review examines key aspects such as data collection methods, settings, feature representations, and system performance, with a focus on recognition accuracy and robustness. Notably, limitations such as the lack of robust multimodal frameworks and challenges in continuous PD recognition are identified, underscoring the need for innovative solutions. The review highlights significant advancements in PD recognition, particularly the transition from handcrafted feature engineering to DL-based methods. Furthermore, it explores promising trends in multimodal approaches, offering insights into overcoming existing challenges such as data scarcity, model generalization, and computational efficiency. The study emphasizes the importance of integrating diverse data sources to achieve higher diagnostic accuracy and reliability. This survey aims to serve as a comprehensive resource for researchers, providing actionable guidance for the development of next-generation PD recognition systems. By leveraging diverse data modalities and cutting-edge machine learning paradigms, this work contributes to advancing the state of PD diagnostics and improving patient care through innovative, multimodal approaches.
\end{abstract}

\titlepgskip=-15pt
\maketitle
\section{Introduction}
\label{sec1}
 Parkinson's disease (PD) is a progressive neurodegenerative disorder affecting dopaminergic neurons, which is crucial for movement control \cite{kulkarni2025parkinson,dauer2003parkinson,perlmutter2009assessment,rizzo2016accuracy}. According to the World Health Organization (WHO), 7–10 million people worldwide are diagnosed with PD, with a male-to-female ratio of approximately 3:2. The disease commonly affects individuals over 50 years old \cite{lamba2022hybrid_who,adewale2025patient}. Symptoms include handwriting difficulties, tremors, freezing gait, falls, and emotional changes. Patients often experience sadness, sleep disturbances, fatigue, digestive issues, and slowed speech or responses. Symptoms worsen over time, making daily life challenging for patients and their families. While there is no cure, care and awareness are essential for managing the disease.
PD presents with both motor and non-motor symptoms that progressively worsen, greatly affecting daily activities. The motor symptoms, crucial for diagnosis, typically include a triad of muscle rigidity, tremors, and disturbances in postural reflexes. Individuals with PD may experience a range of symptoms, such as bradykinesia (slowness in movement), dysarthria (difficulty in speaking), anxiety, depression, sleep disorders, and cognitive challenges. By the time the symptoms of Parkinson's disease become evident, a significant loss of dopamine-producing neurons in the substantia nigra has already occurred, with about 60\% to 80\% of these vital cells being depleted. This depletion greatly impacts the brain's ability to regulate movement and coordination, leading to the characteristic motor symptoms of the disease \cite{dauer2003parkinson}. 
Early diagnosis of Parkinson's disease (PD) is critical for assessing disease progression and improving the quality of life for patients and their caregivers. Medical specialists utilise various clinical methods such as the Schwab and England rating, Hoehn-Yahr staging, and the Unified Parkinson’s Disease Rating Scale (UPDRS) \cite{fahn1987unified_updrs,perlmutter2009assessment}. Revised by the Movement Disorder Society (MDS) in 2008, the UPDRS includes 42 items across four subscales. It involves patients performing motor tasks, such as finger tapping, which physicians visually evaluate on a 5-point scale (0–4) based on parameters like speed, size, and rhythm to detect akinesia. However, this method relies heavily on the physician's expertise, leading to potential variability and subjectivity. Studies reveal that even movement disorder experts achieve a diagnostic accuracy of only 79.6\% at initial assessments and 83.9\% during follow-up evaluations \cite{rizzo2016accuracy}.

\subsection{Background}
A psychiatrist or physician must deeply understand PD to analyze a patient's data and symptoms. Unfortunately, many countries, specifically developing countries, lack a sufficient number of qualified doctors, making it challenging to identify or diagnose PD, which leads to difficulties in detection. While medications can be prescribed, their effectiveness diminishes as the disease progresses from its early stages. Therefore, early detection of Parkinson's disease is crucial for taking timely measures to help individuals maintain their independence for as long as possible. This need has driven healthcare providers to create decision-support systems based on computer-aided diagnosis to aid clinicians in diagnosing PD \cite{loh2021application}. 
Recently, many researchers have been working to develop computer-aided early PD recognition systems using machine learning and deep learning. Their main goal is to develop an automated PD detection or recognition system using various data modalities, including  Magnetic Resonance Imaging (MRI) \cite{Mukherjee1996_F-DMFP_PET_dataset,Grunder2003_F-DMFP_PET_dataset,Pahuja2016,Pahuja2022,Liu2020,yasaka2021parkinson, Wang2023, Talai2021,  Shinde2019, Ramirez2020, miriad2024, tagaris2018machine_NTUA,ds004471:1.0.1}, Gesture-based pose and sensory \cite{Moon2020, elMaachi2019, Zeng2016, Muniz2010, Pfister2020, Eskofier2016, Ricci2019, Talitckii2020, Wahid2015, Tucker2015, Prochazka2015}
,  handwriting exams \cite{margolin1983agraphia,phillips1991handwriting,Drotar2014, Pereira2015_handpd, shaban2021automated, Naseer2020, Kamran2021}, speech patterns \cite{Frid2016, Tsanas2013, Rasheed2020, Gunduz2019, Karabayir2020, Zhang2017, Hirschauer2015},   Electroencephalogram
phy (EEG) \cite{vanegas2018machine, oh2018deep, wagh2020eeg, koch2019automated, shi2019hybrid, lee2019deep, khare2021detection, khare2021pdcnnet, loh2021gaborpdnet, shaban2021automated, Connolly2015}
 and multi-modal fusion data modalities \cite{taleb2020parkinson_thesis}. Although the above-mentioned various modalities have recently been explored for applying ML and DL to detecting Parkinson's disease, clinical diagnosis still primarily relies on observing motor system abnormalities. However, this approach remains subjective and susceptible to human error. Additionally, there are currently no specific or widely recognised clinical biomarkers for Parkinson's disease or its associated complications.
\subsection{Existing PD Detection Survey Papers}
 As we discussed in the above section, there are no specific or widely recognised clinical biomarkers for Parkinson's disease, and we still need to encourage young researchers in the field. To do this, many researchers have been working to combine existing PD detection work to help young researchers understand the current and future trends in PD recognition work. 
  In 2016, Chahine et al. reviewed the PD detection research paper published between 2005-2015, which focused on the significant sleep disorders associated with Parkinson's disease (PD), such as REM sleep behaviour disorder, insomnia, nocturia, restless legs syndrome, sleep-disordered breathing, excessive daytime sleepiness, and circadian rhythm disorders \cite{chahine2017systematic}.  In 2017, Pereira et al. \cite{pereira2019survey} reviewed the papers published between 2015 and 2016, focusing on signal-based data, virtual reality, and e-health monitoring systems. They also suggested future research to improve the understanding and management of sleep dysfunction in PD, but they did not discuss the smartphone-based PD system. In 2020, Anju et al. reviewed smartphones to monitor movements and speech, utilising a deep multi-layer perceptron classifier without direct physician interaction, allowing for early detection even with incomplete data \cite{anju2020recent}. In 2021, More et al. reviewed the only speech data modality-based PD recognition system. They discuss mainly phonatory and articulatory aspects of speech, highlighting methodological issues and suggesting that while these aspects are crucial, further research is needed to establish validated methodologies and identify new biomarkers \cite{moro2021advances}. To overcome the modality-oriented knowledge lacking in 2021, Loh et al. reviewed the PD detection research paper published between 2011 and 2021, where they included  63 studies on deep learning models for automated Parkinson's disease (PD) diagnosis, covering various modalities like brain imaging and motion symptoms \cite{loh2021application,kumari2025exploiting}. They also highlight the various advantages of the existing work in summaries. In 2022, Guo et al. reviewed a research paper that makes the relationship between human gait and PD, emphasising the role of gait analysis systems, both vision-based and wearable sensor-based, in detecting and managing PD with smart systems and treatment \cite{guo2022detection_review}. In 2022, Rana et al. reviewed 112 research papers published until 2022 to provide an overview of machine learning (ML) techniques and data modalities used in PD analysis by highlighting key challenges and offering future recommendations for research \cite{rana2022imperative_review}. In 2023, Shaban et al. \cite{shaban2023deep} reviewed Parkinson's disease (PD) detection research published between 2016 and 2022. Their comprehensive review focused on machine learning and deep learning-based systems for PD recognition, disease staging, and identifying biomarkers using four different data modalities. In 2024, Desai et al. \cite{desai2024enhancing} reviewed existing deep learning (DL) algorithms for PD diagnosis and introduced a CNN model that utilizes 3D brain MRI \cite{zhang2025identification,ye2025individualized} images from the PPMI dataset. More recently, some researchers conducted a survey on gait pose, sensor data, speech, and multimodal approaches, incorporating few data modalities \cite{alkhanbouli2025role, zhao2025artificial,xie2025optimal,wu2025boostering,dudek2025analysis,boyle2025activity,singh2025review}. However, we did not find any survey papers that address PD detection using six different modalities, despite recent publications in the field.
    
  
\begin{figure}[htp]
    \centering
    \includegraphics[width=9cm]{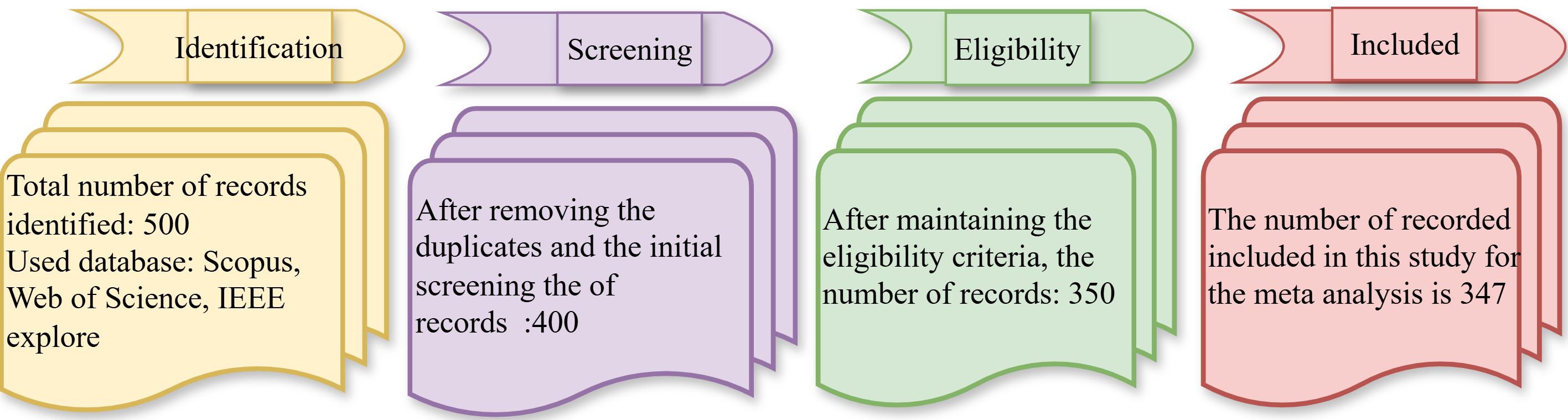}
    \caption{Article selection process procedure.}
    \label{fig:article_slection_pro}
    \end{figure}

\begin{figure}[htp]
    \centering
    \includegraphics[width=8cm]{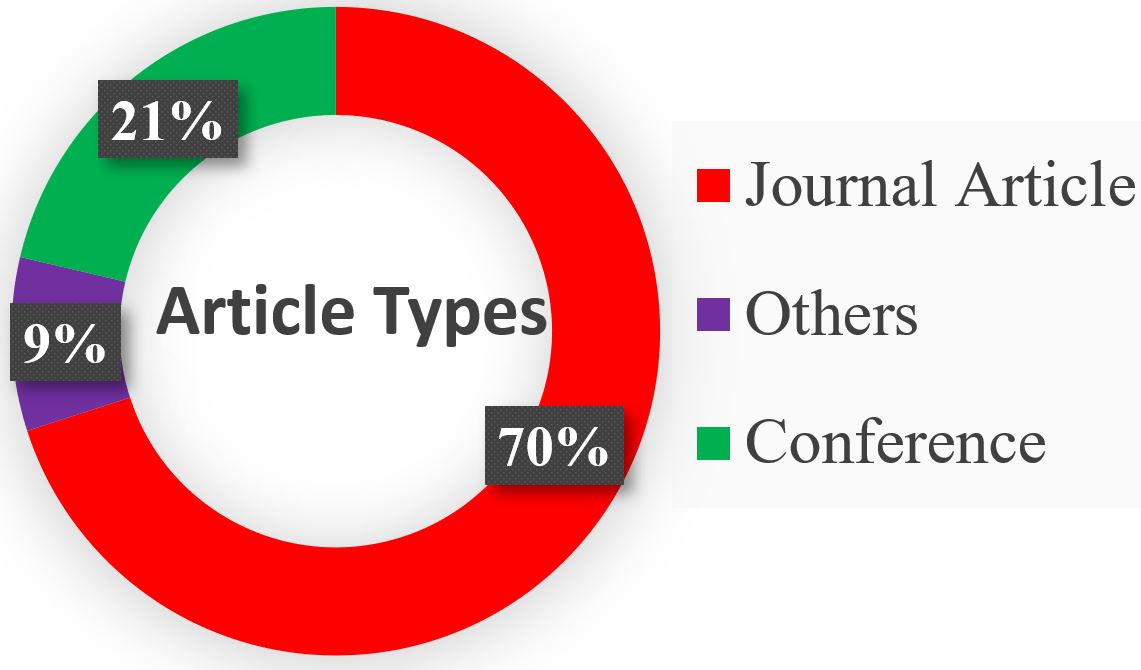}
    \caption{Article types: journal, conference, and others.}
    \label{fig:intro_reference}
\end{figure}

\begin{figure}[htp]
    \centering
    \includegraphics[width=9cm]{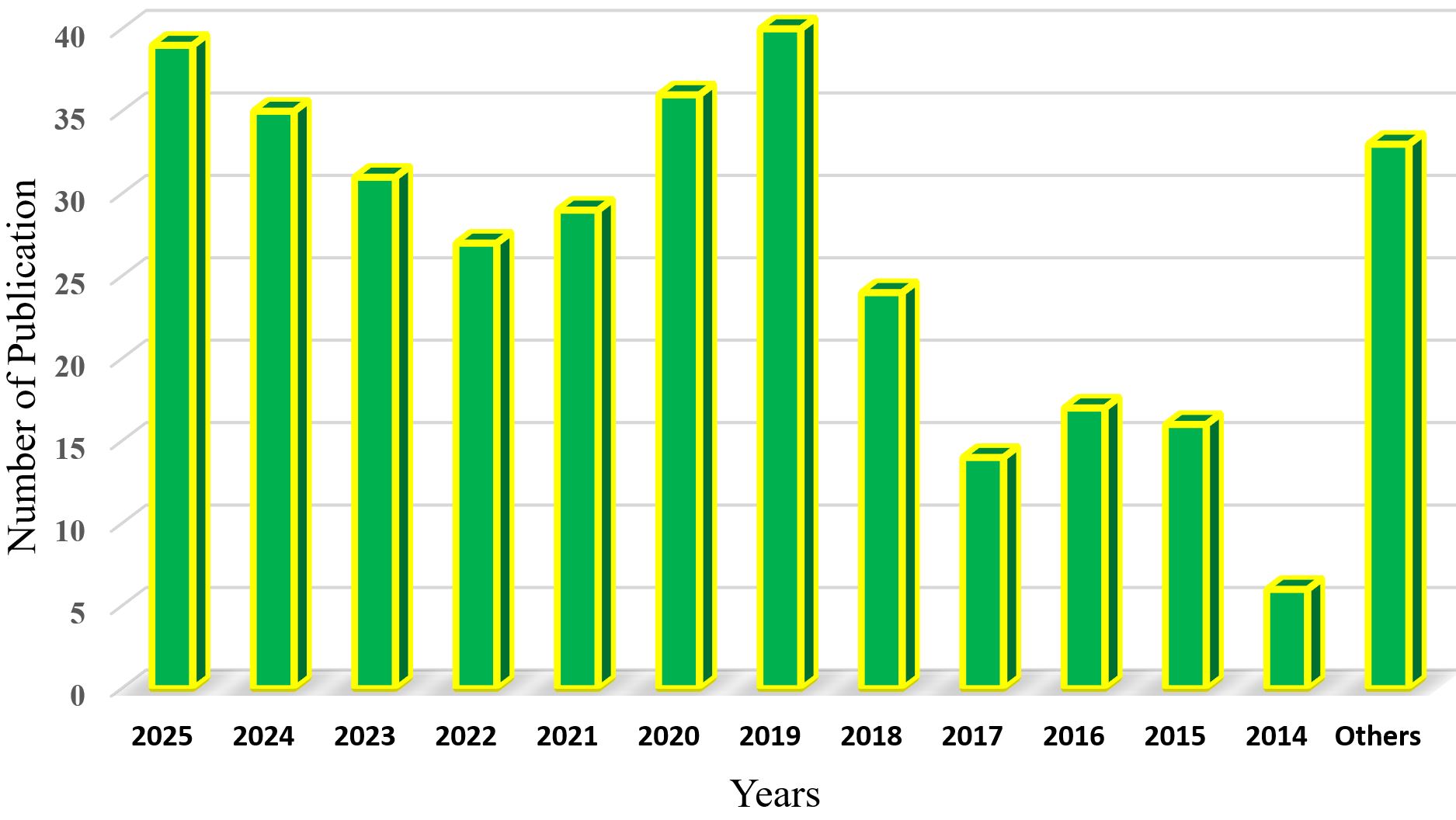}
    \caption{Year-wise peer-reviewed publications used in the study.}
    \label{fig:year_wise_reference}
\end{figure}
 \begin{figure}[htp]
    \centering
    \includegraphics[width=9cm]{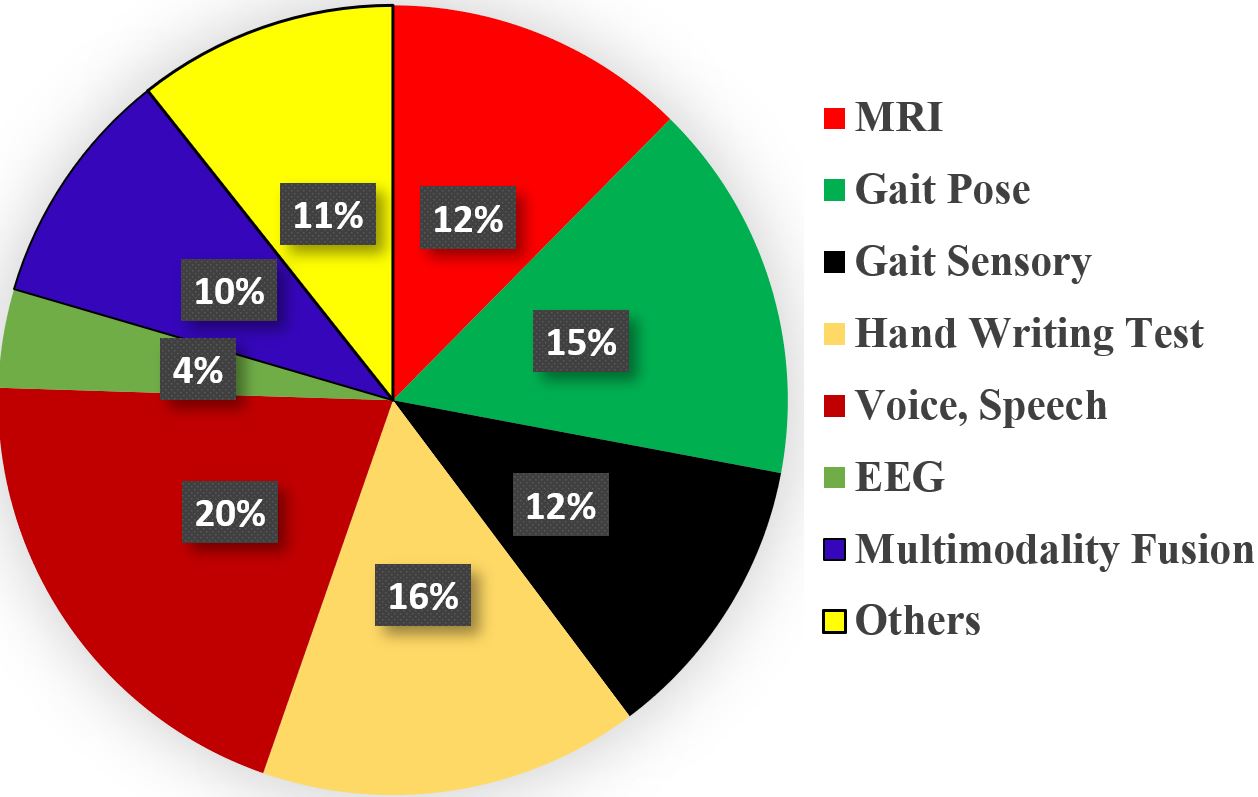}
    \caption{Various data modalities based references.}
    \label{fig:data_modalities_reference}
\end{figure}

\begin{figure}[htp]
    \centering
    \includegraphics[width=9cm]{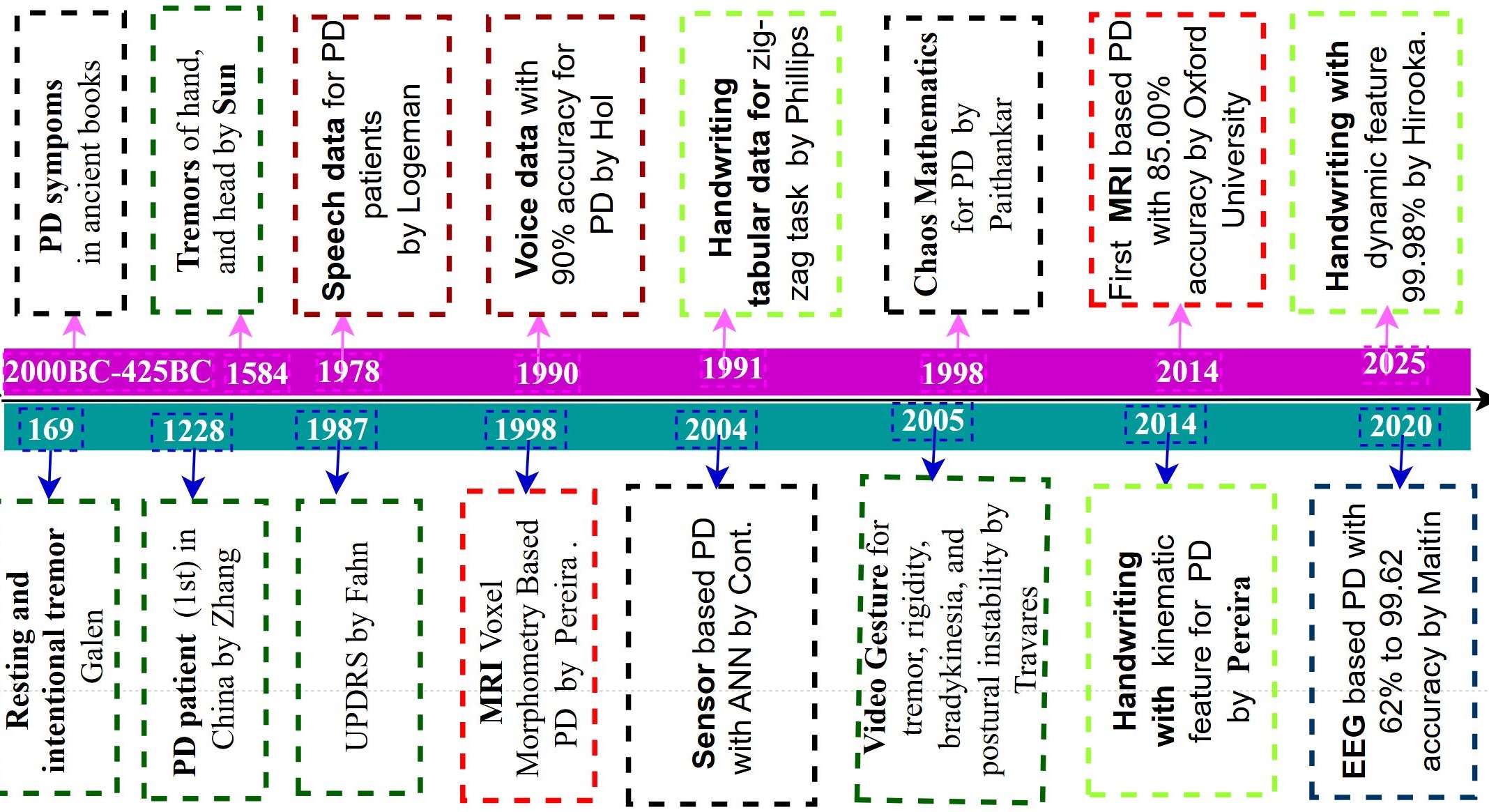}
    \caption{Example of the PD evaluation across various data modalities.}
    \label{fig:domain_various_modalities}
\end{figure}

\subsection{Article Search Strategy and Survey Methodology}\label{sec3.1}
To identify relevant articles on multimodality-based Parkinson's Disease (PD) recognition, we employed a targeted search strategy using specific keywords. The focus was on the following terms:
\begin{itemize}
    \item MRI-based PD recognition systems.
    \item Pose and sensory gait or gesture analysis-based PD recognition systems.
    \item Speech and handwriting-based PD recognition systems.
    \item Multimodal dataset fusion-based PD recognition.
\end{itemize}
We sourced articles from esteemed databases to ensure a comprehensive review of the pertinent literature. The databases included the Web of Science database, Science Direct, Scopus, IEEE Xplore Digital Library, MDPI, Springer Link, ResearchGate, and Google Scholar. These article sources have conceptual data and abstracts from various research publications. The Scopus search engine provides a standardised and homogeneous search approach, which has a strong relationship with various diseases, specifically PD. 

To refine and ensure relevance in our initial search results, we applied the following criteria:
\begin{itemize}
    \item Publications dated between 2014 and 2024.
    \item Inclusion of journals, proceedings, and book chapters.
    \item Focus on various data modality-based PD systems, including MRI, Gait-based Pose, Gait-based sensory, handwriting, speech, EEG, other single modality and multimodal fusion. 
   \item Exclude papers that only mention PD briefly or indirectly.
    \item Exclude literature that mainly reviews other researchers' work.
    \item Exclude studies that do not provide detailed information about their experiments.
    \item Exclude research articles where the full text is not available in either print or online.
    \item Exclude research articles that contain opinions, keynote speeches, discussions, editorials, tutorials, comments, introductions, personal views, or slide presentations.
\end{itemize}

We identified 347 articles that met our inclusion and exclusion criteria using the search keywords outlined in this methodology. Figure \ref{fig:article_slection_pro} demonstrates the article selection procedure.
Figures \ref{fig:intro_reference}, \ref{fig:year_wise_reference}, and \ref{fig:data_modalities_reference} display the types of references, year-wise references, and data modalities based on the number of references, respectively. Figure \ref{fig:domain_various_modalities} shows the evaluation of PD research across various data dimensions. In our survey methodology, each article was reviewed through a structured process involving abstract review, methodology analysis, discussion, and result evaluation. Most of the papers were sourced from the IEEE Xplore Digital Library, ensuring a high standard of research quality. The different modalities used in PD recognition have unique features, each with its own set of advantages and disadvantages.

\subsection{Research Gaps and Emerging Challenges}
Despite the extensive reviews of Parkinson's Disease (PD) detection techniques, existing surveys have several limitations that create research gaps in the field. Firstly, most reviews focus on single-modal detection approaches, such as MRI, speech, or gait analysis, without providing an integrated perspective on the advancements and challenges in multimodal PD detection systems. These surveys often neglect the complexities involved in fusing multiple data modalities, such as combining EEG, MRI, speech, and handwriting data, which could improve diagnostic accuracy. Furthermore, current surveys often fail to address the specific technological and computational hurdles in developing efficient and real-time multimodal PD detection systems, such as data heterogeneity, processing speed, and the scalability of machine learning models. Many existing reviews lack a forward-looking analysis of the potential integration of new data modalities and AI techniques, including deep learning and multimodal fusion strategies. 
Our research aims to address these gaps by thoroughly examining the literature on multimodality-based PD systems, emphasising recent advancements, unique challenges, and future directions. By critically analysing the progress in this area, we seek to provide actionable insights and highlight new research challenges, such as optimising data fusion techniques, improving computational efficiency, and enhancing clinical applicability, to guide future research efforts in developing robust and efficient multimodal PD detection systems.

\subsection{Contribution}
\begin{figure*}[htp]
    \centering
    \includegraphics[width=18cm]{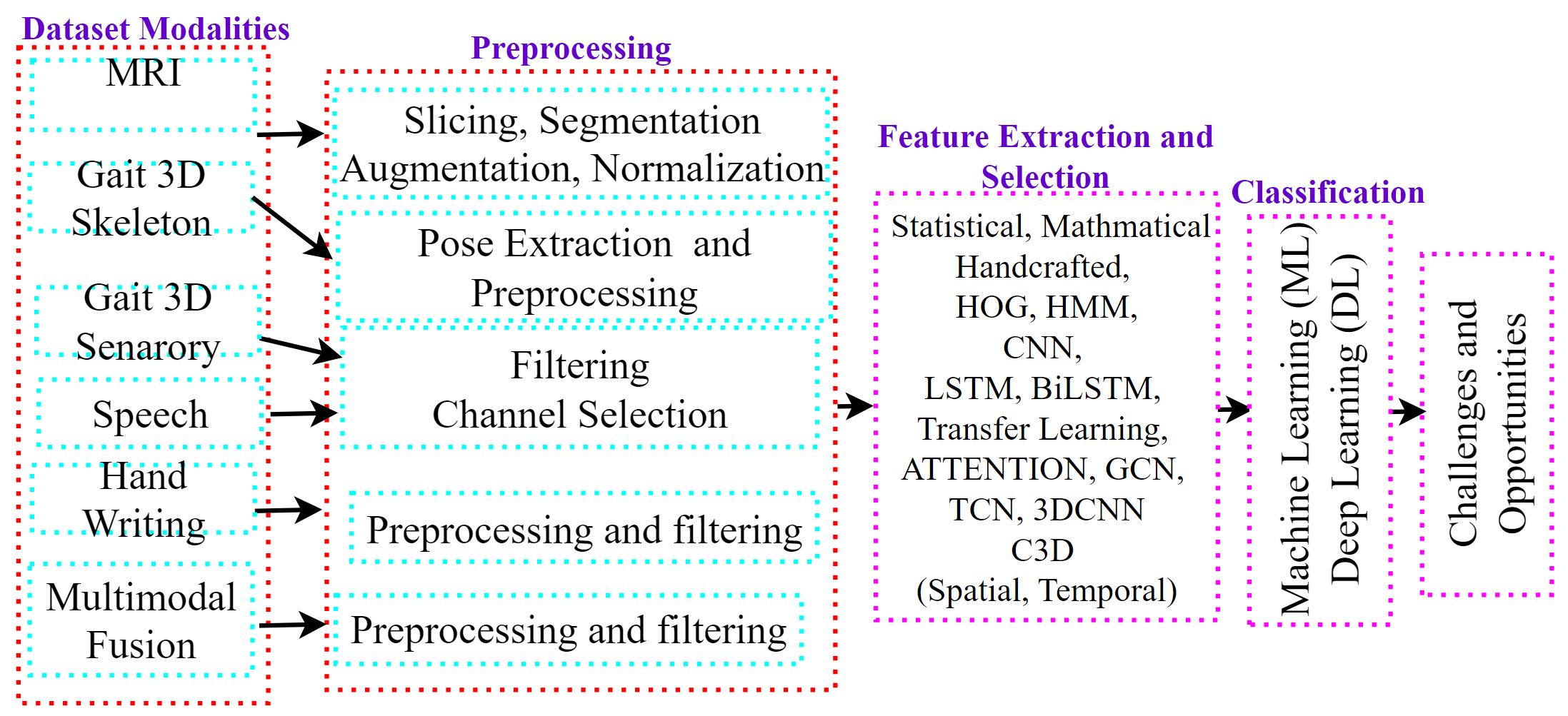}
    \caption{The basic flowgraph of the PD research work.}
    \label{fig:basic_PD_recognition_model}
\end{figure*}

\begin{figure}[htp]
    \centering
    \includegraphics[width=9cm]{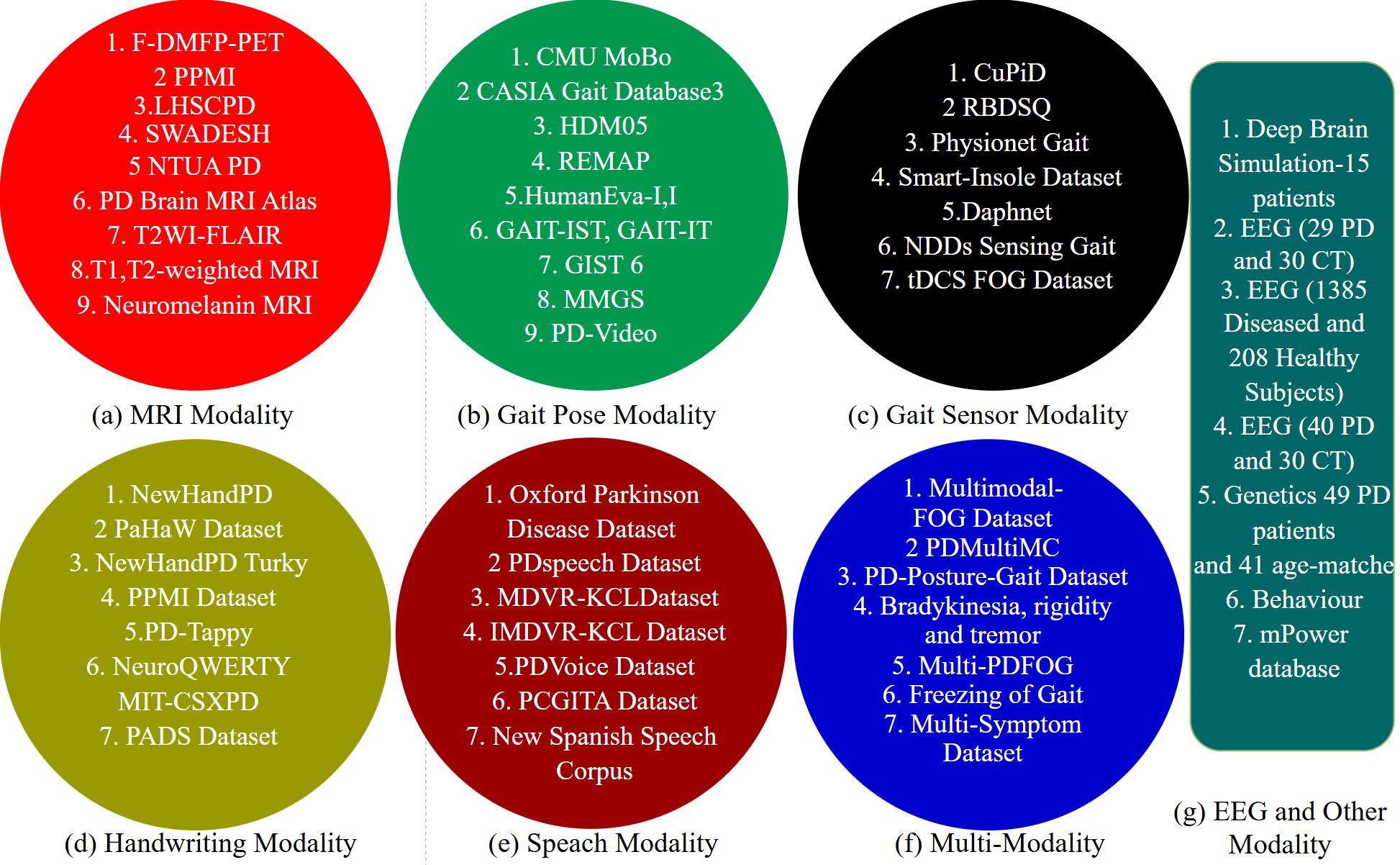}
    \caption{Modality-based dataset names included in the study.}
    \label{fig:modality_based_dataset}
\end{figure}

Figures \ref{fig:basic_PD_recognition_model} illustrate the proposed methodology flowchart and the overall structure of this study, respectively. The key contributions of this research are summarised as follows:
\begin{itemize}
    \item \textbf{Comprehensive Review:} This work provides an extensive and detailed review of Parkinson's Disease (PD) recognition systems, emphasising the evolution of data acquisition methods, data environments, and PD characterisation from 2014 to 2025. It covers various modalities, ensuring a holistic understanding of the field's progress over the last decade.
    \item \textbf{Multimodal Analysis:} For the first time, this study systematically examines the advancements in multiple data modalities used in PD detection systems, including RGB, skeleton, depth, audio, EMG, EEG, and multimodal fusion. By analysing these diverse modalities, our study provides a broad perspective on the state-of-the-art techniques and their applications in PD recognition.
    \item \textbf{Creation of Benchmark Dataset Tables:} This study compiles and presents benchmark tables for various data modalities, including detailed information on datasets and the latest performance metrics, with citations for each modality (MRI, gait pose, gait sensor, speech, EEG, and multimodal fusion). These tables serve as a valuable reference for researchers to compare and assess different PD detection methods. Figure \ref{fig:modality_based_dataset} shows the list of modality-based benchmark datasets. 
    \item \textbf{Identification of Gaps and Future Directions:} By critically reviewing existing literature, we identify significant gaps in current research and propose potential future research directions. Our study also highlights key challenges associated with each data modality and suggests areas for further investigation to enhance the effectiveness and efficiency of PD recognition systems.
    \item \textbf{Evaluation of System Efficacy:} We evaluate the effectiveness of current PD detection systems by analysing recognition accuracy and other performance metrics. This provides a benchmark for future studies, helping researchers understand the relative strengths and weaknesses of different approaches.
    \item \textbf{Guidance for Researchers and Practitioners:} The findings and insights from this comprehensive review offer practical guidance for developing more robust and accurate PD detection systems. The study is a valuable resource because it summarises state-of-the-art techniques, identifies challenges, and suggests potential future research directions. This facilitates informed decision-making and fosters innovation in developing PD detection technologies.
\end{itemize}
By addressing these contributions, our paper not only fills existing gaps in the literature but also paves the way for future advancements in PD recognition and management.
\subsection{Research Questions and Our Approach}
To guide our investigation, we have formulated the following primary research questions:

\begin{itemize}
    \item 
    \textbf{Research Question 1 (RQ1):} What advancements have been made in the development of PD recognition systems using MRI, vision-based gait analysis, sensory gait information, speech, EEG, and handwriting data from 2014 to 2025, particularly in terms of data acquisition methods, data environments, and disease characterization?
    \item 
    \textbf{Research Question 2 (RQ2):} How effective are current PD recognition systems that utilize these various data modalities, and what are the potential future directions for enhancing their diagnostic accuracy and clinical applicability?
\end{itemize}

\subsection{Organization of the Paper}
To organise the paper, we follow the sequence below: Section \ref{sec1} MRI modality. Section \ref{sec:video} described the gait-based pose recognition for PD, and then the sensory data modality-based PD is described in Section \ref{sec:sensor}. Section \ref{sec:handwriting} described the handwriting data modalities based on PD. Section \ref{sec:voice} describes the speech modality-based PD. Section \ref{sec:eeg} demonstrated the EEG signal-modality-based PD,  and Section \ref{sec:multimodal} demonstrates the multimodal  fusion-based PD. Conclusion described in the Section \ref{sect8}.
\section{MRI  Data Modality Based PD Recognition System} \label{sec:mri}
Magnetic Resonance Imaging (MRI) data are widely used in the diagnosis of neurological disorders. Recent advancements in machine learning (ML) and deep learning (DL) have opened up new possibilities to make new advancements in MRI-based PD recognition. Several researchers have successfully applied ML and DL techniques to MRI data, enabling the detection of subtle patterns associated with PD that would otherwise go unnoticed. These techniques can analyze complex data and extract features that improve diagnostic accuracy, making them valuable tools for early and more precise PD detection.
\subsection{Dataset}
Table \ref{tab:MRI_pd_metadata} summarizes MRI-based Parkinson's disease (PD) datasets. Below, we detail key datasets utilized by researchers for PD detection and progression analysis. Pahuja et al. \cite{Pahuja2016} utilised the Parkinson’s Progression Marker Initiative (PPMI) dataset, which includes 150 3d T1-weighted MRI images split equally between healthy individuals and PD patients. Segmentation errors led to the replacement of 72 images with 60 new ones. Liu et al. \cite{Liu2020} employed a dataset from Jinzhou Medical University, featuring T2-weighted, axial T1-weighted, and FLAIR images from 138 participants. Yasaka et al. \cite{yasaka2021parkinson} collected MRI data from 115 PD patients and 115 healthy controls using a 3T Siemens scanner for T1-weighted and magnetization transfer saturation images. Wang et al. \cite{Wang2023} studied PD using Quantitative Susceptibility Mapping (QSM) and T1-weighted MRI, collecting data from 92 PD patients and 287 healthy individuals at Ruijin Hospital, along with a dataset from Zhengzhou University comprising 83 PD patients and 72 controls. Talai et al. \cite{Talai2021} analyzed MRI data from 45 PD patients, 20 individuals with progressive supranuclear palsy, and 38 healthy controls using DTI and T1-weighted images from a 3T Siemens Skyra scanner. Other notable datasets include those by Shinde et al. \cite{Shinde2019}, who used neuromelanin-sensitive MRI from 55 subjects, and Ramírez et al. \cite{Ramirez2020}, who utilized the PPMI dataset for subcortical volume analysis and fractional anisotropy (FA) mapping. These datasets provide diverse neuroimaging modalities, aiding PD diagnosis and progression analysis. Additional details are covered in the following subsections.

\begin{table*}[h]
\centering
\caption{Well-known MRI based PD datasets. }
\label{tab:MRI_pd_metadata}
\begin{adjustwidth}{0cm}{0cm}
\setlength{\tabcolsep}{3pt}
\begin{tabular}{|l|c|c|c|l|l|l|l|l|}
\hline
\textbf{Author} & \textbf{Year} & \begin{tabular}[c]{@{}l@{}}\textbf{Dataset} \\ \textbf{Names}\end{tabular} & \textbf{Classes} & \begin{tabular}[c]{@{}l@{}}\textbf{Dataset} \\ \textbf{Takens}\end{tabular} & \begin{tabular}[c]{@{}l@{}}\textbf{Type} \\ \textbf{of Scan}\end{tabular} & \begin{tabular}[c]{@{}l@{}}\textbf{No. of} \\ \textbf{Sub.}\end{tabular} & \begin{tabular}[c]{@{}l@{}}\textbf{Tot.} \\ \textbf{Sample}\end{tabular} & \begin{tabular}[c]{@{}l@{}}\textbf{Latest} \\ \textbf{Accuracy}\end{tabular} \\ \hline
Mukhar et al. \cite{Mukherjee1996_F-DMFP_PET_dataset, Grunder2003_F-DMFP_PET_dataset} & 1980s & F-DMFP-PET & PD, CN & \begin{tabular}[c]{@{}l@{}}Fluorine-18 \\as a radioisotope,\\ Desmethoxyfallypride\end{tabular} & PET & 87 &- & -\\ \hline
Pahuja et al. \cite{Pahuja2016} & 2016 & - & 2 & PPMI & \begin{tabular}[c]{@{}l@{}}T1-weighted \\ MRI\end{tabular} & \begin{tabular}[c]{@{}l@{}}75 PD, \\ 75 HC\end{tabular} & - &-  \\ \hline
Ramírez et al. \cite{Ramirez2020} & 2020 & - & 2 & \begin{tabular}[c]{@{}l@{}}Open Access \\ Database\end{tabular} & DTI & \begin{tabular}[c]{@{}l@{}}129 PD, \\ 57 HC\end{tabular} &  &  \\ \hline
Miriad et al. \cite{miriad2024} & 2024 & PPMI & 2 (PD, CN) & MRI & Parkinson’s Disease & - & - & 95.29 \\ \hline
DS et al \cite{ds004471:1.0.1} & - & LHSCPD & 2 (PD) & MRI & Parkinson’s Disease & - & - & - \\ \hline
DS et al \cite{ds004471:1.0.1} & - & SWADESH & 2 (PD) & MRI & Parkinson’s Disease & - & - & - \\ \hline
Tagaris et al. \cite{tagaris2018machine_NTUA} & 2018 & NTUA PD \footnote{\url{https://github.com/ails-lab/ntua-parkinson-dataset/tree/master}} & PD, CN & \begin{tabular}[c]{@{}l@{}}MRI and DaT scans, \\ Demographics, \\ Clinical Information\end{tabular} & \begin{tabular}[c]{@{}l@{}}MRI, \\ DaT\end{tabular} & 78 & 43,087 & - \\ 
\hline
 Xia et al. \cite{xiao2019bridging_brain_mri} & 2024 & PD Brain MRI Atlas & PD & \begin{tabular}[c]{@{}l@{}}Multi-contrast \\ brain atlas\end{tabular} & \begin{tabular}[c]{@{}l@{}}T1w (FLASH, MPRAGE), \\ T2*w, T1-T2* fusion, \\ phase, R2* map\end{tabular} & 25 & - & 87.21 \cite{islam2024review_sruvey_handwriting_voice_data}\\ \hline
 PPMI et al. \cite{wu2017exosomes_ppmi,mathur2015rising_ppmi} & 2016 & PPMI & 2 (PD, CN) & \begin{tabular}[c]{@{}l@{}}Longitudinal \\ Observational \\ Study\end{tabular} & \begin{tabular}[c]{@{}l@{}}MRI, \\ Clinical, \\ Biologic, \\ SPECT\end{tabular} & \begin{tabular}[c]{@{}l@{}}402 PD, \\ 184 CN\end{tabular} & Multiple & 97.68 \cite{islam2024review_sruvey_handwriting_voice_data} \\ \hline
 Liu et al. \cite{liu2023large_TWw1-FLAIR} & 2016 & T2WI-FLAIR & 2 (PD, CN) & \begin{tabular}[c]{@{}l@{}}T2WI-FLAIR \\ Study\end{tabular} & \begin{tabular}[c]{@{}l@{}} BIDS\end{tabular} & \begin{tabular}[c]{@{}l@{}}2,888\end{tabular} & Multiple & 98.6 \cite{li2024parkinson} \\ \hline
Yasaka et al. \cite{yasaka2021parkinson} & 2021 & T1-weighted MRI & - & \begin{tabular}[c]{@{}l@{}}Hospital \\ data\end{tabular} & \begin{tabular}[c]{@{}l@{}}T1-weighted MRI, \\ dMRI, MT \\ saturation \\ images\end{tabular} & \begin{tabular}[c]{@{}l@{}}115 PD, \\ 115 HC\end{tabular} & - & - \\ \hline
Wang et al. \cite{Wang2023} & 2023 & T1-weighted MRI & -& \begin{tabular}[c]{@{}l@{}}Hospital \\ data\end{tabular} & \begin{tabular}[c]{@{}l@{}}T1-weighted MRI, \\ QSM\end{tabular} & \begin{tabular}[c]{@{}l@{}}92 PD, 287 HC \\ \& 83 PD, \\ 72 HC\end{tabular} & - &-  \\ \hline
Talai et al. \cite{Talai2021} & 2021 & T1,T2-weighted MRI & - & \begin{tabular}[c]{@{}l@{}}Medical \\ Center\end{tabular} & \begin{tabular}[c]{@{}l@{}}T1-weighted MRI, \\ T2-weighted \\ MRI, DTI\end{tabular} & \begin{tabular}[c]{@{}l@{}}45 PD, 20 \\ Progressive \\ supranuclear \\ palsy, 38 HC\end{tabular} & - & - \\ \hline
Pahuja et al. \cite{Pahuja2022} & 2022 &T1-weighted MRI  & - & Heterogeneous & \begin{tabular}[c]{@{}l@{}}T1-weighted \\ MRI\end{tabular} & \begin{tabular}[c]{@{}l@{}}73 PD, \\ 59 HC\end{tabular} &-  &  -\\ \hline
Shinde et al. \cite{Shinde2019} & 2019 & Neuromelanin MRI & - & \begin{tabular}[c]{@{}l@{}}Medical \\ Center\end{tabular} & \begin{tabular}[c]{@{}l@{}}Neuromelanin \\ sensitive MRI\end{tabular} & \begin{tabular}[c]{@{}l@{}}45 PD, 20 \\ Atypical \\ Parkinsonian \\ Syndromes, \\ 35 HC\end{tabular} & - & - \\ \hline
\end{tabular}
\end{adjustwidth}
\end{table*}

\subsubsection{F-DMFP-PET Dataset}

Fluorine-18, a radioisotope emitting positrons, is used as a radiotracer in medical research, particularly in positron emission tomography (PET) studies \cite{Mukherjee1996_F-DMFP_PET_dataset}. Desmethoxyfallypride, a moderate-affinity dopamine D2/D3 receptor antagonist, is commonly used in these studies \cite{Grunder2003_F-DMFP_PET_dataset}. Initially introduced in the 1980s for oncology, PET is now applied in neurodegenerative disease research, including Parkinson's Disease (PD). The 18F-DMFP-PET dataset focuses on striatal dopamine analysis and brain glucose metabolism, with data from 87 individuals diagnosed with Parkinsonism. Images are taken one hour after the radiopharmaceutical is injected, making this dataset valuable for PD analysis.

\subsubsection{PPMI dataset}
The Parkinson’s Progression Markers Initiative (PPMI) dataset is a comprehensive resource dedicated to studying Parkinson’s Disease (PD) progression \cite{wu2017exosomes_ppmi,mathur2015rising_ppmi}. This dataset comprises 600 instances, including 400 PD patients and 200 healthy controls, with 159 attributes and no missing values. Since its release in December 2011, the PPMI dataset has been widely utilized in PD-related research. For this study, baseline T1-weighted and DTI images were selected and acquired using 3T scanners from GE and Siemens. The dataset features 140 early-stage, drug-naive PD patients and 70 healthy controls (HCS). PD patients are further categorized into Postural Instability Gait Difficulty (PIGD, n=70) and Tremor-Dominant (TD, n=70) groups. To ensure balanced comparisons, age and sex were matched across groups, facilitating both three-label (HC vs. PIGD vs. TD) and binary (PIGD vs. TD) classifications.
This dataset provides valuable insights into PD progression and is available for access at the following link \url{(www.ppmi- info. org)}

\subsection{MRI Preprocessing Techniques for PD Detection}
Preprocessing is crucial and challenging when working with MRI images. Many young researchers use preprocessed or sliced data available on platforms like Kaggle or GitHub to avoid the complexities of MRI preprocessing. Table \ref{tab:mri_cleaning_transformation} presents various preprocessing techniques that assist in converting MRI.nii files into slice images.
\begin{table*}[!htp]
\centering
\caption{MRI Data cleaning and transformation for early PD detection.}
\label{tab:mri_cleaning_transformation}
\begin{adjustwidth}{0cm}{0cm}
\setlength{\tabcolsep}{3pt}
\begin{tabular}{|p{2.5cm}|p{.7cm}|p{2cm}|p{4cm}|p{.7cm}|p{2cm}|p{2cm}|p{2cm}|}
\hline
Author & Year & MRI Modality Datasets & Process & Output & Dataset Name & Classification & Performance \\ \hline
Bhan et al. \cite{Bhan2021} & 2021 & T1 Weighted & MRIcro tool One Hot Encoding & Binary & PPMI & LeNet-5 & 97.62 \\ \hline
Sangeetha et al. \cite{Sangeetha2023} & 2023 & T1 Weighted & Median Filtering & Image & PPMI & CNN & 95.20 \\ \hline
Zhang et al. \cite{Zhang2019} & 2019 & T1 Weighted & WGAN Technology ImageDataGenerator API & Image & PPMI & ResNet & 76.46 \\ \hline
Bhan et al. \cite{Bhan2021a} & 2021 & T1 Weighted & CLAHE Gaussian blur Histogram Equalizer & Image & - & ResNet-50 & 98.91 \\ \hline
Camacho et al. \cite{Camacho2023} & 2023 & T1 Weighted & HDBET ANTs log-Jacobian maps & Image & MNI PD25 atlas & CNN & 85.00 \\ \hline
Rubbert et al. \cite{Rubbert2019} & 2019 & T1 Weighted & FMRIB BET2 FIX & Image & rs-fMRI & LR & 76.2 \\ \hline
Pereira et al. \cite{pereira2016deep_newhandpd} & 2016 & T1 Weighted & Visual Rhythm approaches & Image & HandPD & CNN & 77.92 \\ \hline
Noor et al. \cite{Noor2020} & 2020 & T1 Weighted & Motion correction Spatial normalization Scaling & Image & ADNI, OASIS, MIRIAD & Deep Learning & - \\ \hline
Veetil et al. \cite{Veetil2021} & 2021 & T1 Weighted & Skull stripping Bias field correction Normalization Data augmentation & Image & PPMI & Vgg19 & 92.6 \\ \hline
Erdaş et al. \cite{Erdas2023} & 2023 & T1 Weighted & Image registration using python code & Image & PPMI & 3DCNN & 95.49 \\ \hline
Chakraborty et al. \cite{chakraborty2020detection} & 2020 & T1 Weighted & Image registration & Image & PPMI & 3DCNN & 95.29 \\ \hline
\end{tabular}
\end{adjustwidth}
\end{table*}

Pereira et al. \cite{pereira2016deep_newhandpd} used a CNN to extract features for PD detection, applying normalization to input signals before representation learning. Rubbert et al. \cite{Rubbert2019} preprocessed whole-brain rs-fMRI data for PD detection using FMRIB Software Library (FSL), which included brain extraction, Gaussian smoothing, normalization, and denoising via ICA Xnoisifier. Zhang et al. \cite{Zhang2019} enhanced PPMI datasets with WGAN-based data augmentation and the Keras ImageDataGenerator API to increase image diversity through rotations and inversions. Chakraborty et al. \cite{chakraborty2020detection} aligned MRI data from the PPMI dataset to the MNI atlas using ANTsPy for image registration, ensuring multi-centre consistency. Noor et al. \cite{Noor2020} applied motion correction, spatial normalization, and scaling to improve MRI data quality for PD diagnosis. Bhan et al. \cite{Bhan2021, Bhan2021a} used MRIcro to preprocess PPMI T1-weighted images, removed irrelevant portions, and employed Gaussian blur and CLAHE for image enhancement, converting images to YUV channels for further processing. Veetil et al. \cite{Veetil2021} applied skull stripping, bias field correction, and data augmentation, splitting data into 80\% training and 20\% testing and normalizing image intensities. Sangeetha et al. \cite{Sangeetha2023} used median filtering on PPMI MRI images to reduce noise while preserving edges, enhancing diagnostic accuracy. Camacho et al. \cite{Camacho2023} processed T1-weighted images using HD-BET for brain extraction, bias field correction with ANTs, and registered data to the MNI PD25-T1-MPRAGE atlas.
 Erdaş et al. \cite{Erdas2023} utilized image registration to align the PPMI dataset with the MNI atlas using Python-based FLIRT tools for precise alignment.


\subsection{MRI Based PD Recognition State of the Art ML Approaches}
Machine learning (ML) techniques excel in MRI-based Parkinson's Disease (PD) recognition by detecting subtle, imperceptible patterns in MRI scans that are often missed by the human eye. These approaches hold significant promise for early and accurate PD diagnosis, offering insights beyond traditional clinical assessments. For instance, Shinde et al. \cite{Shinde2019} developed a CNN-based model leveraging NeuroMelanin-sensitive (NMS) MRI biomarkers, achieving an accuracy of 83.6\%, surpassing traditional regression analysis methods (81.8\%). Built on the efficient ResNet architecture, this model highlighted the potential of deep learning for medical image classification. Conversely, Prasuhn et al. \cite{Prasuhn2020} utilized a binary Support Vector Machine (bSVM) and Multiple-Kernel Learning (MKL) on DTI MRI data but achieved a limited AUC of 60\%, indicating diffusion metrics alone may be insufficient for reliable PD detection. Panahi et al. \cite{Panahi_Hosseini_2024} employed Diffusion Tensor Imaging (DTI) MRI data to classify tremor-dominant, postural instability gait difficulty, and healthy control groups. Using Pyradiomics for feature extraction and a Light Gradient Boosting Machine (LGBM), they achieved a notable accuracy and AUC of 95\%, though accuracy decreased to 85\% for distinguishing healthy controls. Similarly, Li et al. \cite{li2024automatic} applied T2WI-FLAIR sequences with a 3D VB-Net and SVM for classification, reaching 95\% accuracy. Khanna et al. \cite{khanna2023novel} explored ML models, including SVM, LR, and random forest, on MRI data from the PPMI and SWADESH datasets. Their models achieved impressive accuracies of 92.45\% and 90.57\%, respectively, emphasizing the effectiveness of combining multiple ML techniques for robust PD detection. 

\begin{table*}[htp!]
\centering
\caption{ MRI dataset modality for PD Recognition using ML and DL based Approaches.}
\label{tab:MRI_ML_methods}
\begin{adjustwidth}{-.5cm}{0cm}
\setlength{\tabcolsep}{4pt}
\begin{tabular}{|l|l|l|l|l|l|l|l|l|l|}
\hline
\textbf{Method} & \textbf{Year} & \begin{tabular}[c]{@{}l@{}}\textbf{Main} \\ \textbf{Objective}\end{tabular} & \begin{tabular}[c]{@{}l@{}}\textbf{MRI} \\ \textbf{Modality}\\ \textbf{Datasets}\end{tabular} & \begin{tabular}[c]{@{}l@{}}\textbf{No of} \\ \textbf{Class.}\end{tabular} & \begin{tabular}[c]{@{}l@{}}\textbf{No of} \\ \textbf{Samples}\end{tabular} & \begin{tabular}[c]{@{}l@{}}\textbf{Feature} \\ \textbf{Extraction} \end{tabular} & \textbf{Classifier} & \textbf{Perfor.[\%]} & \textbf{Limitations} \\ \hline
Shinde et al. \cite{Shinde2019} & 2019 & \begin{tabular}[c]{@{}l@{}} biomarkers of PD \\ using NMS-MRI\end{tabular} & \begin{tabular}[c]{@{}l@{}}NeuroMelanin-\\ sensitive MRI\end{tabular} & 2 & - & - & \begin{tabular}[c]{@{}l@{}}RA\end{tabular} & \begin{tabular}[c]{@{}l@{}}80; \\ 83.6 cv\end{tabular} & \begin{tabular}[c]{@{}l@{}}Dependent on \\ the testing \\ ability \end{tabular} \\ \hline
Prasuhn et al. \cite{Prasuhn2020} & 2020 & PD Detection & \begin{tabular}[c]{@{}l@{}}DTI MRI \\ (162 PD \\ and 70 CN)\end{tabular} & 2 & 232 & \begin{tabular}[c]{@{}l@{}}bSVM, \\ MKL\end{tabular} & ML& \begin{tabular}[c]{@{}l@{}}AUC: \\58, \\ 60\end{tabular} & Low performance \\ \hline
Panahi et al. \cite{Panahi_Hosseini_2024} & 2024 & PD Detection & \begin{tabular}[c]{@{}l@{}}DTI MRI \\ (70 TD, \\ 70 PIGD\\ and 70 HC)\end{tabular} & 2 and 3 & - & Pyradiomics & LGBM & \begin{tabular}[c]{@{}l@{}}Accuracy: \\ 95, 85 \\ AUC:\\ 95, 94 \end{tabular} & - \\ \hline
Li et al. \cite{li2024automatic} & 2024 & PD Detection & T2WI-FLAIR & - & - & 3D VB-Net & SVM & 95.0 & - \\ \hline
Khanna et al. \cite{khanna2023novel} & 2023 & PD Detection & \begin{tabular}[c]{@{}l@{}}MRI \\ (PPMI and \\ SWADESH)\end{tabular} & 2 & - & FDR & \begin{tabular}[c]{@{}l@{}}SVM, L \\ Regression, \\ RF, KNN\end{tabular} & \begin{tabular}[c]{@{}l@{}}92.45, \\ 90.57\end{tabular} & - \\ \hline
Kaplan et al. \cite{Kaplan2022} & 2022 & \begin{tabular}[c]{@{}l@{}}Early \\ diagnosis of PD\end{tabular} & MRI & - & - & - & \begin{tabular}[c]{@{}l@{}}kNN, \\ SVM\end{tabular} & \begin{tabular}[c]{@{}l@{}}99.53, \\ 99.22, \\ 98.70 \end{tabular} & \begin{tabular}[c]{@{}l@{}}Accuracy may vary \\for number and \\ quality of  features\end{tabular} \\ \hline
Sahu et al. \cite{Sahu2022} & 2022 & \begin{tabular}[c]{@{}l@{}}PD Detection\\ using DL\end{tabular} & \begin{tabular}[c]{@{}l@{}}Medical \\ Data\end{tabular} & 2 & - & \begin{tabular}[c]{@{}l@{}}RA, ANN\end{tabular} & \begin{tabular}[c]{@{}l@{}}ANN\end{tabular} & \begin{tabular}[c]{@{}l@{}}93.46 \end{tabular} & \begin{tabular}[c]{@{}l@{}} Used DL tools\end{tabular} \\ \hline
Chowdhary et al. & 2021 & \begin{tabular}[c]{@{}l@{}}PD Detection\\ using DL\end{tabular} & \begin{tabular}[c]{@{}l@{}}MRI\end{tabular} & 2 & - & \begin{tabular}[c]{@{}l@{}}HOG\end{tabular} & \begin{tabular}[c]{@{}l@{}}CNN\end{tabular} & \begin{tabular}[c]{@{}l@{}}94.00 \end{tabular} & \begin{tabular}[c]{@{}l@{}} Used DL tools\end{tabular} \\ \hline
Li et al. \cite{li2024parkinson} & 2024 & PD Detection & T2WI-FLAIR  & - & 582 & CNN & YOLOv5 & \begin{tabular}[c]{@{}l@{}}p=96.1, \\ r=97.4, \\ mAP=98.6\end{tabular} & - \\ \hline
\begin{tabular}[c]{@{}l@{}}Chakr et al. \cite{chakraborty2020detection} \end{tabular}   & 2020 & PD Detection & PPMI PD & 2 & - & CNN & CNN & 95.29 & - \\ \hline
Zhang et al. \cite{Zhang2019} & 2019 & \begin{tabular}[c]{@{}l@{}}Prodromal \\ PD Detection\end{tabular} & \begin{tabular}[c]{@{}l@{}}102 \\ AXI/SAG \\ MRI\end{tabular} & 2 & 102 & \begin{tabular}[c]{@{}l@{}}WGAN/ \\ ResNet\end{tabular} &CNN & \begin{tabular}[c]{@{}l@{}}76.5\end{tabular} & \begin{tabular}[c]{@{}l@{}}low performance, \\ high complexity \end{tabular} \\ \hline
Ramirez et al. \cite{Ramirez2020} & 2020 & \begin{tabular}[c]{@{}l@{}}De Novo \\ PD Detection\end{tabular} & \begin{tabular}[c]{@{}l@{}} PPMI, DTI \\ (129 PD \\ and 57 CN)\end{tabular} & 2 & 186 & \begin{tabular}[c]{@{}l@{}}Convol. \\ Autoencoder\end{tabular} & CNN& \begin{tabular}[c]{@{}l@{}}AUC: \\ 83\end{tabular} & \begin{tabular}[c]{@{}l@{}}Limited \\performance, \\ small dataset\end{tabular} \\ \hline
Islam et al. \cite{islam2024advanced} & 2024 & PD Detection & PPMI MRI & - & 7500 & DenseNet169 & \begin{tabular}[c]{@{}l@{}} CNN\end{tabular} & 85.08 & - \\ \hline
\begin{tabular}[c]{@{}l@{}}  Camacho \\et al. \cite{camacho2023explainable}\end{tabular} & 2023 & PD Detection & MRI & - & 2041 & - & CNN & 79.3 & - \\ \hline
Erda et al. \cite{erdacs2023fully} & 2023 & PD Detection & \begin{tabular}[c]{@{}l@{}}MRI \\ (PPMI)\end{tabular} & 2 & 1130 & CNN & CNN & 96.2 & - \\ \hline
Desai et al. \cite{desai2024enhancing} & 2023 & PD Detection & \begin{tabular}[c]{@{}l@{}}MRI \\ (PPMI)\end{tabular} & 2 & 1180 & - & DL & 90.13 & - \\ \hline

Pahuja et al. \cite{Pahuja2022} & 2022 & PD Detection & \begin{tabular}[c]{@{}l@{}}MRI \\ (PPMI)\end{tabular} & 2 & - & ReliefF & CNN & \begin{tabular}[c]{@{}l@{}}93.33, \\ 92.38\end{tabular} & - \\ \hline
Basnin et al. \cite{basnin2021deep} & 2021 & PD Detection & MRI & - & - & DenseNet & LSTM & 93.75 & - \\ \hline
Talai et al. \cite{talai2021utility} & 2021 & PD Detection & \begin{tabular}[c]{@{}l@{}}MRI (T1, \\ T2-weighted, \\ DTI)\end{tabular} & - & - & - & Multi-modal & 95.1 & - \\ \hline
\begin{tabular}[c]{@{}l@{}} Ahmadlou\\ et al. \cite{Ahmadlou2010}\end{tabular} & 2015 & PD Detection & \begin{tabular}[c]{@{}l@{}}MRI \\ (PPMI)\end{tabular} & - & - & - & \begin{tabular}[c]{@{}l@{}} Gaussian\\ Kernel\end{tabular}  & 98.6 & - \\ \hline

Kuma et al. \cite{Kumaran2022} & 2022 & \begin{tabular}[c]{@{}l@{}}Early and \\ accurate \\ detection of PD\end{tabular} & \begin{tabular}[c]{@{}l@{}}Brain \\ MRI\end{tabular} & 2 & - & - & \begin{tabular}[c]{@{}l@{}}Modified \\ VGG Net\end{tabular} & \begin{tabular}[c]{@{}l@{}}93 \\ accuracy\end{tabular} & \begin{tabular}[c]{@{}l@{}}Cannot detect \\ the stages of PD\end{tabular} \\ \hline

Bhan et al. \cite{Bhan2021} & 2021 & \begin{tabular}[c]{@{}l@{}}Early \\ detection of PD\end{tabular} & \begin{tabular}[c]{@{}l@{}}T1-weighted \\ MRI\end{tabular} & 2 & - & - & \begin{tabular}[c]{@{}l@{}}LeNet-5\end{tabular} & \begin{tabular}[c]{@{}l@{}}97.63 \\ accuracy\end{tabular} & \begin{tabular}[c]{@{}l@{}}Can't analyze \\ medical or \\ neuro images\end{tabular} \\ \hline

Sange et al. \cite{Sangeetha2023} & 2023 & \begin{tabular}[c]{@{}l@{}}Precise \\ diagnosis of PD \\ using CNN\end{tabular} & \begin{tabular}[c]{@{}l@{}}Brain \\ MRI\end{tabular} & 2 & - & - & \begin{tabular}[c]{@{}l@{}}CNN\end{tabular} & \begin{tabular}[c]{@{}l@{}}95 \\ accuracy\end{tabular} & \begin{tabular}[c]{@{}l@{}}Medical picture \\ analysis less \\ efficient\end{tabular} \\ \hline

Kollia et al. \cite{Kollia2019} & 2019 & \begin{tabular}[c]{@{}l@{}}Enhance PD \\ detection\end{tabular} & MRI & 2 & - & - & \begin{tabular}[c]{@{}l@{}}CNN-RNN\end{tabular} & \begin{tabular}[c]{@{}l@{}}98 \\ accuracy\end{tabular} & \begin{tabular}[c]{@{}l@{}}Absence of \\ diagnosing \\ neurodegenerative\end{tabular} \\ \hline

Sivaran et al. \cite{Sivaranjini2019} & 2019 & \begin{tabular}[c]{@{}l@{}}Diagnosis of PD\end{tabular} & \begin{tabular}[c]{@{}l@{}}T2-weighted \\ MRI\end{tabular} & 2 & - & - & \begin{tabular}[c]{@{}l@{}}AlexNet\end{tabular} & \begin{tabular}[c]{@{}l@{}}88.9  \end{tabular} & \begin{tabular}[c]{@{}l@{}}Low performance\end{tabular} \\ \hline

Choi et al. \cite{Choi2017} & 2017 & \begin{tabular}[c]{@{}l@{}} PD with DAT \\ imaging\end{tabular} & \begin{tabular}[c]{@{}l@{}}SPECT\end{tabular} & 2 & - & - & \begin{tabular}[c]{@{}l@{}}CNN\end{tabular} & \begin{tabular}[c]{@{}l@{}} 98\end{tabular} & \begin{tabular}[c]{@{}l@{}}Needed careful\\ analysis, evaluation \end{tabular} \\ \hline

Yasaka et al. \cite{yasaka2021parkinson} & 2021 & \begin{tabular}[c]{@{}l@{}}Differentiate PD\\ from HC \end{tabular} & dMRI & - & - & \begin{tabular}[c]{@{}l@{}}Grad-CAM\end{tabular} & \begin{tabular}[c]{@{}l@{}}CNN\end{tabular} & \begin{tabular}[c]{@{}l@{}}67-89 \\  0.895, \\ 0.800, 0.761\end{tabular} & \begin{tabular}[c]{@{}l@{}}Small sample,\\ lack of \\ external validation\end{tabular} \\ \hline

Khairnar et al. \cite{Khairnar2023} & 2023 & \begin{tabular}[c]{@{}l@{}} Early PD \\ detection\end{tabular} & \begin{tabular}[c]{@{}l@{}}MRI, \\ SPECT\end{tabular} & 2 & - & - & \begin{tabular}[c]{@{}l@{}}CNN, ANN\end{tabular} & \begin{tabular}[c]{@{}l@{}}-\end{tabular} & \begin{tabular}[c]{@{}l@{}}not cost-effective \end{tabular} \\ \hline

Sangeetha et al. \cite{sangeetha2025mri} & 2025 & \begin{tabular}[c]{@{}l@{}} Early PD \\ detection\end{tabular} & \begin{tabular}[c]{@{}l@{}}MRI, 3 axis\end{tabular} & 2 & -  & \begin{tabular}[c]{@{}l@{}}DMFEN\\ EfficientNet-B3-Attn-2\end{tabular} & \begin{tabular}[c]{@{}l@{}}-\end{tabular} & \begin{tabular}[c]{@{}l@{}}92.6 \end{tabular} & Not 3D \\ \hline

\end{tabular}
\end{adjustwidth}
\end{table*}

Other approaches also highlight the role of ML in PD detection, such as Sahu et al. \cite{Sahu2022}, which combined regression analysis (RA) and artificial neural networks (ANN) to detect PD using vocal recognition and biological data, achieving 93.46\% accuracy. Kaplan et al. \cite{Kaplan2022} employed k-nearest neighbours (kNN) for PD staging and motor symptom classification, while SVM proved effective for high-dimensional data in dementia status categorization. Despite the high accuracies demonstrated, limitations persist. Many ML models rely heavily on large datasets and fine-tuned hyperparameters, which may not generalize well across diverse populations or clinical settings. Additionally, some models, like the one used by Prasuhn et al., show poor performance with diffusion-based MRI techniques, highlighting that not all MRI modalities may be equally effective for PD detection \cite{Prasuhn2020}. Moreover, the sensitivity of certain algorithms to noise and hyperparameters can hinder the consistent performance needed for robust clinical applications. Thus, while ML-based MRI approaches show promise, further improvements are necessary to optimise these models for widespread use.


\subsection{MRI Based PD Recognition State of the Art DL Approach}
The limitation of the ML model due to facing difficulties in large size datasets or the 3D MRI classification can be solved by the deep learning (DL) model. DL models have demonstrated significant potential in utilizing MRI data for PD detection. These models employ advanced architectures capable of extracting complex features from high-dimensional imaging data, offering valuable insights for early diagnosis and screening. Table \ref{tab:MRI_ML_methods} provides a detailed of the machine and deep learning techniques applied to MRI datasets for PD detection. However, the performance of these methods remains relatively lower when compared to EEG-based approaches, primarily due to the limited resolution and nuanced features present in MRI data. Additionally, the validation of these models has often relied on small MRI datasets, which restricts their generalizability and applicability to diverse clinical scenarios. Addressing these challenges is essential for advancing MRI-based PD detection methodologies.

Sivaranjini et al. \cite{Sivaranjini2020} applied a pre-trained CNN model, AlexNet, to T2-weighted MRI images for PD detection, achieving 88.9\% accuracy, 89.3\% sensitivity, and 88.4\% specificity by leveraging deep features learned through multiple convolutional layers.
Khairnar et al. \cite{Khairnar2023} combined CNN and ANN models to analyze MRI and SPECT data, applying preprocessing techniques such as image resizing, augmentation, and noise removal. Their approach enhanced the model’s performance, demonstrating that proper preprocessing can significantly impact accuracy. Similarly, Kollia et al. \cite{Kollia2019} employed a hybrid CNN-RNN architecture, achieving around 98\% accuracy by integrating convolutional feature extraction with recurrent layers for temporal data processing. Other models, like that of Sangeetha et al. \cite{Sangeetha2023}, utilized a spatial CNN model that reduced the number of hyperparameters, achieving 95\% accuracy by incorporating multiple convolutional and max-pooling layers. Bhan et al. \cite{Bhan2021} improved early PD detection with the LeNet-5 architecture, achieving 97.6\% accuracy on test data, further proving that deep CNN architectures are effective for MRI-based PD diagnosis. In addition, methods such as VGG Net \cite{Kumaran2022} and the lightweight CNN model by Chowdhary et al.  which used Histogram of Oriented Gradients (HoG) as a feature extraction method, contributed significantly to improving detection accuracy, with up to 94\% accuracy reported. Choi et al. \cite{Choi2017} leveraged CNN for SPECT images, demonstrating that deep learning could overcome the limitations of manual interpretation. Shivangi et al. applied CNNs to spectrograms for early-stage PD detection with accuracy improvements through multimodal data fusion. Additionally, Yasaka et al. \cite{yasaka2021parkinson} used connectome matrices derived from dMRI data and employed Grad-CAM to visualize important areas during CNN-based classification, showing promise in diagnostic accuracy. Ahmadlou et al. \cite{Ahmadlou2010} introduced an Enhanced Probabilistic Neural Network (EPNN) for PD classification, using local decision circles surrounding training samples to regulate the spread of the Gaussian kernel. Applying their model to the Parkinson's Progression Markers Initiative (PPMI) dataset, they achieved 98.6\% accuracy in distinguishing PD patients from healthy controls and a recognition rate of 92.5\% based on clinical exams and neuroimaging data from two brain regions.
\subsection{MRI Based PD Recognition Current Limitation and Challenges}
Despite these advances, MRI-based PD detection systems face certain limitations. One major issue is the relatively small size of available MRI datasets, which restricts the models' ability to generalize effectively. Slicing MRI into a fixed number of sliced images is also responsible for the loss of information.  Additionally, MRI is less efficient in capturing fine-grained temporal patterns of PD progression compared to EEG-based methods. These models are also computationally expensive and require high-quality preprocessing, which can be time-consuming and resource-intensive.

\subsection{MRI Based PD Recognition Future Trends and Recommendation}
To overcome these limitations, future work could focus on larger, multi-institutional datasets to enhance model robustness and generalizability. Working with high-dimensional MRI can overcome the slicing MRI or a fixed number of sliced images, which is related to the loss of information problems. In this situation, 3DCNN or 3D transformer and 3d Transfer learning can be used to handle the high-dimensional work. Moreover, hybrid systems incorporating multimodal data such as EEG, clinical records, and patient demographics could provide a more comprehensive diagnosis. Transfer learning and advanced augmentation techniques like Generative Adversarial Networks (GANs) could also address the dataset size issue by generating synthetic but realistic training data, improving model performance without additional MRI acquisitions. 

\begin{table*}[h]
\centering
\caption{Well-known Gait Pose-based PD datasets. }
\label{tab:video_pose_gait_datasaet}
\begin{adjustwidth}{0cm}{0cm}
\setlength{\tabcolsep}{3pt}
\begin{tabular}{|l|c|c|c|l|l|l|l|l|}
\hline
\textbf{Author} & \textbf{Year} & \begin{tabular}[c]{@{}l@{}}\textbf{Dataset} \\ \textbf{Names}\end{tabular} & \textbf{Classes} & \begin{tabular}[c]{@{}l@{}}\textbf{Dataset} \\ \textbf{Takens}\end{tabular} & \begin{tabular}[c]{@{}l@{}}\textbf{Type} \\ \textbf{of Scan}\end{tabular} & \begin{tabular}[c]{@{}l@{}}\textbf{No. of} \\ \textbf{Sub.}\end{tabular} & \begin{tabular}[c]{@{}l@{}}\textbf{Tot.} \\ \textbf{Sample}\end{tabular} & \begin{tabular}[c]{@{}l@{}}\textbf{Latest} \\ \textbf{Accuracy}\end{tabular} \\ \hline

Gross et al. \cite{gross2001mobo,zhang_posthoc_mobo} & 2001 & CMU MoBo \footnote{\url{http://www.hid.ri.cmu.edu}} & PD, CN & \begin{tabular}[c]{@{}l@{}}6 Camera\end{tabular} & Video & 25 & 25*8k& \\ \hline
Yu et al. \cite{yu2006framework_CASIO_GAIT} & 2006 & CASIA Gait Database\footnote{\url{http://www.cbsr.ia.ac.cn/english/Gait\%20Databases.asp}}   & N/A & \begin{tabular}[c]{@{}l@{}}Camera and \\12 Image \\ Sequences \\ per Person\end{tabular} & Image & 20 & 19139 images &89 \cite{zheng2011robust_casio} \\ \hline
\begin{tabular}[c]{@{}l@{}}Müller et al. \\ \cite{muller2007mocap}\end{tabular} & 2005 & HDM05 \footnote{\url{https://resources.mpi-inf.mpg.de/HDM05/}} & \begin{tabular}[c]{@{}l@{}}More than \\ 70 motion \\ classes\end{tabular} & \begin{tabular}[c]{@{}l@{}} Camera, \\ 10 to 50 \end{tabular} & \begin{tabular}[c]{@{}l@{}}C3D, \\ ASF/AMC\end{tabular} & \begin{tabular}[c]{@{}l@{}}5 Various \\  actors\end{tabular} & \begin{tabular}[c]{@{}l@{}} 3 hours of \\ data\end{tabular} & N/A \\ \hline
\begin{tabular}[c]{@{}l@{}}Human et al. \\ \cite{sigal2010humaneva}\end{tabular} & 2009 & HumanEva-I,I \footnote{\url{https://humaneva.is.tue.mpg.de/}} & \begin{tabular}[c]{@{}l@{}}Multiple \\ subjects \\ performing \\ predefined \\ actions\end{tabular} & \begin{tabular}[c]{@{}l@{}} Synchronized \\ Motion Capture, \\ Multi-view Video\end{tabular} & \begin{tabular}[c]{@{}l@{}}Video, \\ 3D Motion\end{tabular} & \begin{tabular}[c]{@{}l@{}}Multiple \\ Subjects\end{tabular} & \begin{tabular}[c]{@{}l@{}} Over 40,000 \\ frames\end{tabular} & N/A \\ \hline
\begin{tabular}[c]{@{}l@{}}REMAP \\ Dataset \cite{morgan2023multimodal}\end{tabular} & 2023 & REMAP \footnote{\url{https://data.bris.ac.uk/data/dataset/21h9f9e30v9cl2fapjggz4q1x7}} & \begin{tabular}[c]{@{}l@{}}PD, CN\end{tabular} & \begin{tabular}[c]{@{}l@{}}Sit-to-stand \\ transitions, \\ turns in gait\end{tabular} & \begin{tabular}[c]{@{}l@{}}Skeleton Pose, \\ Accelerometry\end{tabular} & \begin{tabular}[c]{@{}l@{}}Multiple \\ Subjects\end{tabular} & \begin{tabular}[c]{@{}l@{}}Free-living \\ and clinical \\ assessments\end{tabular} & N/A \\ \hline

\begin{tabular}[c]{@{}l@{}}GAIT-IST \\ Dataset \cite{loureiro2020skeleton_GAIT-IST}\end{tabular} & 2023 & GAIT-IST  & \begin{tabular}[c]{@{}l@{}}PD, CN\end{tabular} & \begin{tabular}[c]{@{}l@{}} 2 Walking\\ Direction\end{tabular} & \begin{tabular}[c]{@{}l@{}}Energy\\ Silhuttie\end{tabular} &  \begin{tabular}[c]{@{}l@{}}10\end{tabular} & \begin{tabular}[c]{@{}l@{}}-\end{tabular} & 98.00\cite{chen2023fuselgnet} \\ \hline

\begin{tabular}[c]{@{}l@{}}GAIT-IT \\ Dataset \cite{albuquerque2021remote_GAIT-IT}\end{tabular} & 2023 & GAIT-IT  & \begin{tabular}[c]{@{}l@{}}2\end{tabular} & \begin{tabular}[c]{@{}l@{}} 4 Gait \\ Sequence\end{tabular} & \begin{tabular}[c]{@{}l@{}}Energy\\ Silhuttie\end{tabular} & \begin{tabular}[c]{@{}l@{}}21\end{tabular} & \begin{tabular}[c]{@{}l@{}}-\end{tabular} & 95.00\cite{chen2023fuselgnet} \\ \hline

\textbf{GIST \cite{jun2020pathological_GIST}} & 2020 & GIST & 6 Gait Pathologies & \begin{tabular}[c]{@{}l@{}}Captured by 6 \\ Kinect V2 RGB-D \\ Cameras\end{tabular} & \begin{tabular}[c]{@{}l@{}}Skeleton Pose \\ (25 points)\end{tabular} & 10 & 7200 Frames & 98.78 \cite{zhao2024fp_FPGCN}\\ \hline

\textbf{MMGS \cite{khokhlova2019normal_MMGS}} & 2019 & MMGS & 6 Gait Pathologies & \begin{tabular}[c]{@{}l@{}}Captured by 6 \\ Kinect V2 RGB-D \\ Cameras\end{tabular} & \begin{tabular}[c]{@{}l@{}}Skeleton Pose \\ (25 points)\end{tabular} & 10 & 475 Frames & N/A \\ \hline

Sun et al. \cite{sun2022transformer} & 2019 &PD-Video &  \begin{tabular}[c]{@{}l@{}} The subjects walked\\ a 4-
meter straight\\ line on the floor\end{tabular} & \begin{tabular}[c]{@{}l@{}} Cameras\end{tabular} & \begin{tabular}[c]{@{}l@{}}25 fps.\end{tabular} & 43 & \begin{tabular}[c]{@{}l@{}} 115 Video\\ 61:54 (HC:PD)\end{tabular} & N/A \\ \hline
\end{tabular}
\end{adjustwidth}
\end{table*}

\begin{table*}[!htp]
\centering
\setlength{\tabcolsep}{4pt}
\caption{A methodological review of the Gait pose datasets modality-based PD diagnosis using ML.}
\label{tab:vidoe_pose_gai_ml_methodology}
\begin{adjustwidth}{-.5cm}{0cm}
\setlength{\tabcolsep}{3pt}
\begin{tabular}{|p{2cm}|p{.7cm}|p{2cm}|p{2cm}|p{.7cm}|p{.8cm}|p{3cm}|p{2cm}|p{1cm}|p{2.5cm}|}
\hline
Method & Year & Main Objective & Dataset & No of Classes & No of Samples & Feature Extraction & Classifier & Perfor. & Limitations \\ \hline

Chen et al. \cite{chen2021pd} & 2021 & Pose or Video Based PD Recognition & 149 PD patients Hand Gesture videos & 5 & 149 & OpenPose library for pose detection & Custom PD-Net system & 87.6 C. Kappa: 0.82 & Inferior 3D spatial recognition with monocular camera. \\ \hline

Guo et al. \cite{guo2022vision} & 2022 & 3D hand pose based PD Recognition & 112 videos of 59 subjects (48 PD, 11 controls) performing FT & 2 & 112 & tsfresh package for feature extraction & SVM with RBF kernel & 81.2 & Limited to FT task; moderate accuracy \\ \hline

Butt et al. \cite{butt2018objective} & 2018 & LMD based PD recognition & 28 subjects (16 PD patients) hand gesture data from LMD & 2 & - & Time-domain and frequency-domain feature extraction & NB SVM, LR & NB: 81.4 & Small sample size, moderate accuracy \\ \hline

Zhen et al. \cite{zheng2011robust_casio} & 2023 & View Transformation Model for Gait Recognition & CASIA Gait A, B, C & 3 & 19139 & Gait Energy Image (GEI), Robust PCA, Partial Least Square & SVM & 89.00 & Limited to the CASIA dataset \\ \hline

Shin et al. \cite{shin2024classification} & 2024 & Hand-movement Based PD Recognition & 45 PD patients hand movement data & 5 & 2139 & RF, Sequential Forward Floating Selection & SVM & FT 87, OC 93.2, PS 92.2 & Limited to hand movement data, imbalance in data, focused only on akinesia. \\ \hline

Kim et al. \cite{kim2021gait} & 2021 & Gait-based PD & 5: 1
HC gait and 4 disease gaits & 2 & 500 & Integrated gait features & KNN, SVM & 96.52 & Lack of real patients. \\ \hline

Khan et al. \cite{khan2021parkinson} & 2021 & Gait-based PD & 19-PD subjects & 3-5 & 156 & Silhouette, temporal features & SVM & ROC:\newline 80.8 & - \\ \hline

Shetty and Rao \cite{shetty2016svm} & 2016 & Gait-based PD & 15-PD, 16-HC, 20-HD, 13-ALS subjects & 4 & - & Stride, swing, and stance & SVM(RBF) & 83.33 & - \\ \hline
Abdulhay et al. \cite{abdulhay2018gait} & 2018 & Gait-based PD & 93-PD, 73-HC subjects & 2 & - &  Gait, tremor, kinetic, and temporal features & SVM & 92.7 & - \\ \hline
Chavez and Tang \cite{chavez2022gait} & 2022 & Gait-based PD & Youtube video & - & - & Stride, swing, and double support & SVM, KNN, GB & 99.00 & - \\ \hline
Muñoz-Ospina et al. \cite{munozospina2022gait} & 2022 & Gait-based PD & 30-PD, 30-HC & 2 & 60 & Gait spatiotemporal features, swing magnitude, asymmetry & RF, SVM, DT, NB, LR & 84.5 & - \\ \hline
Gong et al. \cite{gong2020gait} & 2020 & Gait-based PD & 30 mix subjects & - & - & Gait energy images & OSVM & 97.33 & - \\ \hline

Jhapate. \cite{jhapate2025gait} & 2025 & Gait-based PD & GAIT-IT&2&30 &Local Features (OCNN), Global Features (Swin Transformer) & AGSO-Optimized CNN, MK-SVM & - & Single dataset \\ \hline

\hline

\end{tabular}
\end{adjustwidth}
\end{table*}

\begin{table*}[!htp]
\centering
\setlength{\tabcolsep}{4pt}
\caption{A methodological review of the Gait pose datasets modality-based PD diagnosis using DL.}
\label{tab:vidoe_pose_gai_ml_methodology}
\begin{adjustwidth}{-.5cm}{0cm}
\setlength{\tabcolsep}{3pt}
\begin{tabular}{|p{2.5cm}|p{.7cm}|p{2cm}|p{2cm}|p{.7cm}|p{.8cm}|p{3cm}|p{2cm}|p{1cm}|p{2cm}|}
\hline
Method & Year & Main Objective & Dataset & No of Classes & No of Samples & Feature Extraction & Classifier & Perfor. & Limitations \\ \hline

Adashzadeh et al. \cite{dadashzadeh2020exploring} & 2020 & Gait and Hand movement based PD recognition & 1058 videos from 25 PD patients & 2 & 1058 & End-to-end DL-based feature extraction & Custom DL framework & Gait: 77.1, Hand: 72.3 & Lower accuracy, limited to gait and hand movements \\ \hline

Chen et al. \cite{chen2021pd} & 2024 & PD Recognition & Gait-IST Energy front, back, Silhouette front, back, IT Energy front, back, Silhouette front, back & 2 & 21 10 & FuseNet & CNN &  98.89, 97.62,  99.78, 99.63 & - \\ \hline

FP-GCN \cite{zhao2024fp_FPGCN} & 2024 & Pathological Gait Classification & GIST NewData & 2 & N/A & Spectral Decomposition Pyramidal Feature & N/A & 98.78 96.54 & Temporal and Spatial Dependency Extraction \\ \hline

Chen et al. \cite{chen2022gait} & 2022 & Gait-based PD & 7 subjects & 3 & 750 & Gait cycle, joint angles & SVM, KNN, LSTM, CNN & 94.9 & - \\
 \hline
 Zhao et al. \cite{zhao2018dual}
 & 2018 & Gait-based PD & 15-PD, 16-HC subjects & 2 & - & Dual Channel LSTM & Softmax & 97.33 & - \\ \hline
Tian et al. \cite{tian2022skeleton} & 2022 & Gait-based PD & 9 subjects & 2 & - & AGS-GCN & GAP and Softmax & 100 & - \\
\hline
Ajay et al. \cite{ajay2018pervasive} & 2018 & Gait-based PD & 26-PD, 23-HC subjects & 2 & - & Analysis Video & DT & 93.75 & - \\\hline
Zou et al. \cite{zou2020deep} & 2018 & Gait-based PD & 18 subject & 2 & - & CNN-LSTM & FC & 93.7 & - \\ \hline
Zhao et al. \cite{zhao2021multimodal} & 2021 & Gait-based PD & 15-PD, 16-HC subjects & 2 & - & SFE, CorrMNN & Multi-Switch Discriminator & 99.86 & - \\ \hline

Balaji et al. \cite{balaji2021automatic} & 2021 & Gait-based PD & 93-PD, 73-HC subjects & 2 & - & LSTM & FC and Softmax & 98.6 & -\\ \hline
Kumar et al. \cite{kumar2023parkinson} & 2023 & Gait-based PD & 93-PD, 73-HC subjects & 2 & - & CNN, CNN-LSTM & CNN, CNN-LSTM & 95 & - \\ \hline
Nguyen et al. \cite{nguyen2022transformers} & 2022 & Gait-based PD & 93-PD, 73-HC subjects & 2 & - & Transformer & FC & 95.2 &- \\ \hline
Cheriet et al. \cite{cheriet2023multi} & 2023 & Beside Gait pd dataset & 43 subjects & 2 & - & Transformer & Multi-Speed Transformer & 96.9 & - \\ \hline
Sun and Zhang \cite{sun2022transformer} & 2022 & Gait-based PD & - & 2 & - & Transformer, GAU & Enhanced Transformer & 97.4 & - \\ \hline
Naimi et al. \cite{naimi2023hct} & 2023 & Gait-based PD & 93-PD, 73-HC subjects & 2 & - & 1D-ConvNets, Transformer & ConvNet-Transformer & 97 & - \\ \hline
Wu and Zhao \cite{wu2022multilevel} & 2022 & Gait-based PD & 52 subjects & 2 & - & Transformer & Multi-Level Fine-Tuned Transformer & 99.83 & - \\ \hline
Wu et al. \cite{wu2023attention} & 2023 & Gait-based PD & 93-PD, 73-HC subjects & 2 & - & Attention-based temporal network & FC and Softmax & 98.86 &  - \\ 
\hline

\end{tabular}
\end{adjustwidth}
\end{table*}

\begin{table*}[h]
\centering
\caption{Well-known Sensor-based PD datasets.}
\label{tab:sensor_gait_dataset}
\begin{adjustwidth}{0cm}{0cm}
\setlength{\tabcolsep}{3pt}
\begin{tabular}{|p{2cm}|p{1cm}|p{2cm}|p{1cm}|p{3.5cm}|p{3.5cm}|p{1cm}|p{2cm}|p{.7cm}|}
\hline
\textbf{Author} & \textbf{Year} & \textbf{Dataset Names} & \textbf{Classes} & \textbf{Dataset Takens} & \textbf{Type of Scan} & \textbf{No. of Sub.} & \textbf{Tot. Sample} & \textbf{Latest Accuracy} \\ \hline
Mazilu et al. \cite{Mazilu2013a_cupid, Mazilu2013b_cupid, Harms2010_cupd} & 2013 & CuPiD & PD & Inertial Measurement Units (IMUs), ECG, GSR, NIRS Sensors & Sensor & 18 & 24 hours of sensing data & N/A \\ \hline
Goldberger et al. \cite{Goldberger2000PhysioBank} & 2007 & RBDSQ, Physionet, Parkinson's Disease Gait Dataset\footnote{\url{https://physionet.org/content/gaitpdb/1.0.0/}} & PD, CN & Vertical ground reaction force measurements & Sensor (Force Plate) & 166 & Gait data with 100 samples/sec & 96.8 \cite{ye2018} \\ \hline
Chartzaki et al. \cite{chatzaki2021smart_Insole} & 2021 & Smart-Insole Dataset\footnote{\url{https://bmi.hmu.gr}} & PD, CN & Pressure sensor insoles & Sensor (Pressure Sensor) & 29 & - & 96.8 \cite{ye2018} \\ \hline
Rogger et al. \cite{roggen2010daphnet} & 2010 & Daphnet\footnote{\url{https://archive.ics.uci.edu/dataset/245/daphnet+freezing+of+gait}} &- &Annotated readings from 3 accelerometers placed on hips and legs during various walking tasks or freezing of gait & Accelerometer  (straight walk, turning, ADL like fetching coffee)& 237 & Multiple trials during lab-based tasks & 92.00 \\ \hline
PDgait Dataset \cite{physionet_pdgait_2024} & 2024 & PhysioNet PD Gait Dataset & 2 (PD, HC) & Vertical Ground Reaction Force (VGRF) signals from 16 sensors per foot & Foot sensor: Subjects walked ~2 mins with 8 force sensors under each foot; raw data includes timestamp, 16 VGRF values, and 2 total foot forces per sample; lacks deep feature extraction or classifier in baseline & 3 & 166 (93 PD, 73 HC) & - \\ \hline
NDDs Dataset \cite{physionet_ndds_2024} & 2024 & NDDs Sensing Gait & 4 (ALS, HD, PD, 
CO) & Temporal gait events and force variation via foot switch system & Foot switch system with force sensors and a resistor to estimate stride intervals & 48 & 64 (48 NDD + 16 HC) & - \\ \hline

tDCS FOG Dataset \cite{salomon2024mlfog} & 2024 & tDCS FOG Dataset& 3 (Start-FOG, Complete-FOG, No-FOG) &  Lower-back 3D accelerometer during video-recorded walking trials &  Sham-controlled, double-blind, multi-site clinical trial data; & 71 PD patients  & 62 train, 14 public test, 7 private test& \\ \hline

Bari et al. \cite{shida2023public,boari2021dataset} & 2021 & Parkinson's Disease Gait Dataset & 2 & Full-body kinematics and kinetics signals, processed in c3d and csv formats, metadata (xls), anthropometrics, model (mdh), and pipeline (v3s) files & 3D Motion Capture (12 cameras), Force Platforms (5), Anatomical Markers (44) & 26 & &\\ \hline

\end{tabular}
\end{adjustwidth}
\end{table*}

\begin{table*}[!htp]
\centering
\setlength{\tabcolsep}{4pt}
\caption{Gait Sensory data modality based PD recognition using ML approach}
\label{tab:Gait_sensory_ML_methods}
\begin{adjustwidth}{-1cm}{0cm}
\setlength{\tabcolsep}{2pt}
\begin{tabular}{|p{2cm}|p{1cm}|p{2cm}|p{2.5cm}|p{.7cm}|p{2cm}|p{2cm}|p{2cm}|p{2cm}|p{2cm}|}
\hline
Method & Year & Main Objective & Dataset & No of Classes & No of Samples & Feature Extraction & Classifier & Perfor [\%]. & Limitations \\ \hline
Ye, Q. et al. \cite{ye2018} & 2018 & PD, ALS, HD from HC & Physionet & 4 & 64, 15 PD + 16 HC + 13 ALS + 20 HD & Not mentioned & LS-SVM, PSO & PD-90.32, HD -94.44, ALS-93.10 & - \\ \hline
Wahid et al. \cite{Wahid2015} & 2015 & PD from HC & Collected from participants & 2 & 49, 23 PD + 26 HC & - & RF, SVM, KFD & 92.6, 80.4, 86.2 & - \\ \hline
Pham et al. \cite{pham2018} & 2018 & PD from HC & Physionet & 2 & 166, 93 PD + 73 HC & Not mentioned & LS-SVM & Sen—100, Sp—100 & - \\ \hline
Mittra et al. \cite{mittra2018} & 2018 & PD from HC & Collected from participants & 2 & 49, 23 PD + 26 HC & LR, DR, SVM (Linear, Pol), KNN & SVM-RBF, RF & 90.39 & - \\ \hline
Klomsae et al. \cite{klomsae2018} & 2018 & PD, ALS, HD from HC & Physionet & 4 & 64, 15 PD + 20 HD + 13 ALS + 16 HC & Not mentioned & Fuzzy KNN & PD—96.43, HD—97.22, ALS—96.88 & - \\ \hline
Milica et al. \cite{djuric2017} & 2017 & PD from HC & Institute of Neurology CCS, School of Medicine, University of Belgrade & 2 & 80, 40 PD + 40 HC & Not mentioned & SVM-RBF & 85 & - \\ \hline
Cuzzolin et al. \cite{cuzzolin2017} & 2017 & PD from HC & Physionet & 2 & 424, 156 PD + 268 HC & - & HMM & 85.51 & - \\ \hline
Felix et al. \cite{felix2019} & 2019 & PD from HC & Neurology Outpatient Clinic at Massachusetts General Hospital, Boston, MA, USA & 2 & 31, 15 PD + 16 HC & Not mentioned & SVM, KNN, NB, LDA, DT & 96.8 & - \\ \hline
Andrei et al. \cite{andrei2019} & 2019 & PD from HC & Laboratory for Gait and Neurodynamics & 2 & 166, 93 PD + 73 HC & - & SVM & 100 & - \\ \hline
Gao et al. \cite{gao2018} & 2018 & PD from HC & University of Michigan & 2 & 80, 40 PD + 40 HC & LR, SVM, XGBoost & RF & 79.6 & - \\ \hline
Rehman et al. \cite{rehman2019} & 2019 & PD from HC & Not mentioned & 2 & 303, 119 PD + 184 HC & Not mentioned & SVM, LR & 97 & - \\ \hline
Kleanthous Natasa et al. \cite{kleanthous2020new} & 2020 & PD from HC & Collected from participants & 2 & 10 PD &  RF, NN, XGBoosting, gradient boosting, SVM(RBF) & Best SVM(RBF) & 72.34, 91.49, 75.00 &-\\ \hline
Zeng et al. \cite{Zeng2016} & 2016 & PD Detection & Gait Features & 2 & 166 & RBF, NN & RBF & 96.4 & Small dataset \\ \hline
Muniz et al. \cite{Muniz2010} & 2010 & PD Detection & Gait Features & 2 & 45 & Logistic Regression, PNN, SVM & SVM & Maximum 94.6 & Small dataset \\ \hline
Trucker et al. \cite{Tucker2015} & 2015 & PD Versus Neurological Disorders & Composed of non-wearable multimodal sensors & 2 & 56 & RF, Logistic Regression, SVM, Light GBM, Stacked Ensemble Model & Manual Measure & 78.00 & Limited performance, small dataset \\ \hline
Procházka et al. \cite{Prochazka2015} & 2015 & PD Stages & Spatial Modeling Kinect sensor & 2 & - & NB & NB & 94.10 & Limited performance, small dataset \\ \hline
Ricci et al. \cite{Ricci2019} & 2019 & De Novo PD Detection & 35 Motor Features & 2 & 60 & Naïve Bayes, SVM, KNN & SVM & Maximum 95 & Small dataset \\ \hline
Talicki et al. \cite{Talitckii2020} & 2020 & PD vs Neurological Dis. & Sensory Data & 2 & 56 & RF, LR, SVM, Light GBM, Stacked Ensemble & Stacked Ensemble Model & Bradykinesia Feat: 85 & Limited performance, small dataset \\ \hline
Alkhatib et al. \cite{alkhatib2020gait} & 2020 & Gait-based PD & 29-PD, 18-HC & 2 & 47 & COP, load distribution & LDA, QDA & 95.00 & - \\ \hline
\end{tabular}
\end{adjustwidth}
\end{table*}

\begin{table*}[!htp]
\centering
\caption{Gait sensory datasets modality-based PD diagnosis using DL approach}
\label{tab:Gait_sensory_DL_methods}
\begin{adjustwidth}{0cm}{0cm}
\setlength{\tabcolsep}{3pt}
\begin{tabular}{|p{2cm}|p{1cm}|p{2cm}|p{2cm}|p{.5cm}|p{1cm}|p{2cm}|p{2cm}|p{2cm}|p{2cm}|}
\hline
\textbf{Method} & \textbf{Year} & \textbf{Main Objective} & \textbf{Dataset} & \textbf{No. of Cl} & \textbf{No of Samples} & \textbf{Feature Extraction} & \textbf{Classifier} & \textbf{Perfor.[\%]} & \textbf{Limitations} \\ \hline

Priya et al. \cite{priya2021} & 2021 & PD from HC & Laboratory for Gait and Neurodynamics & 2 & 166, 93 PD + 73 HC & - & ANN & 96.82 & - \\ \hline

Baby et al. \cite{baby2017} & 2017 & PD from HC & Laboratory for Gait and Neurodynamics & 2 & 166, 93 PD + 73 HC & - & ANN & 86.75 & - \\ \hline

Moon et al. \cite{Moon2020} & 2020 & ET  Versus PD & 48 Balance and Gait Features & 2 & 567 & ANN, SVM, KNN, Decision Tree, RF, Gradient Boosting & ANN & F1-Score 61 & Low performance, unbalanced dataset \\ \hline

Maachi et al. \cite{elMaachi2019} & 2019 & PD Detection and Staging & Sensory Data & 2 & 166 & 18 parallel 1D-CNNs, Fully Connected Network & CNN & 98.7, 85.3 & Small dataset, subjective UPDRS staging \\ \hline

Pfister et al. \cite{Pfister2020} & 2020 & PD Diagnosis & Sensory Data & 3 & 30 & CNN & CNN & 65.4 & Limited performance, small dataset \\ \hline

Eskoifer et al. \cite{Eskofier2016} & 2016 & Bradykinesia Detection & Sensory Data & 2 & 10 & SVM, KNN, 7-Layer CNN & CNN & Maximum 91 & Small dataset \\ \hline

Perumal et al. \cite{perumal2016} & 2016 & PD from HC & Laboratory for Gait and Neurodynamics & 2 & 166, 93 PD + 73 HC & Not mentioned & SVM, ANN & 86.9 & - \\ \hline
Orphanidou et al. \cite{orphanidou2018gait} & 2018 & Gait-based PD & 93-PD, 73 HC & 2 & - & Time, frequency domain  & MLP, RF, XGB, SVML, KNN, NB, SVMP & 91 & - \\ \hline
Borzì et al. \cite{borzi2023gait} & 2023 & Gait-based PD & ADL,Rempark,Daphnet & - &  - & Temporal and spectral features & LR, RF, CNN & 80-96.00 sen & - \\ \hline

Balaji et al. \cite{balaji2020gait} & 2020 & Gait-based PD & 93-PD, 73-HC  & 2 & - & RCNN Spa-Temp & DT, SVM, EC & 97.00 & - \\ \hline
Zhao et al. \cite{zhao2018gait} & 2018 & PD, HC & Physionet & 2 & 93-PD, 73-HC & CNN, LSTM (S,T) & Deep learning & 98.88 & - \\ \hline
Wahid et al. \cite{Wahid2015} & 2015 & PD Stages Disorders & 23 PD patients and 26 aged matched control & 2 & - & Spatial-Temporal gait features & KFD, NB, KNN, SVM, RF & 87.40, 82.00, 84.40, 86.00, 80.00 & Limited performance, small dataset \\ \hline
Li et al. \cite{li2020} & 2020 & PD from HC & Lcal & 2 & 20, 10 PD + 10 HC & Not mentioned & Deep CNN & 91.9 & - \\ \hline

Odonga et al. \cite{odonga2025bias} & 2024 & PD from HC &Freezing of Gait Detection Dataset &  2 & - & HAR Features (e.g., accelerometer, gyroscope) & Convlstm & DPR +0.027, EOR +0.039 F1-score +0.026, +0.018) &-\\ \hline

\end{tabular}
\end{adjustwidth}
\end{table*}

\section{Video and Pose Data Modality Based PD recognition}\label{sec:video}
Parkinson's Disease (PD), in particular, is a challenge to diagnose using traditional MRI methods, as the characteristic features of the disease are often not visible to the naked eye. Gait analysis using visual data is a significant approach for detecting movement disorders, including \cite{taylor2005quantitative} Parkinson's Disease (PD), and can be applied to the mentioned MRI-related problems \cite{parsay2025gait}. By leveraging machine learning (ML) algorithms, researchers have investigated the distinct gait patterns exhibited by individuals with PD compared to healthy controls (HC). Gait is highly individualistic, but PD patients show notable deviations in their walking patterns, making it a valuable diagnostic indicator. Visual data, such as videos and pose information, provide a non-invasive and accessible way to analyze these gait transformations, enabling early detection and monitoring of PD progression. This approach underscores the potential of video and pose data modalities in improving PD diagnosis and treatment planning.


\subsection{Video and Pose Modalitity Dataset}
\begin{figure}[htp]
    \centering
    \includegraphics[width=9cm]{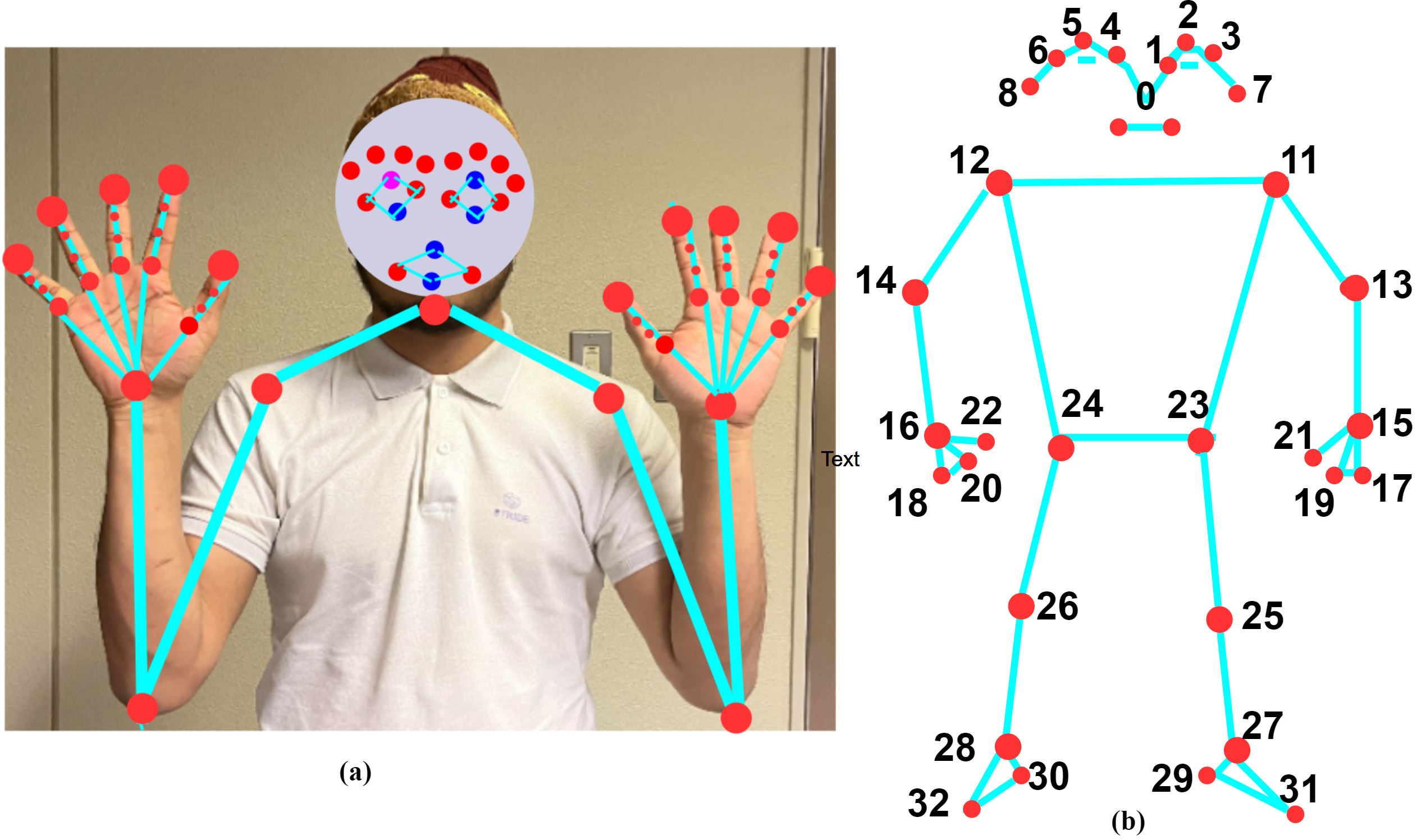}
    \caption{Pose extraction (a) hand and face skeleton joints (b) full body skeleton joints \cite{miah2024sign_largescale}.}
    \label{fig:pose_skeleton}
\end{figure}

Video and pose modalities are critical in the detection and monitoring of Parkinson's disease (PD) and other movement disorders. PD is characterized by motor symptoms like tremors, bradykinesia, and postural instability, which are evident through gait and movement patterns. Table \ref{tab:video_pose_gait_datasaet} shows the benchmark video or pose-based PD dataset, and most used benchmark datasets are also described in the below subsection. Video-based and pose estimation techniques allow for detailed analysis of these motor symptoms, providing valuable biomarkers for early detection and continuous monitoring. Figure \ref{fig:pose_skeleton} shows the extraction process of the pose or skeleton information from the RGB-based image or video. This non-invasive approach is particularly useful for capturing real-time movement data in clinical and natural settings, making it essential for developing accurate diagnostic tools.

In terms of datasets, the CMU MoBo dataset, developed by Gross et al. \cite{gross2001mobo}, includes gait recordings from multiple cameras, capturing movement data from 25 subjects. Yu et al. \cite{yu2006framework_CASIO_GAIT} introduced the CASIA Gait Database, comprising images from 12 sequences per person. The HDM05 dataset by Müller et al. \cite{muller2007mocap} provides motion data across over 70 classes. HumanEva-I, by Sigal et al. \cite{sigal2010humaneva}, contains synchronized motion capture and multi-view video data. More recently, the REMAP dataset \cite{morgan2023multimodal} includes sit-to-stand transitions and turns in gait, using skeleton pose and accelerometry data for both free-living and clinical assessments.

\subsubsection{MoBo Dataset}
The CMU Motion of Body (MoBo) database, introduced in March 2001, is a well-known gait-based dataset for PD recognition and other movement studies. It features 25 individuals walking on a treadmill in CMU's 3D room, performing four distinct walking patterns: slow, fast, incline, and while carrying a ball. These actions were captured using six high-resolution Sony DXC 9000 cameras, providing detailed motion data. Each subject's walking sequence consists of 340 frames, recorded at 30 frames per second (fps), typically covering 10 gait cycles. The dataset contains 8160 images per individual, with the uncompressed data totalling 175 GB (PPM format), while the compressed JPG version reduces this to 9.9 GB. MoBo offers a rich source of video-based gait data, with body silhouettes extracted, making it highly valuable for gait analysis in detecting movement disorders like Parkinson's disease \cite{gross2001mobo}.

\subsection{Preprocessing}
Figure \ref{fig:pose_skeleton} shows the skeleton points we can extract from the RGB image or video using various pose extraction techniques, including Mediapipe, OpenPose, MMPose, etc. We can extract only hand skeletons, only faces or full bodies, including all organs \cite{miah2023dynamic_graph_general,egawa2023dynamic_fall_miah,miah2023skeleton_euvip,miah2024hand_multiculture,miah2024sign_largescale}. 
 
\subsection{Machine Learning (ML) Based Parkinson's Disease (PD) Recognition Using Video}
In recent years, machine learning (ML) techniques have shown great promise in recognizing Parkinson’s Disease (PD) based on various physiological signals and movements. Chen et al. \cite{chen2021pd} developed a system called PD-Net, which used video-based inputs from the OpenPose library. The system focused on detecting 21 key locations on the hand in RGB video recordings. These hand gesture signals were used to identify motor signs of PD during specific tasks, such as finger tapping (FT), open-close (OC), and pronation-supination (PS). Their approach achieved an impressive 87.6\% accuracy, with a Cohen’s kappa coefficient of 0.82, indicating strong agreement with clinical evaluations. However, one limitation of their approach was the use of a monocular camera, which restricted its 3D spatial recognition abilities.
Similarly, Guo et al. \cite{guo2022vision} utilized a more advanced 3D hand pose estimation method based on a depth camera to detect PD symptoms. They collected 112 videos from 59 subjects, using the fresh package in Python to extract motion-based features such as amplitude, velocity, and rhythm. Their system, employing various ML models, including Support Vector Machines (SVM) with a radial basis function (RBF) kernel, achieved a classification accuracy of 81.2\%. Another notable ML-based approach was introduced by Butt et al. \cite{butt2018objective}, who focused on linear movement data (LMD) to differentiate PD patients from healthy controls. They gathered hand gesture signals from 28 subjects, applying time-domain and frequency-domain feature extraction. Their use of NB for classification reached an accuracy of 81.4\%. Shin et al. \cite{shin2024classification} further explored the use of hand movement data for PD recognition. They collected data from 45 PD patients, focusing on various hand movements. Using Random Forest (RF) for feature selection and Support Vector Machines (SVM) for classification, they achieved classification accuracies of 87\% for FT, 93.2\% for OC, and 92.2\% for PS tasks. However, their model was limited to hand movement data, and it faced challenges such as data imbalance and a focus primarily on akinesia, which may limit its generalization.

\subsection{Deep Learning (DL) Based Parkinson's Disease (PD) Recognition Using Video}

Deep learning (DL) techniques have also been increasingly explored for PD recognition due to their ability to automatically extract relevant features from raw data \cite{razzouki2025deep}. Dadashzadeh et al. \cite{dadashzadeh2020exploring} proposed an end-to-end DL-based framework that evaluated PD symptoms based on both gait and hand movements. They collected 1058 videos from 25 PD patients, achieving a classification accuracy of 77.1\% for gait analysis and 72.3\% for hand movement recognition. Despite the potential of deep learning, their model's relatively lower accuracy suggests a need for more diverse data and more refined feature extraction techniques. FP-GCN \cite{zhao2024fp_FPGCN} is proposed for the classification of pathological gait patterns, focusing on both temporal and spatial feature extraction. The model utilizes spectral decomposition and a pyramidal feature extraction approach to enhance the detection of gait abnormalities. FP-GCN demonstrates superior performance, achieving 98.78\% accuracy on public datasets and 96.54\% on proprietary data, surpassing existing methods for pathological gait classification.

\subsection{Video-Based PD Recognition: Current Limitations and Challenges}
Current pose and gait-based PD recognition systems face several challenges. Many systems rely on monocular or depth cameras, which limit the accuracy of 3D spatial recognition, making it difficult to capture subtle motor impairments, such as fine tremors or slight rigidity, that may not be fully visible. Existing datasets are often small and task-specific, typically focusing on specific movements or stages of PD, which makes it challenging to generalize across the full spectrum of symptoms. For example, while gait analysis can reveal bradykinesia through slow or reduced movement, it may not effectively detect postural instability, which doesn’t always manifest in visible movement changes. Furthermore, data imbalances, such as unequal representation of PD patients at various disease stages, can skew results and limit model accuracy, especially when assessing early-stage or mild symptoms like subtle tremors or postural instability.

\subsection{Video-Based PD Recognition: Future Trends and Recommendations}
Future research could address these limitations by integrating multi-sensor data, such as inertial measurement units (IMUs) and electromyography (EMG) sensors, alongside video data to capture a broader range of movements. For instance, combining video data with motion sensors can provide a more comprehensive understanding of tremor severity and rigidity, which are harder to detect through video alone. Expanding datasets to include diverse PD symptoms—like tremor, rigidity, bradykinesia, and postural instability—will improve model generalizability and clinical relevance. Future systems could also explore hybrid models that combine traditional machine learning (ML) techniques with advanced deep learning (DL) architectures for better feature extraction and robust classification. This would enhance the ability to recognize and track PD symptoms across different stages of the disease, ultimately aiding in early diagnosis, personalized treatment, and continuous monitoring of disease progression.

\section{Sensor Data Modality Based PD Recognition }\label{sec:sensor}
\begin{figure}[htp]
    \centering
    \includegraphics[width=9cm]{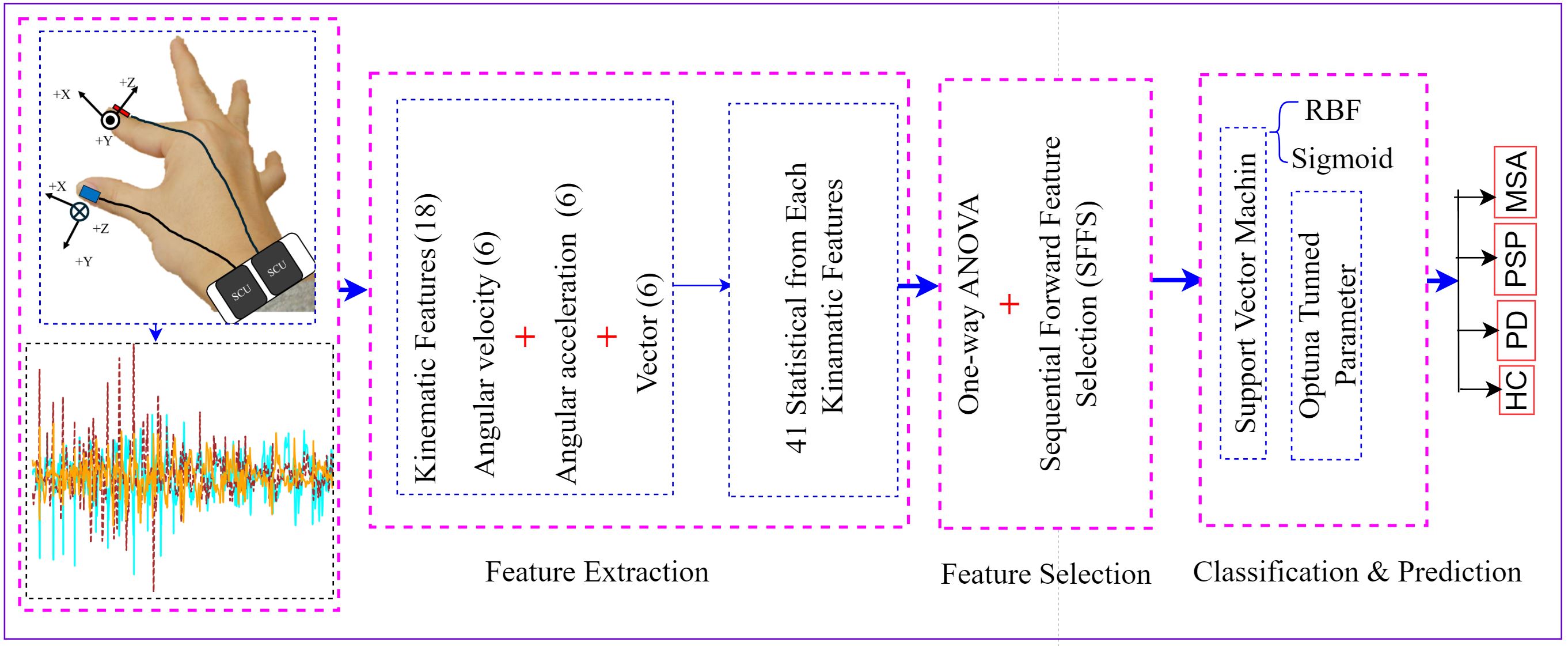}
    \caption{Example of sensor based pd recognition working architecture \cite{10937116_Matsumoto_shin_pd}.}
    \label{fig:sensor_pd_workflow}
\end{figure}

Sensor-based data, such as from Inertial Measurement Units (IMUs), accelerometers, or foot pressure sensors \cite{miah2024sensor}, is highly effective for recognizing Parkinson's Disease (PD) compared to MRI. These sensors capture detailed movement and gait data, which are crucial in detecting the motor symptoms of PD, including tremors, rigidity, and freezing of gait. In 2004 Con et al, applied ANN to recognition sensor signal based PD where they used a flexure sensor and magnetic sensor to record the data from the ankle and near the shoulder \cite{cont2004real}. Unlike MRI or video-based approaches, sensor-based systems offer continuous, real-time monitoring, allowing for long-term tracking of subtle changes in movement patterns. Figure \ref{fig:sensor_pd_workflow} shows the example of the sensor based PD recognition architecture \cite{10937116_Matsumoto_shin_pd}.
    
This makes them more practical for daily use and less invasive than medical imaging like MRI.

\subsection{Sensor Data Modality Based Benchmark Datasets}
The datasets reviewed provide clear evidence of the benefits of sensor-based PD recognition, as shown in Table \ref{tab:sensor_gait_dataset}. For instance, the CuPiD dataset by Mazilu et al. \cite{Mazilu2013a_cupid, Mazilu2013b_cupid, Harms2010_cupd} used IMUs, ECG, and other sensors to gather 24 hours of continuous data from PD patients, providing a comprehensive view of patient activity. The RBDSQ Physionet Parkinson's Disease Gait Dataset \cite{Goldberger2000PhysioBank} used force plates to measure ground reaction forces with a high accuracy of 96.8\%. Similarly, Chartzaki et al.'s \cite{chatzaki2021smart_Insole} Smart-Insole dataset recorded pressure sensor data from PD patients, also achieving a 96.8\% accuracy. Lastly, the Daphnet dataset by Roggen et al. \cite{roggen2010daphnet} utilized accelerometers to capture freezing of gait episodes with a 92\% accuracy.
These datasets demonstrate that sensor-based data can be more accurate and practical than traditional methods, providing real-time and long-term movement analysis, which is critical for early PD detection and ongoing monitoring.

\subsubsection{CuPiD Dataset}
The CuPiD dataset, one of the most widely used sensor-based datasets for Parkinson's Disease (PD) recognition, contains 24 hours of sensing data collected from 18 PD patients. Inertial Measurement Units (IMUs) were attached to both wrists, capturing hand movements during different walking protocols designed to provoke freezing of gait (FoG) episodes. These walking tasks included 360- and 180-degree turns, straight-line walking, navigating narrow corridors, and walking through crowded hospital halls \cite{Mazilu2013a_cupid, Mazilu2013b_cupid}. The dataset also includes additional sensors, such as electrocardiograms (ECG), galvanic skin response (GSR), and near-infrared spectroscopy (NIRS) sensors, to provide a comprehensive view of physiological responses during movement \cite{Harms2010_cupd}. The rich data from multiple body parts makes the CuPiD dataset highly valuable for analyzing the correlation between hand movements and FoG, offering insights for more accurate PD detection and symptom tracking.

\subsubsection{Daphnet Gait Parkinson Dataset}
The Daphnet Gait Parkinson Dataset is a key resource for studying freezing of gait (FoG) in Parkinson's Disease (PD) patients \cite{roggen2010daphnet}. It consists of motion data collected using three accelerometers placed on the hips and legs of PD patients during walking tasks. The dataset captures detailed information about gait patterns, focusing on identifying FoG episodes \cite{roggen2010daphnet}. This data is invaluable for developing algorithms aimed at detecting and predicting FoG, contributing to improved diagnostic tools and therapeutic strategies for managing gait disturbances in PD.

\subsection{Sensor Data Modality Signal Preprocessing}
Sensor signal preprocessing is crucial for Parkinson's Disease (PD) recognition using sensor-based modalities, as it ensures raw data is cleaned, standardized, and optimized for machine learning (ML) and deep learning (DL) models. Various methods address noisy, imbalanced, or incomplete data to improve model performance. Moon et al. applied the Synthetic Minority Oversampling Technique (SMOTE) to counter dataset imbalances in gait analysis, preventing models from being biased toward dominant classes \cite{Moon2020}. Wahid et al. normalized spatial-temporal gait features using multiple regression to minimize variance from patient-specific factors, enhancing data reliability \cite{Wahid2015}. Talitckii et al. incorporated tremor and bradykinesia features to ensure ML classifiers captured relevant motor patterns \cite{Talitckii2020}. Noise filtering techniques are widely used to enhance sensor data quality. Zeng et al. utilized radial basis function networks for gait dynamics modelling, with preprocessing playing a vital role in improving feature extraction \cite{Zeng2016}. Eskofier et al. employed standard signal processing methods to filter accelerometer noise, significantly improving PD detection accuracy \cite{Eskofier2016}.
Key preprocessing techniques, including data normalization, noise filtering, and feature extraction, are essential for improving sensor data quality. These methods ensure the data is robust and reliable for ML and DL models, enhancing the accuracy of PD recognition systems.

\subsection{Sensor Based PD Recognition using ML Approach}
Table \ref{tab:Gait_sensory_ML_methods} summarizes 18 studies employing machine learning (ML) techniques to analyze sensor-based gait data for Parkinson's Disease (PD) recognition. Below are refined descriptions of key studies, their methodologies, advantages, and limitations. Ye et al. \cite{ye2018} utilized the Physionet dataset to classify PD, Amyotrophic Lateral Sclerosis (ALS), and Huntington's Disease (HD) from healthy controls (HC). Using Least Squares Support Vector Machines (LS-SVM) and Particle Swarm Optimization (PSO), they achieved classification accuracies of 90.32\% for PD, 94.44\% for HD, and 93.10\% for ALS. Despite high accuracy, the study lacked detailed feature extraction steps, reducing transparency. Wahid et al. \cite{Wahid2015} analyzed gait data from 49 participants, comparing RF, Support Vector Machine (SVM), and Kernel Fisher Discriminant (KFD). RF achieved the highest accuracy of 92.6\%, emphasizing its effectiveness for small datasets. Similarly, Pham et al. \cite{pham2018} used LS-SVM on the Physionet dataset, achieving 100\% sensitivity and specificity, but provided minimal insight into feature extraction methods. Mittra et al. \cite{mittra2018} combined LR, Decision Trees (DT), and SVM, achieving 90.39\% accuracy with an SVM radial basis function (RBF) kernel. However, the small sample size limited result generalizability. Klomsae et al. \cite{klomsae2018} expanded classifications to include ALS and HD using Fuzzy KNN, achieving 96.43\% accuracy for PD, though fuzzy methods complicate interpretability. Milica et al. \cite{djuric2017} achieved 85\% accuracy with SVM and an RBF kernel, slightly underperforming compared to others. Cuzzolin et al. \cite{cuzzolin2017} employed Hidden Markov Models (HMM) for time-series data, reaching 85.51\% accuracy—lower than RF or SVM. Felix et al. \cite{felix2019} achieved 96.8\% accuracy with SVM, KNN, and Naive Bayes on a small dataset (31 participants), highlighting the need for larger samples. Andrei et al. \cite{andrei2019} reported 100\% accuracy using SVM on a similarly limited dataset, raising concerns about model generalizability. Gao et al. \cite{gao2018} applied LR, SVM, and XGBoost to 80 participants but achieved only 79.6\% accuracy with RF, reflecting dataset limitations. Larger datasets, as used by Rehman et al. \cite{rehman2019}, yielded 97\% accuracy with SVM and LR on 303 participants, although feature extraction details were omitted. Overall, these studies demonstrate promising advancements in sensor-based PD detection but emphasize the need for standardized feature extraction, larger datasets, and transparent methodologies to enhance reproducibility and generalization.

\subsection{Sensor Based PD Recognition using DL Approach}
Table \ref{tab:Gait_sensory_DL_methods} summarizes the application of deep learning (DL) techniques in sensor-based gait data to diagnose Parkinson's Disease (PD) across 11 studies. Over the years, significant advancements have been made in this domain. Early studies laid the groundwork for DL-based PD recognition. Baby et al. \cite{baby2017} applied an Artificial Neural Network (ANN) to gait data from the Laboratory for Gait and Neurodynamics, achieving 86.75\% accuracy. Similarly, Perumal et al. \cite{perumal2016} used the same dataset with ANN and Support Vector Machine (SVM), attaining 86.9\% accuracy. These efforts demonstrated the potential of neural networks but were constrained by small datasets, limiting their generalization. Building on this, Priya et al. \cite{priya2021} enhanced ANN performance, achieving a significantly improved accuracy of 96.82\%. Oğul et al.  further advanced this approach, achieving 98.3\% accuracy with ANN, but dataset size remained a significant limitation. Researchers shifted to advanced deep learning architectures to overcome these challenges. Maachi et al. \cite{elMaachi2019} utilized 18 parallel 1D-CNNs to detect PD and stage its severity, achieving an accuracy of 98.7\% for detection and 85.3\% for staging. Despite its success, the small dataset size and subjective UPDRS staging hindered generalization. Similarly, Pfister et al. \cite{Pfister2020} applied CNNs to a dataset of 30 samples but achieved only 65.4\% accuracy, emphasizing the necessity of larger datasets for effective deep learning training. Focused studies on specific PD symptoms also demonstrated the potential of CNNs. Eskofier et al. \cite{Eskofier2016} applied SVM, KNN, and a 7-layer CNN for bradykinesia detection, achieving a maximum accuracy of 91\% on a dataset of 10 samples. Moon et al. \cite{Moon2020} aimed to differentiate between PD and essential tremor using balance and gait features, but an unbalanced dataset limited their model’s F1-score to 61\%, highlighting the need for balanced and comprehensive datasets. The evolution of DL methods in sensor-based PD diagnosis underscores the growing efficacy of CNNs in analyzing complex gait data. However, challenges such as small and imbalanced datasets persist, restricting model generalization. Future research should prioritize larger, more diverse datasets and explore hybrid models, combining CNNs with temporal architectures like Recurrent Neural Networks (RNNs) to better capture sequential gait patterns, improving PD diagnosis and staging.

\subsection{Sensor-Based PD Recognition: Current Limitations and Challenges}
These studies collectively show the promise of machine learning (ML)-based approaches to PD recognition using sensor-based gait data. The advantage of methods like Support Vector Machines (SVM) and RF lies in their high accuracy, with studies like Pham et al. (2018) and Andrei et al. (2019) achieving near-perfect performance. However, many ML-based studies suffer from small dataset sizes, which limits the generalization of the models. Additionally, certain methods, like Fuzzy KNN or Hidden Markov Models (HMM), while effective, may introduce complexity in interpretation and deployment in clinical settings. Deep learning (DL) approaches, while offering automated feature extraction, face challenges as well. One significant issue is the need for large, labeled datasets, which are often scarce in sensor-based PD recognition. DL models are also computationally intensive, making them difficult to implement in real-time or resource-constrained environments, such as wearable devices. Furthermore, interpretability remains a challenge with DL models, as their "black-box" nature can limit clinical trust and understanding. 

\subsection{Sensor-Based PD Recognition: Future Trends and Recommendations}
The trend across these works points to the importance of using larger datasets and diverse classifiers. Future research should focus on improving generalization by employing larger, more varied datasets and combining traditional ML approaches with deep learning to enhance both feature extraction and classification performance. Deep learning, with its ability to automatically extract meaningful features from raw sensor data, can provide significant advancements in sensor-based PD recognition. However, to fully leverage its potential, future research should address some of the current challenges. Effective channel selection is also a crucial point in this case. To enhance DL models, researchers should focus on creating larger, labelled datasets through collaboration between clinical institutions, which can also help balance data and reduce bias. Semi-supervised learning and transfer learning can be explored to improve model training when labelled data is limited. Furthermore, optimizing DL models to work efficiently on low-power devices, such as wearables, will be key for real-time PD monitoring. Finally, interpretability techniques, such as attention mechanisms or explainable AI (XAI) methods, can be integrated into DL models to make them more transparent and acceptable in clinical settings. Additionally, combining wearable sensors for real-time monitoring will enhance the practicality of PD detection systems in everyday clinical practice.

\section{Hand Writing Data Modality Based PD Research} \label{sec:handwriting}
Handwriting analysis has become a valuable tool for early Parkinson's Disease (PD) detection, utilizing motor skill impairments associated with the condition. Micrographia, a handwriting abnormality common in PD patients, serves as a significant diagnostic indicator. The concept of the graphics tablet based handwriting data for pD recognition used by Margolin et al \cite{margolin1983agraphia} and philips \cite{phillips1991handwriting} in 1991. Philips \cite{phillips1991handwriting} proposed a simple zig-zag drawing for PD recognition an dthe revealed some difficulites in smooth movement generation compare to controlling stroke length. Teulings et al, updated that work by modifing the speed and motion of the hand writing \cite{teulings1991control}. Later in 2013-2016, Rosenblum \cite{rosenblum2013handwriting} and Droter et al \cite{Drotar2016} used graphics tablets and made pen tip pressure-based features from a handwriting dataset, and they also opened their dataset. Based on the previous work of handwriting tabular data  Raudmann et al \cite{raudmann2014handwriting} in 2014 reported that kinematic features are more effective for the pd compared to size of the tabular information. 

\begin{figure}[htp]
    \centering
    \includegraphics[width=9cm]{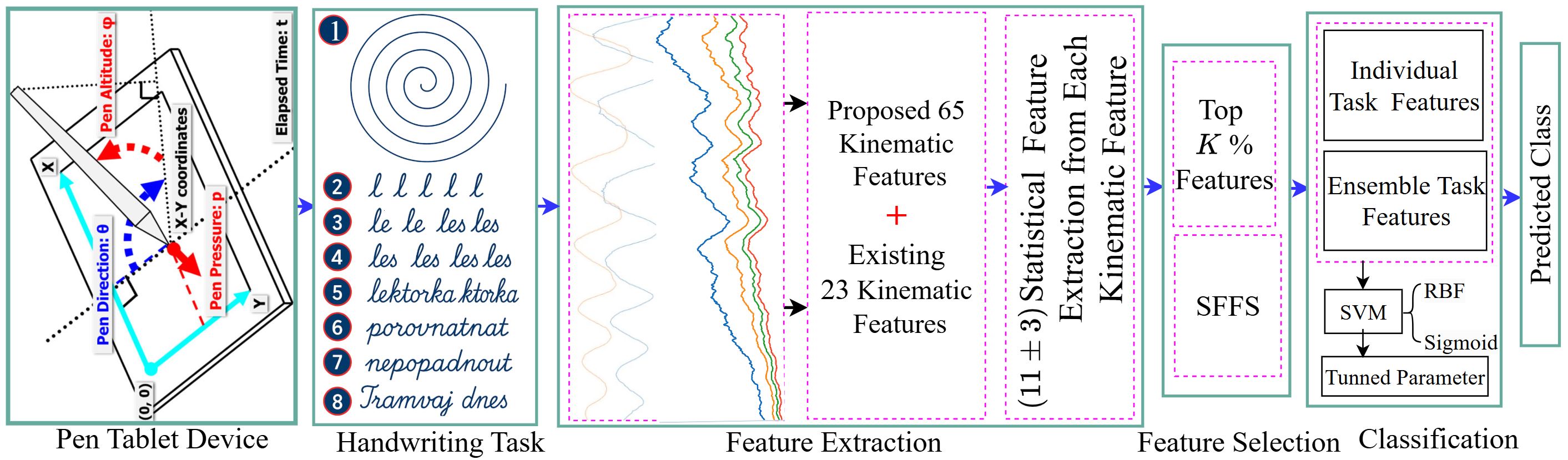}
    \caption{Handwriting based pd recognition flowgrap \cite{shin2024parkinson}.}
    \label{fig:handwriting_pd_workflow}
\end{figure}

Machine learning (ML) techniques have been applied extensively to analyze handwriting patterns, using datasets such as the UCI Machine Learning Repository, Parkinson’s Progression Markers Initiative (PPMI), Parkinson’s Disease Handwriting (PaHaW) database, and Picture Archiving and Communication System (PACS). Figure \ref{fig:handwriting_pd_workflow} shows the example of the handwriting based pd recognition system \cite{shin2024parkinson}. A review of 18 studies highlights the effectiveness of ML algorithms like Support Vector Machines (SVM), RF, and k-Nearest Neighbours (KNN). Among 961 subjects analyzed, 657 were from the PPMI database, while 304 came from local datasets. SVM with a linear kernel achieved the highest accuracy of 97.9\% for 652 subjects (443 PD patients and 209 healthy controls). Accuracy for local datasets ranged from 88\% to 92\%, while PPMI data yielded 95\% to 97\%, emphasizing handwriting analysis as a promising method for early PD detection. While ML algorithms have shown success, challenges such as the need for larger datasets and clinical integration remain. Future research should aim to enhance algorithm robustness and generalizability, enabling handwriting analysis to become a standard, non-invasive tool for early PD diagnosis and timely intervention.

\subsection{Handwriting Modality Datasets}
Handwriting analysis has emerged as a valuable tool for recognizing Parkinson's Disease (PD), offering several advantages over other modalities such as MRI, video, pose, and sensor data. Handwriting tests can capture early motor skill deterioration, specifically micrographia, which is a hallmark of PD. Unlike MRI or video, handwriting data is easier and more cost-effective to collect, making it highly practical for large-scale testing and continuous monitoring. Additionally, handwriting data is often collected using simple tools like tablets or biometric pens, allowing for real-time analysis, unlike MRI, which requires costly, sophisticated equipment. Furthermore, handwriting data directly captures the fine motor control impairments associated with PD, making it more focused and relevant for early diagnosis compared to generalized motion data from video or sensors.
The Table \ref{tab:handwriting_datasets} summarizes several well-known handwriting-based PD datasets. For instance, Pereira et al. (2015) introduced the HandPD dataset, which consists of 736 handwriting images from 92 subjects, achieving 95\% accuracy. Similarly, Droter and Pavlovic et al. (2015) developed the PaHaW dataset, featuring handwriting samples from 75 subjects, yielding an accuracy of 94\%. Isenkul et al. (2014) provided the New HandPD dataset, incorporating both static and dynamic spiral tests, along with stability tests, capturing images and sensor data from 77 subjects with an impressive accuracy of 99\%. Perai (2016) enhanced the NewHandPD dataset with multimodal handwriting signals and images, utilizing a biometric smart pen, achieving accuracies of 89.4\% and 97.5\% on different test sets. Additionally, the Kaggle Handwriting Drawings dataset (2021) includes spiral and wave drawings from 55 subjects, with an accuracy of 89\%.
These datasets underscore the potential of handwriting-based data in achieving high diagnostic accuracy for PD while being less invasive and more accessible than other diagnostic modalities. We also described the well known benchmark dataset below subsection. 
\begin{table*}[h]
\centering
\caption{Well-known handwriting-based PD datasets.}
\label{tab:handwriting_datasets}
\begin{adjustwidth}{0cm}{0cm}
\setlength{\tabcolsep}{3pt}
\begin{tabular}{|p{2.5cm}|p{.8cm}|p{1.5cm}|p{1cm}|p{3.5cm}|p{2cm}|p{1cm}|p{3cm}|p{1.5cm}|}
\hline
Author & Year & Dataset Names & Classes & Dataset Takens & Type of Scan & No. of Sub. & Tot. Sample & Latest Accuracy [\%] \\ \hline

Pereira et al. \cite{Pereira2015_handpd} & 2015 & NewHandPD & PD, CN & Images from handwriting exams & Image & 92 & 736 images (368 spirals, 368 meanders) & 95.00\cite{islam2024review_sruvey_handwriting_voice_data} \\ \hline
Droter and Pavlovic et al. \cite{Drotar2016} & 2015 & PaHaW & PD, CN & Handwriting samples & Tablet & 75 & Multiple handwriting signals & 94.00 \cite{allebawi2024parkinson} \\ \hline

Isenkul et al. \cite{Isenkul2014Improved} & 2014 & New HandPD & PD, CN & Static Spiral Test, Dynamic Spiral Test, Stability Test & Image, Sensor & 77 & 3 handwriting recordings per subject & 99.00 \\ \hline

Perai \cite{pereira2016deep_newhandpd} & 2016 & NewHandPD & PD, CN & Multimodal (Handwriting signals, images) & Biometric Smart Pen & 66 & 264 & 89.4, 97.5 \cite{islam2024review_sruvey_handwriting_voice_data} \\ \hline

Kaggle. \cite{mader2019parkinsons_kaggle_handwriting,zham2017distinguishing_kaggle_handwriting} & 2021 & Handwriting Drawings & 2 (PD, CN) & Kaggle Handwriting dataset \footnote{\url{https://www.kaggle.com/datasets/kmader/parkinsons-drawings}} & Spiral and Wave Drawings & 55 & 204 & 89\cite{islam2024review_sruvey_handwriting_voice_data} \\ \hline

Prashanth et al. \cite{prashanth2016} & 2016 & PPMI & 2 (PD, CN) & - & - & - & 584 & 96.4 \\ \hline

Adams et al. \cite{adams2017high} & - & PD-Tappy Dataset & PD, HC & Handwriting assessments (Static Spiral Test, Dynamic Spiral Test, Stability Test on Certain Point) & Image & 25 PD, 15 HC & Multiple spiral images (for analysis) & - \\ \hline

Giance et al. \cite{giancardo2016computer} & - & NeuroQWERTY MIT-CSXPD Dataset & PD, HC & Keystroke records from typing on a laptop (simulated home typing habits) & Keystroke data & 85 (PD, HC) & Keystroke data with high temporal resolution & - \\ \hline

Vj et al. \cite{vj2024machine} & 2024 & PADS Dataset & 3 (PD, DD, HC) & - & - & 469 & 5159 & - \\ \hline

\end{tabular}
\end{adjustwidth}
\end{table*}

 \begin{figure}[htp]
    \centering
    \includegraphics[width=7cm]{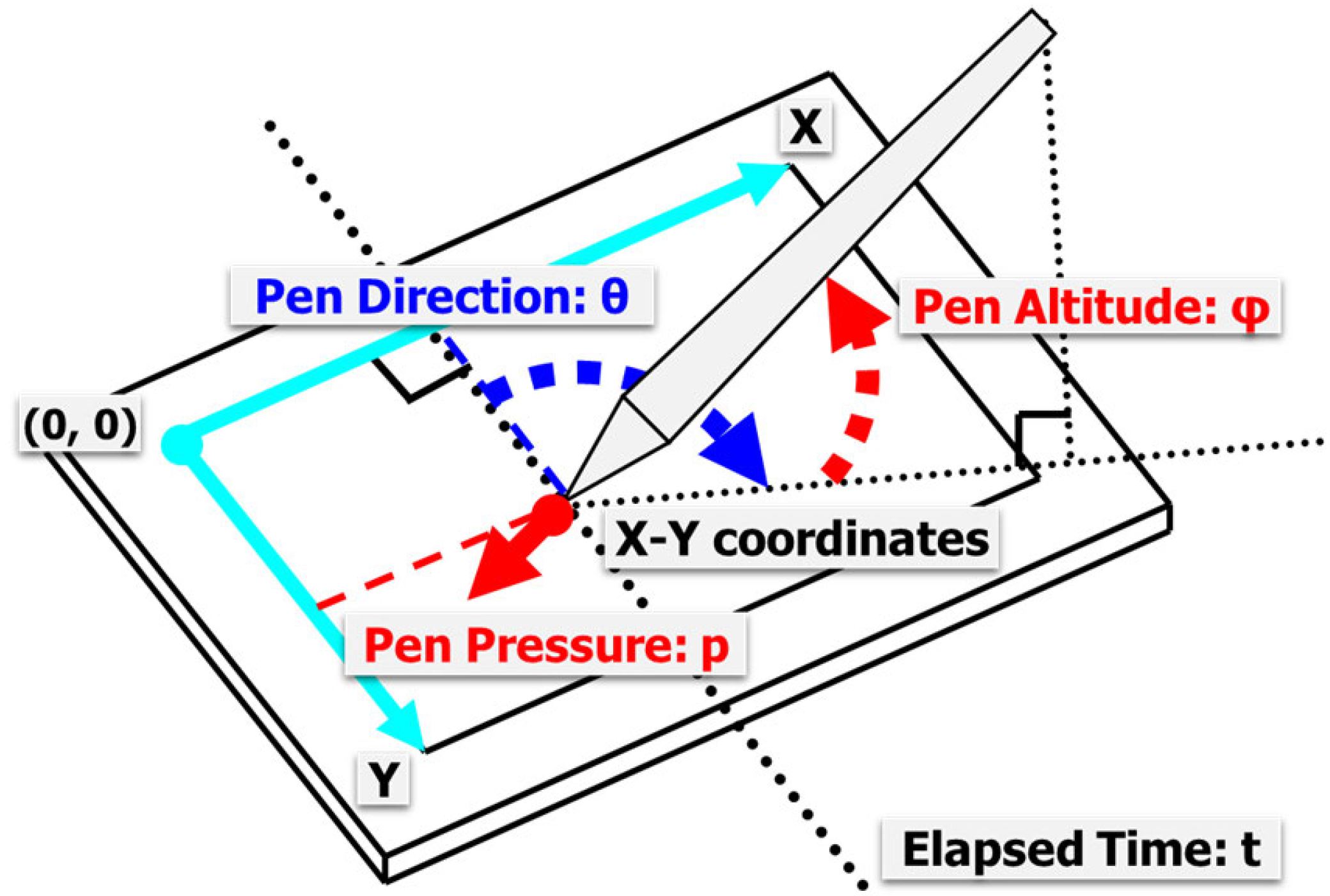}
    \caption{Pen tablet device used to collect hand writing data from PD patient \cite{shin2023handwriting}.}
    \label{fig:handwriting_data_collection}
\end{figure}

We present the primary benchmark datasets utilized in the previously discussed studies and explain how they were specifically developed to address challenges associated with Parkinson's Disease in the below section.

\subsubsection{HandPd Dataset}
Handwriting analysis is a useful tool for Parkinson's Disease (PD) detection, offering advantages over MRI, video, and sensor data. Handwriting tests can detect early motor issues, like micrographia, and are easier and cheaper to collect, making them suitable for large-scale and continuous monitoring. They can be recorded with simple devices, allowing real-time analysis without expensive equipment. Handwriting data specifically targets fine motor impairments relevant to PD, making it more effective for early diagnosis than general motion data.
The table below summarizes key handwriting-based PD datasets. Pereira et al. (2015) introduced the HandPD dataset with 736 samples from 92 subjects, achieving 95\% accuracy  \cite{Pereira2015_handpd}. Droter and Pavlovic (2015) developed the PaHaW dataset with 75 subjects, achieving 94\%. Isenkul et al. (2014) created the New HandPD dataset with spiral tests and sensor data from 77 subjects, achieving 99\%. Perai (2016) expanded this with multimodal signals, reaching up to 97.5\%. The Kaggle Handwriting Drawings dataset (2021) includes 55 subjects with 89\% accuracy.

\subsubsection{NewHandPD Dataset}
The NewHandPD dataset \cite{pereira2016deep_newhandpd} extends the original HandPD dataset by incorporating images from two drawing tasks: the standard spiral cognitive test and the modified spiral (meander) test. It includes both offline images and time-based online handwriting signals, captured using a Biometric Smart Pen (BiSP). The dataset consists of 264 samples from 66 participants, including 31 PD patients and 35 healthy controls, with 39 men and 29 women, mostly right-handed (59 participants). Released in 2016, NewHandPD is useful for classification tasks, though further exploration is needed to unlock its full potential.

\subsubsection{Parkinson's Disease Handwriting Dataset – PaHaW}
The Parkinson's Disease Handwriting Database (PaHaW) (PaHaW\footnote{https://www.researchgate.net/publication/289525377\_Parkinson's\_Disease\\ \_Handwriting\_Database\_PaHaW}) \cite{Drotar2016} contains handwriting samples from 37 Parkinson's patients (19 men, 18 women) and 38 healthy controls (20 men, 18 women). Data collection was conducted in collaboration with Masaryk University's Movement Disorders Center and St. Anne's University Hospital in Brno, Czech Republic. Participants used a standard ink pen on a Wacom Intuos 4M digitizing tablet, recording at 200 Hz. The dataset includes movement signals and perpendicular pressure applied to the tablet, with recordings starting upon pen contact and ending after task completion, enabling real-time tracking of handwriting dynamics.

\subsection{Preprocessing Techniques for Handwriting Data Modality}
Preprocessing handwriting data for Parkinson’s Disease (PD) recognition involves techniques to enhance data quality and optimize machine learning (ML) and deep learning (DL) model performance. Pereira et al. \cite{Pereira2015_handpd} employed feature selection to reduce dimensionality and focus on key handwriting movements. Drotar et al. \cite{Drotar2014} highlighted the significance of in-air trajectory preprocessing between handwriting strokes for pattern differentiation. Kamran et al. \cite{Kamran2021} applied data augmentation, including flipping, rotation, contrast, and illumination adjustments, to address small dataset limitations and improve robustness. Transfer learning techniques, as employed by Naseer et al. \cite{Naseer2020}, involved fine-tuning pre-trained networks on handwriting data, further optimizing the performance of deep learning models. These preprocessing strategies ensure that handwriting data is clean, well-represented, and ready for subsequent feature extraction and classification.
\begin{table*}[htp!]
\centering
\caption{Hand writing dataset modality based PD recognition using ML.}
\label{tab:handwriting_ml_methodology}
\begin{adjustwidth}{-1cm}{0cm}
\setlength{\tabcolsep}{2pt}
\begin{tabular}{|p{2cm}|p{.8cm}|p{1.5cm}|p{3.5cm}|p{.7cm}|p{2cm}|p{2.5cm}|p{2cm}|p{2cm}|p{1cm}|}
\hline
Method & Year & Main Objective & Dataset & No of Classes & No of Samples & Feature Extraction & Classifier & Perfor.[\%] & Limitations \\ \hline

Taylor et al. \cite{Taylor2017} & 2017 & PD from HC & PPMI and local & 2 & PPMI: 657, 448 PD + 209 HC, local: 304,191 PD + 113 HC & - & SVM (10CV) & Local: 88-92 PPMI: 95-97 & - \\ \hline

Oliveira et al. \cite{Oliveira2018} & 2017 & PD from HC & PPMI database & 2 & 652, 443 PD + 209 HC & LR-LOOCV KNN & SVM linear & 97.9 & - \\ \hline

de Souza et al. \cite{desouza2018} & 2018 & PD from HC & HandPD & 2 & 92, 74 PD + 18 HC & OPF, NB & SVM (RBF) (10CV) & 85.4 & - \\ \hline

Drożdż et al. \cite{Drotar2016} & 2016 & PD from HC & PaHaW database & 2 & 75, 37 PD + 38 HC & KNN, Ensemble AdaBoost & SVM & Ac-81.3 sp—80.9 and sen—87.4 & - \\ \hline

Hsu, et al. \cite{hsu2019} & 2019 & PD from HC & PACS & 2 & 202, 94 Severe PD + 102 mild PD + 6 HC & SVM RBF, LR & SVM with RBF & 83.2, sen-82.8, sp-100 & - \\ \hline

Kurt, L. et al. \cite{kurt2019} & 2019 & PD from HC & UCI machine learning repository & 2 & 72, 57 PD + 15 HC & SVM (linear and RBF), KNN & SVM linear & 97.52 & - \\ \hline

Mabrouk et al. \cite{mabrouk2019} & 2019 & PD from HC & PPMI Database & 3 & 550, 342 PD + 157 HC + 51 SCAN (SWEDD) & RF, SVM, MLP, KNN & SVM KNN & SVM—78.4 82.2 & - \\ \hline

Fabian et al. \cite{Maass2020} & 2020 & PD from HC & UCI machine learning repository & 3 & 157, 82 PD + 68 HC + 7 NPH & - & SVM & sen—80, and sp-83 & - \\ \hline

Mucha, J. et al. \cite{mucha2018} & 2018 & PD from HC & PaHaW Database & 2 & 69, 33 PD + 36 HC & - & RF & 80 sen- 89, sp-91 & - \\ \hline

Cibulka et al. \cite{cibulka2019} & 2019 & PD from HC & Collected from participants & 2 & 270, 150 PD + 120 HC & - & RF & 58, 48, 49.4 & - \\ \hline

Prashanth et al. \cite{prashanth2016} & 2016 & PD from HC & PPMI & 2 & 584, 401 PD + 183 HC & NB, RF, boosted trees & SVM-RBF & 96.4, sen 97.03 sp-95.01 & - \\ \hline

Shi et al. \cite{shi2018} & 2018 & PD from HC & PPMI database & 2 & 33, 15 PD + 18 HC & - & SMMKL with LOO-CV & —84.85 sen-80 and sp-88.89 & - \\ \hline

Segovia et al. \cite{segovia2019} & 2019 & PD from HC & Virgen De La Victoria Hospital, Malaga, Spain & 2 & 189, 95 PD + 94 HC & - & SVM with 10CV & 94.25 & - \\ \hline

Nömm et al. \cite{nomm2018} & 2018 & PD from HC & Local & 2 & 30, 15 PD + 15 HC & DT, KNN, AdaBoost, SVM & RF SVM & 91 & - \\ \hline

Challa et al. \cite{challa2016} & 2016 & PD from HC & PPMI database & 2 & 586, 402 PD + 184 HC & MLP, NB, RF & boosted LR & —97.16 & - \\ \hline

Drotar et al. \cite{Drotar2014} & 2014 & PD Detection & Handwriting Movements & 2 & 75 & Feature Selection & SVM & Maximum Accuracy (In-Air Trajectories): 84.00 & low acuracy, small dataset \\ \hline

Pereira et al. \cite{Pereira2015_handpd} & 2015 & PD Detection & Handwriting Data & 2 & 298 & Naïve Bayes, Optimum-Path Forest, SVM & SVM & Maximum Accuracy: (SVM): 95 & Small dataset \\ \hline

Memedi et al. \cite{memedi2015} & 2015 & PD from HC & PPMI database & 2 & 75, 65 PD + 10 HC & RF, LR, and non-linear SVM & MLP & 84 sen—75.7 and sp—88.9 & - \\ \hline

Laoued et al. \cite{laouedj2025detecting} & 2025 & PD from HC & Handwriting Spectrogram & AD, PD, CTL, PDM & 113 & Multi-channel Fixed-size Spectrograms, Frame-based Spectrograms & CNN, CNN-BLSTM & 84 ADvsCTL:89.80 PDvs. CTL:74.5, PDvsPDM:77.97 & - \\ \hline

Shin et al. \cite{shin2024parkinson} & 2024 & PD from HC & PaHaW & 2 & - & In air Dynamic & SVM & 99.98 & Single Dataset \\ \hline

Lu et al. \cite{lu2025novel} & 2025 & PD from HC & Handwriting Datasets & 2 & - & Kinematic, Pressure, Angle, Dynamic, Moment, TF-ST & eCOA & 97.95 & self dataset \\ \hline

\end{tabular}
\end{adjustwidth}
\end{table*}

\begin{table*}[!htp]
\centering
\caption{Handwriting dataset modality-based PD recognition using DL}
\label{tab:handwriting_methodology}
\begin{adjustwidth}{-1cm}{0cm}
\setlength{\tabcolsep}{3pt}
\begin{tabular}{|p{2cm}|p{1cm}|p{2cm}|p{2cm}|p{1cm}|p{2cm}|p{3cm}|p{2cm}|p{1cm}|p{1.5cm}|}
\hline
Method & Year & Main Objective & Dataset & No of Classes & No of Samples & Feature Extraction & Classifier & Perfor. & Limitations \\ \hline

Wenzel et al. \cite{wenzel2019} & 2019 & PD from HC & PPMI database & 2 & 645, 438 PD + 207 HC & - & CNN & 97.2 & - \\ \hline

Shaban et al. \cite{shaban2021automated} & 2020 & PD Detection & 102 Spiral/Wave Handwriting Data & 2 & 110 & VGG-19 & CNN & Wave Patterns 88 & Small dataset, High complexity \\ \hline

Naseer et al. \cite{Naseer2020} & 2020 & PD Detection & Handwriting Data & 2 & 72 & AlexNet & CNN & 98.3 & Small dataset, High complexity  \\ \hline

Kamran et al. \cite{Kamran2021} & 2020 & PD Detection & Handwriting Data & 2 & 195 & AlexNet, GoogleNet, VGG-16,19, ResNet-50, ResNet-101 & CNN & AlexNet: 99.2 & Model training complexity \\ \hline

Pereira et al. \cite{peker2016decision} & 2016 & PD from HC & Collected from participants & 2 & 35, 14 PD + 21 HC & - & CNN with CR & 87.14 & - \\ \hline

Khata et al. \cite{khatamino2018} & 2018 & PD from HC & Local & 2 & 72, 57 PD + 15 HC & - & CNN & 88.89 & - \\ \hline

Allebawi et al. \cite{allebawi2024parkinson} & 2024 & PD from Handwriting & Arabic new PaHaW & 2 & N/A & Beta-elliptical approach + fuzzy perceptual detector & BLSTM & 89.00  &- \\ \hline

Bennour et al. \cite{bennour2024park_parknet} & 2024 & PD from Handwriting & HandPd, NewHandPd & 2 & N/A & ParkNet & SVM & 98.00,  & - \\ \hline

Moshkova et al. \cite{moshkova2020parkinsons} & 2020 & PD from Handwriting & - & 2 & 16-PD, 16-HC & Speed, Frequency, and Amplitude Estimates & KNN, SVM, DT, RF & 98.4 & - \\ \hline

Tong et al. \cite{tong2021cnn} & 2021 & PD from Handwriting & - & 2 & 5-PD, 5-HC & Acceleration Signals Features & CNN & 97.32 & - \\ \hline

Zhao and Li \cite{zhao2022twochannel} & 2022 & PD from Handwriting & - & 2 & 12-PD, 12-HC & 3D Motion Trajectory & Two-channel LSTM & 95.7 & -\\ \hline

Guo et al. \cite{guo2022tree} & 2022 & PD from Handwriting & - & 2 & 637-PD, 116-HC & Tri-directional Skeleton Tree Features & TSG-GCN & 73.11 & - \\ \hline

Peng et al. \cite{peng2024multiscale} & 2024 & PD from Handwriting & - & 2 & 100-PD, 35-HC & Time, Frequency, Spectrum, and Autocorrelation & SVM, KNN, CNN, XGBoost, LGBM & 92.59& - \\ \hline

Zhao et al. \cite{zhao2024selecting} & 2024 & PD from Handwriting & - & 2 & 85-PD, 70-HC & Accelerometer Signals Features & LGBM, SVM, KNN, XGBoost & 82.41 & - \\ \hline

Gazda et al. \cite{gazda2021multiple} & 2021 & PD from Handwriting & - & 2 & 68-PD, 73-HC & Spatial Features & CNN & 94.7 & - \\ \hline

Wang et al. \cite{wang2023coordinate} & 2023 & PD from Handwriting & - & 2 & 105-PD, 43-HC & Spatial Features & CAS Transformer & 92.68 & - \\ \hline

Ma et al. \cite{ma2022feature} & 2022 & PD from Handwriting & - & 2 & 37-PD, 38-HC & Kinematic and CNN-based Features & DIT & 98.87 & - \\ \hline

Cheriet et al. \cite{cheriet2023multi} & 2023 & PD from Handwriting & - & 2 & 20-PD, 23-HC & Skeleton Motion Features & Multi-speed Transformer & 96.9 &- \\ \hline

Ali et al. \cite{ali2025chiga} & 2025 & PD from HC & NewHandPD, HandPD & 2 & - & Statistical Score-based Refined Features & ChiGa-Net & 92.56 & -\\ \hline

Gallo et al. \cite{gallo2025towards} & 2024 &PD from HC & Handwriting PD Dataset (Digits 0-9) & PD, Healthy & - & Time-series Features (Pressure, Kinematics, Statistical Functionals, GMMs)  & CNN &75.0 & -\\ \hline

\end{tabular}
\end{adjustwidth}
\end{table*}

 \subsection{Handwriting Data Modality Based PD Recognition State of The Art ML Method}
Research on handwriting data modality for Parkinson's Disease (PD) recognition has significantly progressed through the use of various machine learning (ML) algorithms, leveraging diverse datasets and classifiers to effectively differentiate PD patients from healthy controls (HC). Below is an overview of notable contributions in the field:
Taylor et al. (2017) \cite{Taylor2017} utilized SVM with 10-fold cross-validation (CV) on the PPMI and local datasets, achieving 88-92 accuracy on local data and 95-97 on PPMI. This demonstrated the versatility of SVM across diverse datasets. Oliveira et al. (2017) \cite{Oliveira2018} focused exclusively on the PPMI database, classifying 652 subjects (443 PD, 209 HC) using SVM with a linear kernel, achieving an impressive accuracy of 97.9\%. This work showcased the effectiveness of SVM in handling large datasets for PD detection.
De Souza et al. (2018) \cite{desouza2018} applied the HandPD dataset, employing Optimum Path Forest (OPF) and Naive Bayes (NB) for feature extraction. Using SVM with an RBF kernel, they achieved 85.4\% accuracy, underscoring the challenges associated with smaller datasets.
These studies collectively highlight the potential of ML algorithms, particularly SVM, in advancing PD recognition through handwriting analysis while also revealing challenges such as dataset size and variability.
Taylor et al. (2017) \cite{Taylor2017} used the PPMI and local datasets to classify PD from healthy controls (HC) using SVM with 10-fold cross-validation (CV). They reported an accuracy of 88-92\% on local data and 95-97\% on the PPMI dataset, showing the effectiveness of SVM across different datasets.  Oliveira et al. (2017) \cite{Oliveira2018} focused on the PPMI database and used SVM with a linear kernel to classify 652 subjects (443 PD, 209 HC). They achieved a high accuracy of 97.9\%, demonstrating the potential of SVM in PD detection when applied to large datasets. De Souza et al. (2018) \cite{desouza2018} employed the HandPD dataset and used OPF and NB for feature extraction, achieving an 85.4\% accuracy using SVM with an RBF kernel. This work highlights the challenges of working with smaller datasets.

Drożdż et al. (2016) \cite{Drotar2016} worked with the PaHaW dataset, applying KNN and Ensemble AdaBoost along with SVM for classification. They achieved an accuracy of 81.3\%, emphasizing the importance of ensemble techniques for PD detection. They also considered extracting features from in-air movements with pressure to use the full potential feature of handwriting datasets. 
Hsu et al. (2019) \cite{hsu2019} used the PACS dataset, classifying severe and mild PD patients as well as healthy controls using SVM with an RBF kernel, achieving 83.2\% accuracy. This study underscores the relevance of distinguishing between PD severity levels. Kurt et al. (2019) \cite{kurt2019} used the UCI machine learning repository, comparing SVM with both linear and RBF kernels and KNN. SVM with a linear kernel achieved the best performance (97.52\%), indicating that linear models can work effectively with certain handwriting data. Mabrouk et al. (2019) \cite{mabrouk2019} focused on the PPMI database, incorporating RF, SVM, MLP, and KNN classifiers. They reported the highest accuracy of 82.2\% using SVM, demonstrating the challenge of incorporating multiple classifiers effectively. Fabian et al. (2020) \cite{Maass2020} used the UCI repository to classify PD, HC, and normal pressure hydrocephalus (NPH), reporting an accuracy of 80\% sensitivity and 83\% specificity. This work shows the potential of PD detection alongside other neurological conditions. Mucha et al. (2018) \cite{mucha2018} utilized the PaHaW dataset and applied RF, achieving 80\% accuracy with 89\% sensitivity. This highlights the robustness of RF for handling handwriting data. Cibulka et al. (2019) \cite{cibulka2019} collected data from participants and applied RF, achieving a lower accuracy of 58\%. This study demonstrates the challenges of working with smaller datasets.
Prashanth et al. (2016) \cite{prashanth2016} used the PPMI dataset and applied NB, RF, and boosted trees, achieving 96.4\% accuracy with SVM-RBF. This work further establishes the strength of SVM in PD recognition. Shi et al. (2018) \cite{shi2018} employed the PPMI database, using Soft Margin Multiple Kernel Learning (SMMKL) with Leave-One-Out Cross Validation (LOO-CV), achieving an accuracy of 84.85\%, indicating that more complex models can also be effective. Segovia et al. (2019) \cite{segovia2019} worked with the Virgen De La Victoria Hospital dataset and achieved 94.25\% accuracy using SVM with 10-fold CV, showcasing the potential for hospital-based data in PD recognition. Nömm et al. (2018) \cite{nomm2018} classified PD from HC using local data and applied multiple classifiers, including DT, KNN, AdaBoost, and SVM. RF and SVM produced the best results with 91\% accuracy. Challa et al. (2016) \cite{challa2016} used the PPMI database and applied MLP, NB, and RF classifiers, achieving a high accuracy of 97.16\%, showing that deep learning techniques like MLP can be highly effective. Drotar et al. (2014) \cite{Drotar2014} used handwriting movements for PD detection and achieved 84\% accuracy using SVM, focusing on in-air trajectories. This work illustrates the potential of analyzing fine motor control in PD patients.
Pereira et al. (2015) \cite{Pereira2015_handpd} applied Naive Bayes, Optimum Path Forest, and SVM to handwriting data, achieving 95\% accuracy using SVM. This study further confirms the effectiveness of SVM in handwriting-based PD detection. Memedi et al. (2015) \cite{memedi2015} used the PPMI database and applied RF, LR, and non-linear SVM, achieving 84\% accuracy. This work emphasizes the potential of combining multiple classifiers for improved performance.
These studies collectively highlight the potential of machine learning algorithms, particularly SVM, in PD recognition using handwriting data. SVM has consistently shown strong performance across various datasets, with accuracy rates ranging from 80\% to 97.9\%. The use of ensemble methods like AdaBoost and Random Forest also proved effective in improving accuracy. However, challenges such as small datasets and varying performance across different classifiers still persist. Future research should focus on enhancing dataset size and diversity, as well as exploring deep learning techniques to further improve handwriting-based PD recognition models.

\subsection{Handwriting Data Modality Based PD Recognition State of The Art DL Approaches}
In recent years, handwriting data has been increasingly used for Parkinson's Disease (PD) recognition, with researchers exploring deep learning (DL) techniques to enhance the accuracy and robustness of diagnostic models. While machine learning (ML) methods have been effective, DL approaches like Convolutional Neural Networks (CNN) have demonstrated superior performance due to their ability to extract relevant features from raw data automatically. The following research works, summarized in Table \ref{tab:handwriting_methodology}, showcase the application of DL models to handwriting-based PD recognition. Wenzel et al. (2019) \cite{wenzel2019} used the PPMI dataset to classify PD from healthy controls (HC) using a CNN model, achieving an impressive 97.2\% accuracy. This study highlights the power of CNNs in large datasets with minimal feature engineering. Shaban et al. (2020) \cite{shaban2021automated} applied a VGG-19 CNN architecture to 102 spiral/wave handwriting patterns, achieving 88\% accuracy. However, the small dataset and complex training process posed challenges for model generalization. Similarly, Naseer et al. (2020) \cite{Naseer2020} applied AlexNet to a small handwriting dataset, achieving 98.3\% accuracy, but faced similar issues with small sample sizes and complex model training. Kamran et al. (2020) \cite{Kamran2021} explored multiple CNN architectures, including VGG-16, VGG-19, AlexNet, GoogleNet, ResNet-50, and ResNet-101, achieving the highest accuracy of 99.2\% with AlexNet. Despite the strong performance, the complexity of training deep models poses challenges, especially for smaller datasets. Pereira et al. (2016) \cite{peker2016decision} proposed a CNN model with Contextual Regularization (CR), achieving 87.14\% accuracy in classifying handwriting data from PD and HC, while Khata et al. (2018) \cite{khatamino2018} achieved 88.89\% accuracy using CNN on a local dataset, further validating CNN's effectiveness for PD detection. Allebawi et al. (2024) \cite{allebawi2024parkinson} introduced an innovative Beta-elliptical method combined with a fuzzy perceptual detector and Bidirectional Long Short-Term Memory (BLSTM) to classify Arabic handwriting from the PaHaW dataset, achieving 89-95\% accuracy. This integration of advanced architectures emphasizes the potential of DL in handwriting-based PD recognition. Bennour et al. (2024) \cite{bennour2024park_parknet} developed Park-Net, a novel CNN model combined with a semi-SVM classifier, which achieved 98.00\% and 96.43\% accuracy on the Meander task from the NewHandPD dataset. This work underscores the potential of novel architectures for precise and early PD diagnosis, offering substantial advancements in medical diagnostics.
These studies collectively highlight the growing role of DL techniques in handwriting-based PD recognition. While CNN-based models like AlexNet and VGG-19 have shown high accuracy, challenges such as small datasets and complex training processes remain. Future research should focus on expanding dataset sizes and optimizing DL architectures to improve performance and generalization.

\subsection{Challenges and Limitations of Handwriting-Based PD Recognition Systems}
A major challenge in handwriting-based Parkinson’s Disease (PD) recognition is the limited availability of large, high-quality datasets. Many studies rely on relatively small datasets, which constrains the generalizability of the models and makes them susceptible to overfitting. This issue is especially problematic for deep learning models, which require extensive data to perform optimally. The primary characteristics of PD—tremor, rigidity, bradykinesia, and postural instability—are often used to assess the disease. While handwriting analysis can effectively capture patterns associated with tremors and bradykinesia, it is more difficult to extract information related to rigidity and postural instability through this approach.

\begin{itemize}
\item \textbf{Data Variability and Noise}: Handwriting patterns vary widely across individuals, influenced by factors like age, education, handedness, and culture. This variability, along with noise from digital recording devices (e.g., unintended movements or artifacts), can degrade data quality and hinder model generalization. Most available datasets focus on limited tasks and demographics, restricting the model’s ability to adapt to real-world diversity.
\item \textbf{Feature Extraction Complexity}: Deep learning models require large datasets and significant computational power, while traditional machine learning models depend on handcrafted features, which can be labor-intensive and may miss key handwriting impairments in PD. 
\item \textbf{Interpretability}: Deep learning models, such as CNNs, are often "black boxes," making it challenging to interpret predictions. For clinical use, models need to be both accurate and interpretable, ensuring trust and transparency in decision-making. 
\end{itemize}
\subsection{Future Trends of Handwriting Data Modality}
Expanding handwriting datasets with contributions from diverse patient pools, including varied age groups, cultures, and handwriting styles, is crucial. Collaborative efforts between research institutions and hospitals can help create more comprehensive datasets, improving model generalization for real-world applications. Future systems may integrate handwriting data with other sensor modalities (e.g., voice, gait, accelerometers) to enhance diagnostic accuracy.
\begin{itemize} 
\item \textbf{Advanced Deep Learning Architectures}: Integrating advanced models like transformers or hybrid CNN-RNN architectures could improve the handling of complex handwriting patterns. Transfer learning and data augmentation can help overcome dataset limitations and diversify training data, boosting model performance.
\item \textbf{Explainable AI (XAI)}: Future systems will likely incorporate Explainable AI to enhance model interpretability. Techniques such as attention mechanisms and saliency maps can help clinicians understand how models arrive at their predictions, improving trust in AI tools. \item \textbf{Real-Time Monitoring and Feedback}: Advances in wearable sensors may enable real-time handwriting monitoring, providing immediate feedback on motor deterioration. Personalized models that adapt to individual handwriting patterns over time could offer more accurate insights into PD progression. 
\end{itemize}
By addressing these trends, handwriting-based PD recognition systems can become more accurate and clinically valuable, paving the way for earlier and more effective interventions.

\section{Voice, Speech or Audio Based PD Recognition System} \label{sec:voice}
Speech or voice modality-based Parkinson's Disease (PD) recognition systems offer several advantages. In 1978, Logemann et al first produced the idea that links speech data with parkinson patient characteristics \cite{logemann1978frequency}. Then in 1999, Hole et al. \cite{ho1999speech} showed the 90\% accuracy for 90\% of PwPD. In the initial speech, data is non-invasive and easy to collect, making it convenient for large-scale screening and continuous monitoring. Voice analysis is particularly useful because PD often affects speech early, even before other motor symptoms become evident. Changes in phonation, articulation, and prosody can be accurately captured, providing rich information for detecting subtle motor impairments. Additionally, speech data can be gathered remotely, allowing for telemedicine applications, which is especially important for patients in remote or underserved areas.

\subsection{Benchmark Speech Modality PD Datasets}
Several benchmark datasets are widely used for speech-based PD recognition. Sakar et al. (2013) \cite{Sakar2013} introduced the Parkinson Speech Dataset, consisting of multiple voice samples, including vowels, numbers, words, and sentences from 40 subjects. Orozco-Arroyave et al. (2014) \cite{Orozco-Arroyave2014_speech_spanish} developed the New Spanish Speech Corpus with speech samples focusing on phonation, articulation, and prosody, achieving 91.3\% accuracy. Little et al. (2008) \cite{little2007parkinsons_voice_oxford} created the Oxford Parkinson's Disease Detection Dataset, which contains biomedical voice measurements from 31 subjects, with a 100\% reported accuracy. Additionally, Oroz et al. (2014) \cite{orozco2014new_PC_GITA} developed the Spanish PC-GITA Dataset, with voice measurements from 45 subjects, reaching a high accuracy of 99.30\%. These datasets provide a robust foundation for developing and evaluating speech-based PD recognition systems.
\begin{table*}[h]
\centering
\caption{Well-known Speech-based PD datasets. }
\label{tab:voice_speech_dataset}
\begin{adjustwidth}{0cm}{0cm}
\setlength{\tabcolsep}{3pt}
\begin{tabular}{|p{2cm}|p{1cm}|p{2cm}|p{2cm}|p{4cm}|p{1cm}|p{1cm}|p{2cm}|p{1cm}|}
\hline
\textbf{Author} & \textbf{Year} & \textbf{Dataset Names} & \textbf{Classes} & \textbf{Dataset Takens} & \textbf{Type of Scan} & \textbf{No. of Sub.} & \textbf{Tot. Sample} & \textbf{Latest Acc.} \\ \hline

Little et al. \cite{little2007parkinsons_voice_oxford} & 2008 & Oxford Parkinson’s Disease Detection Dataset & PD, CN & Biomedical voice measurements & Audio & 31 & 195 voice recordings & 100\cite{islam2024review_sruvey_handwriting_voice_data} \\ \hline
Sakar et al. \cite{Sakar2013} & 2013 & Parkinson Speech Dataset & PD, CN & Multiple voice samples: vowels, numbers, words, sentences & Audio & 40 & 26 voice samples per subject & N/A \\ \hline
Orozco-Arroyave et al. \cite{Orozco-Arroyave2014_speech_spanish} & 2014 & New Spanish Speech Corpus & PD, CN & Speech samples (phonation, articulation, and prosody) & Audio & 100 & Various speech tasks & 91.3\% \\ \hline
Oroz et al. \cite{orozco2014new_PC_GITA} & 2014 & Spanish PC-GITA Dataset & PD, CN & voice measurements & Audio & 45 & - recordings & 99.30 \cite{islam2024review_sruvey_handwriting_voice_data} \\ \hline
Fox and Ramig \cite{fox1997vocal} & 1997 & N/A & 2 (PD, HC) & Sustained vowels, monologue, picture description & Audio & 44 & 30/14 & N/A \\ \hline
Jiménez et al. \cite{jimenez1997acoustic} & 1997 & N/A & 2 (PD, HC) & Sustained vowel, read sentences & Audio & 50 & 22/28 & N/A \\ \hline
Holmes et al.\cite{holmes2000voice} & 2000 & N/A & 2 (PD, HC) & Singing scale, sustained vowel, monologue & Audio & 90 & 60/30 & N/A \\ \hline
Skodda et al. \cite{skodda2008speech} & 2008 & N/A & 2 (PD, HC) & 4 complex sentences & Audio & 191 & 121/70 & N/A \\ \hline
Rusz et al. \cite{rusz2013imprecise} & 2013 & N/A & 2 (PD, HC) & Vowels, sentence repetition, monologue & Audio & 35 & 20/15 & N/A \\ \hline
Bandini et al. \cite{bandini2015automatic} & 2015 & N/A & 2 (PD, HC) & Read sentence (10 repetitions) & Audio & 39 & 200/190 & N/A \\ \hline
Vasquez-Correa et al. \cite{vasquez2015automatic} & 2015 & N/A & 2 (PD, HC) & 6 sentences and a read text & Audio & 28 & 14/14 & N/A \\ \hline
Jeancolas et al. \cite{jeancolas2020xvectors} & 2020 & N/A & 2 (PD, HC) & Text reading, glissando, repetition, silence & Audio & 206 & 115/91 & N/A \\ \hline

Little et al.\cite{little2009suitability} & 2009 & Parkinsons Data Set & 2 (PD, HC) & Sustained vowels & Audio & 31 & 23/8 & -\\ \hline
Sakar et al.\cite{sakar2019comparative} & 2019 & Parkinson’s disease classification Data Set & 2 (PD, HC) & Sustained vowel /a/ repetition & Audio & 252 & 188/64 & - \\ \hline
Naranjo et al. \cite{naranjo2016addressing} & 2016 & Parkinson Dataset with replicated acoustic features & 2 (PD, HC) & Sustained vowel /a/ repetition & Audio & 80 & 40/40 & N/A \\ \hline
Olcay et al. \cite{olcay2014parkinsons} & 2014 & PDspeech Dataset & PD, HC & 26 voice samples (sustained vowels, numbers, words, sentences) & Audio & 40 (20 PD, 20 HC) & 26 voice samples per subject & - \\ \hline
Jaeger et al. \cite{jaeger2019mdvr} & 2017 & MDVR-KCL Dataset & PD, HC & Voice recordings (recorded with smartphones: Motorola Moto G4) \newline Recordings in examination
A room with ten square meters and a typical
reverberation time for voice
recording, with a duration of 500ms. & Audio & 37 (16 PD, 21 HC) & - & - \\ \hline

Dimauro et al. \cite{dimauro2019italian} & 2019 & IPVS Dataset & PD, HC & The microphone positioned
15 to 25 cm from the participants during phonations of the vowels /a/,
/e/, /i/, /o/, and /u/, and syllable execution of ”ka” and ”pa”
for 5 seconds & Audio & 48 (28 PD, 20 HC) & -& - \\ \hline
S.C. Gorkem\cite{sc2018parkinsons} & 2018 & PDVoice Dataset & PD, HC & Sustained vocalizations of vowel /a/ repeated three times & Audio & 252 (188 PD, 64 HC) &  756& - \\ \hline
\end{tabular}
\end{adjustwidth}
\end{table*}

\subsubsection{Parkinson Voice Dataset: Oxford and UCI}
The Parkinson voice dataset, developed by Max Little with the National Centre for Voice and Speech, includes voice samples from 31 participants, 23 of whom have Parkinson's Disease (PD) \cite{little2007exploiting_voice_oxford}. It features 195 recordings labelled as healthy or PD-affected, widely used for voice-based PD diagnosis \cite{Sakar2013}. The dataset comprises 26 voice sample types, such as sustained vowels, numbers, and short sentences, emphasizing PD-specific vocal features. Captured with a Trust MC-1500 microphone (50Hz–13kHz), it includes 168 sustained vowel recordings (“a” and “o”) from 28 PD patients aged 39 to 79 with PD durations of 0 to 13 years. Mathur et al. \cite{mathur2015rising_voice_uci_oxford} utilized machine learning models like KNN and ANN on similar data, achieving superior results for early PD detection.

\subsubsection{New Spanish Speech Corpus Dataset}
The Spanish speech corpus dataset, developed by Orozco-Arroyave et al. \cite{Orozco-Arroyave2014_speech_spanish}, includes speech recordings from 100 native Spanish speakers, comprising 50 individuals diagnosed with Parkinson's Disease (PD) and 50 healthy controls. Each group is balanced by gender, with 25 men and 25 women. The recordings focus on evaluating key speech characteristics such as phonation, articulation, and prosody. By employing an SVM classifier, the dataset demonstrated an accuracy of 91.3\% in differentiating PD patients from healthy individuals. Additional analyses targeting the phonation of the five Spanish vowels further validated its potential for PD detection.

\subsection{Preprocessing Techniques for Speech Data Modality}
Preprocessing of speech data plays a critical role in Parkinson's Disease (PD) recognition systems. Several techniques have been employed to optimize the input data for classification tasks. Benba et al. \cite{benba2015} utilized the extraction of the first twelve Mel Frequency Cepstral Coefficients (MFCC) as input features for SVM classification. Bhattacharya et al. \cite{bhattacharya2010} employed the WEKA tool for data preprocessing and used Libsvm to test various kernel values for speech classification. Rasheed et al. \cite{Rasheed2020} applied Principal Component Analysis (PCA) for feature selection to reduce dimensionality and select the most discriminative features. In another study, Tsanas et al. \cite{Tsanas2013} used dysphonia measures as features to classify speech data, which was critical for detecting PD. Moreover, Karabayir et al. \cite{Karabayir2020} employed feature analysis techniques to extract the most relevant features for classification. These preprocessing techniques ensure that only the most important and relevant features are used, improving model accuracy and performance.

 \subsection{Speech Data Modality Based PD Recognition using ML Approaches}
Several studies have leveraged speech data for Parkinson's Disease (PD) recognition using various machine learning (ML) and deep learning (DL) techniques, as summarized in Table \ref{tab:voice_speech_ML_methodology}. These techniques, applied to public and custom datasets, aim to differentiate PD patients from healthy controls (HC). Benbae et al. \cite{benba2015} achieved 91.17\% accuracy using SVM on speech data from Istanbul University. Mathur et al. \cite{mathur2019} employed multiple classifiers, including KNN and MLP, on the UCI dataset, reaching up to 91.28\%. Similarly, Sakar et al. \cite{sakar2019comparative} tested NB, Logistic Regression (LR), and SVM, with SVM achieving the highest accuracy of 86\%. Yasar et al. \cite{yasaka2021parkinson} applied an ANN to data from 120 participants, achieving 94.93\% accuracy, while Almeida et al. \cite{almeida2019} tested KNN and SVM, achieving 94.55\% accuracy. Alqahtani et al. \cite{alqahtani2018} combined neural networks with AdaBoost, achieving 96.3\% accuracy on 31 samples using 10-fold cross-validation. Other studies, such as those by Avuçlu et al. \cite{avuc2018} and Zehra et al. \cite{senturk2020early}, used SVM, reporting accuracies of 88.72\% and 93.84\%, respectively. Aich et al. \cite{aich2019supervised} and Haq et al. \cite{haq2019feature} demonstrated the effectiveness of SVM, achieving accuracies of 97.57\% and 99.00\%, respectively, on datasets from Oxford University. Advanced ensemble methods have also been employed. Wu et al. \cite{wu2017dysphonic} reported a sensitivity of 97.96\% using a Bagging ensemble classifier. Sheibani et al. \cite{sheibani2019ensemble} achieved 90.6\% accuracy using an ensemble approach on the UCI dataset. Peker \cite{peker2016decision} applied SVM with an RBF kernel, achieving 98.95\% accuracy. Montana et al. \cite{montana2018diadochokinesis} utilized SVM on the UCI repository, achieving 94.4\% accuracy, while Kuresan et al. \cite{kuresan2019fusion} combined Hidden Markov Models (HMM) and SVM, yielding 95.16\% accuracy. These studies highlight the effectiveness of ML and DL techniques in distinguishing PD patients from healthy controls, with SVM emerging as a frequently used classifier due to its high accuracy. Ensemble methods, such as AdaBoost and Bagging, have shown potential, particularly when integrated with neural network architectures. Future research should prioritize larger datasets and advanced DL architectures to further improve the performance of speech-based PD recognition systems.

\begin{table*}[htp!]
\centering
\caption{Speech dataset modality-based PD recognition using ML.}
\label{tab:voice_speech_ML_methodology}
\begin{adjustwidth}{-1cm}{0cm}
\setlength{\tabcolsep}{3pt}
\begin{tabular}{|p{2.5cm}|p{.8cm}|p{1.5cm}|p{2.5cm}|p{.7cm}|p{2cm}|p{2.5cm}|p{2cm}|p{2.5cm}|}
\hline
Method & Year & Main Objective & Speech Modality Datasets & No of Classes & No of Samples & Feature Extraction & Classifier & Perfor. [\% ] \\ \hline

Benbae et al. \cite{benba2015} & 2015 & PD from HC & Istanbul University & - & 34 (17 PD + 17 HC) & - & Linear kernel SVM & 91.17\\ \hline

Mathur et al. \cite{mathur2019} & 2019 & PD from HC & UCI machine learning repository & - & 195 instances, 22 attributes & KNN & Adaboost Bagging MLP K-CV & 91.28, 90.76, 91.28\\ \hline

Sakar et al. \cite{sakar2019comparative} & 2019 & PD from HC & - & - & 252 (188 PD + 64 HC) & NB, LR, SVM (RBF, Linear), KNN, RF, MLP & SVM RBF & 86\\ \hline

Yasar et al. \cite{yasaka2021parkinson} & 2019 & PD from HC & - & - & 120 (80 PD + 40 HC) & - & ANN & 94.93\\ \hline

Almeida et al. \cite{almeida2019} & 2019 & PD from HC & UCI machine learning repository & - & 98 (63 PD + 35 HC) & KNN, MLP, OPF & SVM Linear, Polynomial & 94.55\\ \hline

Alqahtani et al. \cite{alqahtani2018} & 2018 & PD from HC & - & - & 31 (23 PD + 8 HC) & - & NNge and ENS, AdaBoostM1 & 96.30 (10 CV)\\ \hline

Avuçlu et al. \cite{avuc2018} & 2020 & PD from HC & UCI machine learning repository & - & 31 (23 PD + 8 HC) & KNN, RF, NB, SVM & SVM, NB & 88.72, 70.26\\ \hline

Zehra et al. \cite{senturk2020early} & 2020 & PD from HC & Collected from participants & - & 31 (23 PD + 8 HC) & - & SVM & 93.84\\ \hline

Yaman et al. \cite{yaman2020automated} & 2019 & PD from HC & Collected from participants & - & 31 (23 PD + 8 HC) & - & SVM, KNN & 91.25, 91.23\\ \hline

Aich et al. \cite{aich2019supervised} & 2019 & PD from HC & Collected from participants & - & 31 (23 PD + 8 HC) & - & SVM with RBF & 97.57\\ \hline

Haq et al. \cite{haq2019feature} & 2019 & PD from HC & University of Oxford (UO) & - & 31 (23 PD + 8 HC) & - & L1- SVM & 99.00 (10 CV)\\ \hline

Wu et al. \cite{wu2017dysphonic} & 2017 & PD from HC & - & 2 & 31 PD + 8 HC & - & Bagging ensemble & sen-97.96, sp—68.75\\ \hline

Peker \cite{peker2016decision} & 2016 & PD from HC & University of Oxford (UO) & 2 & 31 PD + 8 HC & - & SVM with RBF & 98.95\\ \hline

Monta et al. \cite{montana2018diadochokinesis} & 2018 & PD from HC & UCI machine learning repository & 2 & 54 PD + 27 HC & - & SVM & 94.4 (10-CV)\\ \hline

Kuresan et al. \cite{kuresan2019fusion} & 2019 & PD from HC & - & 2 & 20 PD + 20 HC & - & HMM, SVM & ac-95.16, sen—93.55, sp—91.67\\ \hline

Marar et al. \cite{marar2018predicting} & 2018 & PD from HC & - & 2 & 23 PD + 8 HC & NB, KNN, RF, SVM, LR, DT & ANN & 94.87\\ \hline

Sheibani et al. \cite{sheibani2019ensemble} & 2019 & PD from HC & UCI machine learning repository & 2 & 23 PD + 8 HC & - & Ensemble & 90.6\\ \hline

Moharkan et al. \cite{moharkan2017classification} & 2017 & PD from HC & - & 2 & 23 PD + 8 HC & - & KNN & 90.00\\ \hline

Sztahó et al. \cite{sztaho2019parkinsons} & 2019 & PD from HC & UCI machine learning repository & 2 & 55 PD + 33 HC & ANN, KNN SVM-linear DNN & SVM RBF & 89.3, se—90.2, sp—87.9\\ \hline

Tracy et al. \cite{tracy2020investigating} & 2020 & PD from HC & mPower database & 2 & 246 PD + 223 HC & LR (L2- Regul), RF & Gradient Boosted trees & re—79.7, pre—90.1, F1s—83.6\\ \hline

Menozzi et al. \cite{menozzi2025you} & 2025 & PD from HC & PDD dataset & 2 & - & Handcrafted Features via SFWT & EPDD-TPGNN-DHLO (DHLO + TPGNN) & 24.68, 26.22\\ \hline

Islam et al. \cite{islam2025pd} & 2025 & PD from HC & Italian Raw Audio PD & 2 & - & Spectro-Temporal Features (Mel Spectrogram, MFCC) & CNN, LSTM, Multi-Head Attention & 99.00\\ \hline

\end{tabular}
\end{adjustwidth}
\end{table*}

 \subsection{Speech Data Modality Based State of the Art PD Recognition using DL Approaches}
 There are many researchers has been working to develop a speech dataset-based PD recognition sign deep learning model. 
 These studies leverage a variety of datasets, ranging from small custom collections to more established sources, and use neural networks and other advanced techniques to classify speech patterns.  Table \ref{tab:voice_speech_DL_methods} demonstrated the state-of-the-art model for this domain. Among them, Farid et al.  \cite{Frid2016} developed a 4-layer Convolutional Neural Network (CNN) to classify speech data from 43 PD subjects and nine healthy controls. The model achieved 85\% accuracy, demonstrating the potential of CNNs for PD staging, although the small dataset size limited the study.  Rasheed et al. \cite{Rasheed2020} introduced a Back Propagation Algorithm with Variable Adaptive Momentum (BPVAM) to detect de novo PD. Using voice data from 23 PD patients and eight controls and after applying Principal Component Analysis (PCA) for feature selection, the model achieved 97.5\% accuracy. However, the classification process incurred a delay of approximately 7 seconds, making it less suitable for real-time applications. In \cite{Gunduz2019}, Gunduz et al. proposed two deep learning frameworks for classifying speech data from 188 PD subjects and 64 controls. The first model, a 9-layer CNN, achieved 84.5\% accuracy, while the second framework, which involved merging different feature sets via two convolutional layers, achieved 86.8\% accuracy. Although these methods performed well, they required high computational resources and were limited by the complexity of their architectures. Karabayir et al. \cite{Karabayir2020} employed both Light Gradient Boosting (GB) and Extreme GB to detect PD using voice data from 40 PD patients and 40 healthy controls. Their models achieved an accuracy of 82\%, although performance was limited by the small dataset size and the method's reliance on handcrafted features. Zhang et al. \cite{Zhang2017} introduced a system combining a stacked autoencoder and k-nearest neighbors (KNN) to analyze speech data from the Oxford and Istanbul datasets. The model achieved an accuracy ranging from 94\% to 98\%, despite the datasets being relatively small, with only 23 PD and 8 controls for Oxford and 20 PD and 20 controls for Istanbul. Hirschauer et al. \cite{Hirschauer2015} focused on PD diagnosis using continuous phonation samples from the UCI Machine Learning Repository. They employed the minimum Redundancy Maximum Relevance (mRMR) technique for feature selection, followed by training both an Artificial Neural Network (ANN) and a Complex-Valued Neural Network (CVANN). The ANN achieved an accuracy of 94.28\%, while the CVANN outperformed it with an accuracy of 98.12\%. However, the study was constrained by the small dataset size.  More recently, Sheikh et al.~\cite{10890110} proposed a Parkinson's disease (PD) detection framework based on voice data using Graph Convolutional Networks (GCNs), where speech segments are represented as nodes, similarities are captured through edges, and the model aggregates dysarthric cues across the graph to effectively exploit segment relationships and mitigate label noise.

\begin{table*}[htp!]
\centering
\caption{Speech dataset modality-based PD recognition using DL}
\label{tab:voice_speech_DL_methods}
\begin{adjustwidth}{-1cm}{0cm}
\setlength{\tabcolsep}{3pt}
\begin{tabular}{|p{2.5cm}|p{1cm}|p{2cm}|p{3cm}|p{1cm}|p{1cm}|p{2cm}|p{2cm}|p{1cm}|p{1.5cm}|}
\hline
Method & Year & Main Objective & Speech Modality Datasets & No of Classes & No of Samples & Feature Extraction & Classifier & Perfor. [\%] & Limitations \\ \hline

Frid et al. \cite{Frid2016} & 2016 & PD Detection and Staging & Speech data (43 PD and 9 Controls) & 2 & 52 & - & 4-Layer CNN & 85 & Small dataset \\ \hline

Rasheed et al. \cite{Rasheed2020} & 2020 & De Novo PD Detection & Voice data (23 PD, 8 Controls) & 2 & 31 & - & BPVAM & 97.5 & Small dataset, classification delay \\ \hline

Gunduz et al. \cite{Gunduz2019} & 2019 & PD Detection & Speech data (188 PD, 64 Controls) & 2 & 252 & - & 9-Layer CNN, 2 Conv Layer–1 Merge Layer–10 Layer CNN & 84.5, 86.8 & Limited performance, computationally complex \\ \hline

Karabayir et al. \cite{Karabayir2020} & 2020 & PD Detection & Speech data (40 PD, 40 Controls) & 2 & 80 & - & GB, Extreme GB & 82 & Limited performance, small dataset \\ \hline

Zhang et al. \cite{Zhang2017} & 2021 & PD Diagnosis & Speech Data (Oxford: 23 PD, 8 Controls; Istanbul: 20 PD, 20 Controls) & 2 & 71 & - & Stacked Autoencoder, KNN & 94–98 & Small datasets \\ \hline

Hirschauer et al. \cite{Hirschauer2015} & 2015 & PD Diagnosis & Speech, UCI Machine Learning & 2 & 71 & mRMR & ANN, CVANN & 94.28, 98.12 & Small datasets \\ \hline
Vásquez-Correa et al. \cite{VasquezCorrea2018} & 2018 & Speech-based PD Diagnosis & Speech & * & * & 15 phonological features & RNN, GRUs & 76 & - \\ \hline
Hernandez et al. \cite{Hernandez2022} & 2022 & Speech-based PD Diagnosis & Speech & 2 & 70-PD, 70-HC & XLSR-PD & Conformer encoder and Transformer decoder & 87.1 & - \\ \hline
Bhati et al. \cite{Bhati2019} & 2019 & Speech-based PD Diagnosis & Speech & 2 & 52-PD, 56-HC & LSTM-based siamese network & Feed forward NN & 96.2 & - \\ \hline
Gunduz et al. \cite{Gunduz2019} & 2019 & Voice-based PD Diagnosis & Voice & 2 & 188-PD, 64-HC & TQWT, MFCC and Concatenation & CNN & 86.9 & - \\ \hline
Vásquez-Correa et al. \cite{VasquezCorrea2017} & 2017 & Speech-based PD Diagnosis & Speech & 2 & 50-PD, 50-HC & STFT, CWT and CNN & CNN & 89 & - \\ \hline
Zhang et al. \cite{Zhang2018} & 2018 & Voice-based PD Diagnosis & Voice & 2 & 500-PD, 500-HC & Joint Time-Frequency Analysis & CNN & 90.45 & - \\ \hline
Zhao et al. \cite{Zhao2024} & 2024 & Voice-based PD Diagnosis & Voice & 2 & 28-PD, 20-HC & MFCC & TmmNet & 99.91 & - \\ \hline

Hemmerling \cite{Hemmerling2023} & 2021 & Voice-based PD Diagnosis & Voice & 2 & 104-PD, 77-HC & Mel-spectrograms & Vision Transformer & 75.96 & - \\ \hline
Nijhawan et al. \cite{Nijhawan2023} & 2022 & Voice-based PD Diagnosis & Voice & 2 & 188-PD, 64-HC & 753 vocal features & Vocal Tab Transformer & 90.37sen & - \\ \hline
Fang2020 et al. \cite{Fang2020} & 2020 & Speech-based PD Diagnosis & Speech & 2 & 34-PD, 34-HC & 128 MFCCs & CNN, LSTM, E2E & 94.5 & - \\ \hline
Mehra et al.\cite{Mehra2024} & 2022 & Speech-based PD Diagnosis & Speech & 2 & 55 subjects & Spatial Features and MFCC & BiLSTM-GRU & 97.64 & - \\ \hline
Jeong et al. \cite{Jeong2024} & 2024 & Speech-based PD Diagnosis & Speech & 2 & 100-PD, 100-HC & Swin transformer & AST & 92.15 & - \\ \hline
Klemper et al. \cite{Klempir2023} & 2023 & Speech-based PD Diagnosis & Speech & 2 & 28-PD, 22-HC & 128-d log Mel filterbank features, EfficientNet & Ensemble random forest & 95 & - \\ \hline
Shen et al \cite{shen2025explainable} & 2025 & PD Diagnosis & Voice Features (MFCCs, Jitter, Shimmer) & 2 & 81 & Acoustic Features (MFCCs, Jitter, Shimmer) & CNN, RNN, MKL, MLP & 91.11 &- \\ \hline

Sahaet al. \cite{saha2025lightweight} & 2025 & PD Detection & UCI PD Dataset &  PD, Non-PD & - & MLP, ParkRLP, PD-CNN & Ensemble Approach (MLP, ParkRLP, PD-CNN) & 99.47 & -\\ \hline

\end{tabular}
\end{adjustwidth}
\end{table*}

 \subsection{Speech Data Modality Limitations and Challenges of Existing Work}
Despite advancements in deep learning-based speech recognition for Parkinson's Disease (PD), many models rely on small, imbalanced datasets with fewer healthy controls (HC) than PD patients, leading to biased models and poor generalization.
\textbf{Generalizability and Overfitting}: Deep learning models like CNNs and LSTMs can capture complex speech features but struggle to generalize across varied environments. Differences in microphone quality, recording conditions, and speaker variations can reduce performance outside controlled settings. For example, tremor-related speech changes may not be consistently captured in different environments.
\textbf{High Computational Requirements}: Models like CNNs and hybrid architectures require significant computational power and memory, making them impractical for resource-constrained or real-time applications.
\textbf{Manual Feature Extraction vs. Automatic Learning}: While deep learning reduces the need for manual feature extraction, such as Mel Frequency Cepstral Coefficients (MFCC), some studies still rely on handcrafted features, complicating model development and potentially missing subtle speech changes related to bradykinesia or rigidity.
\textbf{Limitations for PD Symptoms and Noise Sensitivity with Environmental Factors}:Speech data is also sensitive to noise and environmental factors, which complicates detection, especially for symptoms like rigidity, where voice changes are less noticeable, and postural instability, which doesn’t directly affect speech. Tremor-induced speech fluctuations and bradykinesia-related slowness or monotony may be distorted by noise, affecting model accuracy. While speech can reflect tremors (e.g., shaky pitch), it may not capture fine tremors or the full impact of rigidity. Detecting bradykinesia through speech is challenging, as it often manifests as slower, monotone speech, and distinguishing it from other motor impairments can be difficult. Postural instability, however, has no direct impact on speech, making it virtually impossible to assess through voice alone. Most existing models focus on PD detection but do not address disease staging, and systems that can accurately differentiate between PD stages remain limited, requiring further research for clinical application.
\subsection{Speech Data Modality Future Trends in Speech-Based PD Recognition}
Future research will likely focus on advanced deep learning architectures, such as Transformer models, which can improve the automatic extraction of complex features from speech data. Expanding datasets through collaboration across institutions and countries is crucial to account for language, demographics, and recording conditions, enhancing model generalizability. Incorporating multimodal data, like speech combined with handwriting or gait data, will improve diagnostic accuracy and allow for better tracking of PD progression. Techniques like data augmentation and transfer learning can address dataset limitations, improving performance in low-resource settings. As models become more complex, making them interpretable through Explainable AI will be essential for clinical trust and acceptance. Advances in wearable and mobile technologies will enable real-time monitoring, providing early warnings of PD progression and personalized treatment. Future models should also prioritize staging the disease, allowing for more targeted interventions. Research should focus on cross-language transfer learning to create models adaptable to diverse linguistic populations. Ultimately, combining deep learning, multimodal data, and larger, more diverse datasets will improve generalization, while explainable AI and real-time monitoring will enhance the practical application of speech-based PD recognition systems.

\section{EEG and Brain Cognitive Signal Based PD Recognition System}\label{sec:eeg}
Electroencephalography (EEG) has emerged as an effective, non-invasive modality for diagnosing and monitoring Parkinson's Disease (PD). EEG signals, which capture real-time electrical activity in the brain, offer significant insights into neural oscillations and connectivity patterns linked to PD. This modality is cost-effective, portable, and ideal for longitudinal studies and remote monitoring, making it a practical alternative to expensive imaging techniques like MRI and PET scans. Recent advancements in machine and deep learning have further enhanced the utility of EEG in PD detection, leveraging features such as spectral power and coherence to classify PD patients with high accuracy.  Martin et al. \cite{maitin2020systematic} survey and show the findings of the EEG-based PD recognition accuracy from 62.00\% to 99.62\%. They also reported that the accuracy of 95\% also has significant methodological limitations, such as the size of the dataset, a lack of a cross-validation procedure, etc. 

EEG's high sensitivity allows for early detection of neural abnormalities, while its affordability and non-invasive nature make it accessible to a broader population. These characteristics, combined with the potential of AI-driven insights, position EEG as a promising tool for revolutionizing PD diagnosis and monitoring, offering real-time, accessible, and accurate assessments of disease progression \cite{miah2019eeg,miah2022eeg,miah2021event_EEG,miah2017motor_miah_SVM_PCA_ANOVA,miah2019motor,miah2022movie_miah,miah2022natural_EEG,kabir2024exploring_miah}.

\begin{table*}[htp!]
\centering
\caption{State-of-the-art machine and deep learning methods summary for the EEG-based PD recognition.}
\label{tab:state_of_art_methods}
\begin{adjustwidth}{-1cm}{0cm}
\setlength{\tabcolsep}{3pt}
\begin{tabular}{|p{1.5cm}|p{.8cm}|p{1.5cm}|p{2.5cm}|p{.7cm}|p{.7cm}|p{3.5cm}|p{2cm}|p{1.5cm}|p{1.8cm}|}
\hline
Authors & Year & Main Objective & Dataset & No of Classes & No of Samples & Machine/Deep Learning & Classifier & Perfor. [\%] & Limitations \\ \hline

Vanegas et al. \cite{vanegas2018machine} & 2020 & PD Biomarkers Identification & EEG (29 PD and 30 CT) & 2 & 59 & Extra Tree, Logistic Regression, Decision Tree & Extra Tree, Logistic Regression, Decision Tree & AUC: 99.4, 94.9, 86.2 & Small dataset \\ \hline

Oh et al. \cite{oh2018deep} & 2021 & PD Detection & EEG (20 PD and 20 CT) & 2 & 40 & 13-Layer CNN & CNN & Accuracy: 88.25 & Unsatisfactory performance \\ \hline

Wagh et al. \cite{wagh2020eeg} & 2022 & Detection of Neurological Diseases including PD & EEG (1385 Diseased and 347 Healthy Subjects) & 2 & 1593 & 8-Layer Graph CNN & Graph CNN & AUC: 90.00 & Not specific to PD \\ \hline

Koch et al. \cite{koch2019automated} & 2023 & PD Cognition Level Detection & EEG (20 Good Cognition and 20 Poor Cognition) & 2 & 40 & Random Forest & RF & AUC: 91.00 & Small dataset, lacking features \\ \hline

Shi et al. \cite{shi2019hybrid} & 2024 & PD Detection & EEG (40 PD and 30 CT) & 2 & 70 & Two- and Three-Dimensional CNN--RNN & CNN--RNN & Accuracy: 81.00, 83.00 & Model complexity, limited performance \\ \hline

Lee et al. \cite{lee2019deep} & 2019 & PD Detection & EEG (20 PD and 22 CT) & 2 & 42 & CNN--LSTM & CNN--LSTM & Accuracy: 97.00 & High complexity \\ \hline

Khare et al. \cite{khare2021detection,khare2021pdcnnet} & 2021 & PD Detection & EEG (35 PD and 36 CT) & 2 & 71 & Tunable Q-factor Based LSSVM, SPWVD-Based CNN & LSSVM, CNN & Accuracy: 97.70, 99.50 & High complexity \\ \hline

Loh et al. \cite{loh2021gaborpdnet} & 2021 & PD Detection & EEG (15 PD and 16 CT) & 2 & 31 & Gabor-Transform-Based 8-Layer CNN & CNN & Accuracy: 99.50 & High complexity \\ \hline

Shaban et al. \cite{shaban2021automated} & 2021 & PD Detection & EEG (15 PD and 16 CT) & 2 & 31 & 13-Layer ANN, Wavelet-Based 12-Layer CNN & ANN, CNN & Accuracy: 98.00, 99.90 & High complexity \\ \hline

Connolly et al. \cite{Connolly2015} & 2016 & PD Detection & Deep Brain Simulation-15 patients & 2 & - & Spectral & SVM with leave-one-out cross-validation & 70.00 & High complexity \\ \hline

Lensky et al. \cite{lensky2025central} & 2025 & Central-Parietal EEG channel & Rest EEG in PD dataset & 2 & 31 & CNN & Deep learning & 76.00 & Small dataset \\ \hline

Hasib et al. \cite{hasib2025early} & 2025 & PD & EEG PD Dataset & 2 & - & Time-Frequency Features & Deep learning & 94.64 & Small dataset \\ \hline

Afonso et al. \cite{afonso2025optimizing} & 2025 & PD & UC San Diego Resting State EEG Data & 2 & - & ANOVA F-value, Mutual Information, Chi-square, Linear Regression F-value & Deep learning & 97.31 & Small dataset \\ \hline

\end{tabular}
\end{adjustwidth}
\end{table*}

\subsection{Preprocessing Methods for EEG-Based PD Recognition Systems}
Preprocessing EEG data is critical to address its inherent noise and variability, ensuring reliable feature extraction and accurate classification. Filtering techniques, such as band-pass filters, isolate frequency bands (delta, theta, alpha, beta, gamma) relevant for PD detection while removing unwanted noise. Artefact removal methods like Independent Component Analysis (ICA) and Principal Component Analysis (PCA) further eliminate interference from eye blinks and muscle movements. Advanced feature extraction techniques, including Fast Fourier Transform (FFT), wavelet transforms, and tunable Q-factor wavelet transform (TQWT), provide insights into power spectral density (PSD) and time-frequency characteristics essential for detecting PD-related anomalies. Additionally, time-frequency transformation techniques like the Morlet wavelet and Gabor transforms enable the conversion of EEG signals into 2D representations, enhancing deep learning models' ability to capture both temporal and spectral information. These preprocessing steps collectively optimize EEG data for machine and deep learning applications in PD recognition.

\subsection{Existing Methodology Analysis}
Various machine learning and deep learning methods have been applied to EEG data for PD detection. Below is an overview of significant works:

\subsubsection{Machine Learning Approaches}
Few researchers have worked to develop EEG signal-based PD recognition systems. Vanegas et al. explored the use of machine learning techniques to identify Parkinson's Disease (PD) biomarkers from EEG spectral amplitudes \cite{vanegas2018machine}. Their study utilized an Extra Tree classifier, logistic regression, and decision trees to analyze EEG data from 29 PD patients and 30 healthy controls. Among the models, the Extra Tree classifier achieved the highest performance with an AUC of 99.4\%, followed by LR(94.9\%) and decision trees (86.2\%). The study highlighted the significance of specific frequency bins, particularly in the theta, alpha, and beta ranges, as critical for distinguishing PD from healthy controls. While the results were promising, the relatively small dataset limits the generalizability of their findings, necessitating further validation on larger cohorts. Koch et al. \cite{koch2019automated} employed a RF model to differentiate PD patients based on cognitive abilities, specifically classifying individuals with good versus poor cognition. Their model was trained on EEG features derived from 40 PD patients, evenly split between the two cognitive groups, and achieved an impressive AUC of 91\%. This study emphasized the potential of EEG data in assessing cognitive impairments associated with PD, providing valuable insights into non-motor symptoms of the disease. However, the small dataset and reliance on manually extracted features present challenges, limiting the scalability and robustness of their approach in broader applications. Connolly et al.  \cite{Connolly2015} focused on detecting PD using local field potentials recorded from deep brain stimulation (DBS) devices. They applied SVM, Linear Discriminant Analysis (LDA), and k-Nearest Neighbors (k-NN) models, achieving a classification accuracy of 91\% in a dataset comprising 15 advanced PD patients. By leveraging data directly linked to PD pathology, their approach provided an innovative perspective on utilizing EEG-derived features for diagnosis. Despite its high accuracy, the study's applicability is constrained by the small dataset size and the requirement for DBS recordings, which are typically available only in patients undergoing surgical treatment. This limits its potential for early-stage PD detection or widespread clinical use.

\subsubsection{Deep Learning Approaches}
Besides machine learning, some researchers have been working to develop EEG-based PD recognition using various deep learning algorithms. Oh et al. proposed a 13-layer Convolutional Neural Network (CNN) for classifying resting-state EEG data from 20 Parkinson’s Disease (PD) patients and 20 healthy controls, achieving an accuracy of 88.3\%. This approach reduced the need for manual feature extraction, demonstrating the potential of CNNs for EEG-based PD detection. However, the limited dataset size constrained the generalizability of the model \cite{oh2018deep}. Wagh et al. introduced an 8-layer Graph CNN to analyze EEG feature matrices from 1,385 patients with neurological conditions, including PD, and 208 healthy individuals. Their model achieved an AUC of 90\%, showcasing the utility of graph-based convolutional networks in EEG data analysis. Despite these promising results, the method was not specifically designed for PD detection, limiting its targeted applicability \cite{wagh2020eeg}. Shi et al. developed hybrid CNN-RNN models (2D and 3D) to classify PD patients and healthy controls, achieving accuracies of 81\% and 83\%, respectively. These models combined convolutional layers for spatial feature extraction with recurrent layers to capture temporal dependencies in EEG signals. The study demonstrated the potential of hybrid architectures, though accuracy and dataset limitations posed challenges \cite{shi2019hybrid}.
Khare et al. applied a 10-layer CNN to EEG data processed with the smoothed pseudo-Wigner Ville distribution (SPWVD), achieving accuracies of 99.9\% and 100\% on two datasets. This approach highlighted the efficacy of advanced time-frequency representations in improving EEG-based PD detection. However, its success heavily depended on high-quality preprocessing, which may not always be feasible in clinical applications \cite{khare2021detection, khare2021pdcnnet}. Loh et al. proposed an 8-layer CNN utilizing Gabor transforms to classify PD patients, achieving an accuracy of 99.5\%. The Gabor transform effectively captured relevant time-frequency features from EEG signals, contributing to the model’s high accuracy. Nonetheless, reliance on specialized transformations could present scalability challenges \cite{loh2021gaborpdnet}. Shaban et al. developed a wavelet-based 12-layer CNN and a 13-layer Artificial Neural Network (ANN) for PD classification from EEG data, achieving accuracies up to 99.9\%. The Morlet wavelet transform was instrumental in identifying meaningful EEG patterns for PD detection. While highly accurate, the computational intensity of these models may hinder their use in real-time applications \cite{shaban2021automated}.
Lee et al. employed a hybrid CNN-LSTM model to detect PD from EEG data, achieving 97\% accuracy. This architecture combined convolutional layers for spatial feature extraction with LSTM layers to model temporal dependencies, offering a comprehensive analysis of EEG signals. Despite its promising results, the complexity of hybrid models may present challenges in implementation and optimization \cite{lee2019deep}.
The reviewed studies underscore the success of deep learning models, particularly CNNs and hybrid architectures, in leveraging EEG data for PD detection. While these methods demonstrate significant promise in automating feature extraction and enhancing classification accuracy, challenges such as limited datasets, computational complexity, and real-time adaptability remain. Future research should focus on collecting diverse, large-scale datasets and developing efficient, scalable models to improve the practical application of EEG-based PD recognition systems.

 \subsection{EEG Based PD Recognition Current Limitation and Challenges}
The current EEG-based Parkinson's Disease (PD) recognition systems face several limitations:\\
\textbf{Noisy, Small and Imbalanced Datasets:} EEG signals are susceptible to noise, artefacts, and variability due to factors such as subjects' mental states, recording equipment, and environmental conditions. This variability makes it challenging to develop models that are robust and consistently perform well in different clinical environments. Many studies rely on small datasets, often with an imbalance between PD patients and healthy controls (HC). This leads to biased models that struggle to generalize to diverse populations, limiting their effectiveness in real-world settings. The absence of standardized, large-scale EEG datasets hinders the replication of results and cross-study comparisons. Variations in data collection protocols, equipment, and environmental conditions further reduce the models' generalizability.\\
\textbf{Manual Feature Extraction:} Traditional methods frequently rely on handcrafted features such as frequency bins or wavelet transformations. This manual process can miss crucial EEG patterns relevant to PD detection, adding complexity and making it difficult to compare results across studies.\\
\textbf{Model Complexity and Computational Demands:} While deep learning models (e.g., CNNs, RNNs) are powerful, they tend to be computationally expensive. These complex architectures may be unsuitable for real-time applications or resource-constrained environments. In addition, deep learning models, particularly CNNS and RNNS, are often considered "black boxes." The lack of interpretability makes it difficult for clinicians to trust and adopt these systems in real-world healthcare settings.

\subsection{EEG Based PD Recognition Future Trends and Recommendation}
Future research should focus on creating larger, more diverse datasets from multiple institutions and standardizing data collection protocols to improve reproducibility and model reliability. Collaborative data collection efforts can address limitations in small, imbalanced datasets, leading to more robust and generalizable models.

\textbf{Automated Feature Extraction and Explainable AI (XAI)}: Deep learning architectures like CNNs, RNNs, and hybrid models (e.g., CNN-LSTM) can reduce reliance on manual feature extraction, automatically learning temporal and spatial features from raw EEG data. Transfer learning and data augmentation (e.g., noise injection) can overcome dataset limitations. Advanced signal processing techniques like time-frequency analysis and wavelet transforms will enhance model performance. Explainable AI will be essential for transparency, helping clinicians trust and interpret model predictions.

\textbf{Multiband Decomposition with Temporal-Spectral Approaches}: EEG is a low-amplitude signal with high noise and artefacts. The main challenge of EEG-based Parkinson's disease (PD) detection is removing artefacts such as eye blinks and other movement-related signals from the actual signal. In this case, decomposing the signal into multiple bands and then extracting temporal-spectral features from each band can help in identifying effective features. Band-wise filtering can also be useful in removing artefacts. Combining EEG with other modalities, such as gait, handwriting, or voice data, can improve diagnostic accuracy and enable earlier, more precise detection.

\textbf{Cross-Population Generalization and Real-Time Monitoring}: Models should generalize across populations considering age, gender, medication, and disease severity. Mobile and wearable EEG devices will enable real-time monitoring, facilitating continuous tracking and timely interventions for PD patients.

\section{Other Single Modalities Based PD Recognition}
In addition to the more common modalities such as MRI and gait-based analysis, several other single modalities play a crucial role in Parkinson’s Disease (PD) recognition and diagnosis. One such modality is eye movement analysis, which has shown promise in assessing cognitive function and disease severity in PD patients, as shown in Table \ref{tab:other_single_data_modality_performance}. Traditionally, infrared eye-tracking devices were required for such analyses, limiting their widespread use due to cost and technical limitations. However, recent advancements have made eye-tracking more accessible. A novel approach using the embedded camera of an iPad Pro for gaze tracking (Eye-Tracking Neurological Assessment - ETNA™) has demonstrated its potential in providing precise quantification of various eye movement parameters such as saccade velocity, latency, and accuracy. Studies show that combining these eye movement features with machine learning classifiers, such as support vector classifiers (SVC), yields high accuracy (90\%) in distinguishing between mild and moderate PD stages, highlighting the growing potential of mobile eye-tracking technology in clinical settings \cite{koch2024eye}.
Another important modality is electromyography (EMG), particularly surface EMG (sEMG) for hand gesture recognition, which plays a vital role in both diagnosing and assisting PD patients. EMG signals provide direct insights into muscle activity during movement, which can be especially useful for patients with motor symptoms such as tremors or rigidity. By analyzing sEMG signals, hand gestures can be predicted with remarkable accuracy. Recent studies have shown that by using spectrogram images generated via Short-Time Fourier Transform (STFT) and training models like a 50-layer CNN (ResNet), accuracy rates of up to 99.59\% in classifying seven different hand gestures have been achieved \cite{ozdemir2020emg}. This ability to recognize and predict hand gestures is invaluable, particularly in assistive technologies like prostheses and for enhancing human-computer interactions. Some researchers have also applied facial expressions for PD recognition, like Zhang et al. \cite{zhang2025parkinson}, who collected a dataset from 154 people and then calculated AU intensities, AU derivative features, and applied XGBoost and Random Forest, which showed AUC values of 81.39\% and 83.7\%, respectively. Both eye movement and EMG-based systems have emerged as powerful tools, providing new insights into PD progression and offering significant advancements for early and accurate diagnosis of the disease.

\begin{table*}[htp!]
\centering
\caption{Methodological review with the Other Single data modality}
\label{tab:other_single_data_modality_performance}
\begin{tabular}{|p{2cm}|p{1cm}|p{2cm}|p{2cm}|p{1cm}|p{1cm}|p{2.5cm}|p{2cm}|p{1.5cm}|}
\hline

Author    &Year       & Dataset Name & Modality Names & Class & Sample & Feature Extraction & Classifier & Performance Acc \\ \hline
Ryan et al. \cite{ryan2025integrative} & 2024 & PPMI dataset & Behavioral, Clinical & PD, PL or HC & - & DNAm Signatures, SNPs & Flexible Data Integration (Longitudinal and Cross-sectional) & 95.8 (3year) \\ \hline
Smith et al. \cite{Smith2015} &2016 &  49 PD patients and 41 age-matched healthy controls   & Genetics & -  & -     & Motor Function  & Evolutionary Algorithms & 78.00   \\ \hline
Cook et al. \cite{Cook2015} &2015 &  Behaviour   & Behavior & -  & -     & Daily Activity  & Mannual Decision & -   \\ \hline
Shamir et al. \cite{Shamir2015} &2015 &  10 patients and 89 post-DBS    & Behavior & -  & -     & CDSS and patient-specific details on both stimulation and medication.   &  SVM, NB, RF  & 71.00, 64.00, 64.00 \\ \hline
Bot et al. \cite{bot2016mpower} & 2016 & mPower database & Smartphone Sensor Signal & CN, PD, and the PD severity & 8 channel & 898 features & - & -  \\ \hline

Jiang et al. \cite{jiang2025diagnosis}  & 2016 & - & Eye movements Signal & 2 & 239 & Eye-Movement Features & ECA-CNN \cite{miah2024sensor} & 92.73 \\ \hline 
Kock et al. \cite{koch2024eye} & 2024 & Oculomotor PD Dataset & Eye movement tablet & Mild PD, Moderate PD & 50 & Oculomotor Parameters (e.g., Trail Making Test) & SVC & 90.00 \\ \hline

Ozdem et al. \cite{ozdemir2020emg} & 2024 & Hand Gesture & sEMG & 7 Hand Gestures & 30 & Spectrogram Images (STFT) & 50-layer CNN (ResNet) & 99.59  \\ \hline

Kumar et al. \cite{kumar2025parkin} & 2024 & EMG and Accelerometer PD Dataset & EMG, Accelerometer & PD, Non-PD & - & Statistical Features from EMG and Accelerometer & Decision Tree, Random Forest & 93.00, 98.00 \\ \hline
Aborageh et al. \cite{aborageh2025predicting} & 2024 & UK Biobank and PPMI Cohort Dataset & Genetic, Comorbidities, Lifestyle, Environmental Factors & PD, PDD & - & Genetic Predisposition, Comorbidity Interactions, MR & SHAP, NB Network Structure Learning & - \\ \hline

Baba et al. \cite{babaalidata} & 2024 & NeuroQWERTY PD Dataset & Keystroke Dynamics & Early PD, De-novo PD & - & Features (DWT, DTW), Keystroke Metrics & XGB, Self-Attention  & 86.00 ROC-AUC (1st), 91.0 ROC-AUC (2nd) \\ \hline

Droby et al. \cite{droby2025radiological} & 2025 & PPMI SPECT PD Dataset & SPECT Images & PD, CN & - & Gaussian Noise, Feature Representation & CNN, Reconstruction Sub-networks & 97.14 \\ \hline

Momeni et al. \cite{10879429} & 2025 & Mobile Voice PD Dataset & Voice Features & PD, Healthy & - & Voiced and Unvoiced Segments, Pitch Variation, Spectral Flux & Group-Wise Scaling, SHAP & 82.00 \\ \hline

Zhang et al. \cite{zhang2025parkinson} & 2025 & Hypomimia PD Dataset & Faical Video& PD, Healthy & 154 & AUs Intensities, AUs Derivatives & XGBoost, Random Forest & 81.39, 83.7 AUC \\ \hline
\end{tabular}
\end{table*}

\section{Multi Modal Dataset Based PD Recognition System} \label{sec:multimodal}
In the previous section, we analyzed various data modalities used in Parkinson's Disease (PD) recognition systems, including MRI, video, sensor, handwriting, speech, and EEG data. Additionally, we reviewed existing PD detection studies that incorporated behavioural and genetic data, as well as those that combined two or more of these modalities.
In this study, we aimed to cover as many diverse modality-based PD recognition systems as possible to contribute to the broader PD research landscape. This section focuses on multimodal datasets used for PD recognition systems, along with an exploration of behavioural and genetic data modalities to provide a comprehensive understanding of our research work.
The following subsections discuss the preprocessing of these datasets and the state-of-the-art methodologies used for PD recognition, both for multimodal and single-modal datasets.

\begin{table*}[h]
\centering
\caption{List of \textbf{multimodal} databases for PD.}
\label{tab:multimodal_PD_datasets}
\renewcommand{\arraystretch}{1.2}
\begin{tabular}{|p{1.5cm}|p{.7cm}|p{1cm}|p{3.5cm}|p{9.6cm}|}
\hline
\textbf{Reference} & \textbf{Year} & \textbf{Size} & \textbf{Acquisition Device} & \textbf{ Description of Modalities} \\
\hline
Barth et al. \cite{barth2012combined} & 2012 & 18 PD \newline 17 HC & 
- BiSP \newline
- 3D gyroscopes and 3D accelerometers & 
\textbf{Gait:} 10-meter walk, Heel-toe tapping, Circling \newline
\textbf{Handwriting:} Drawing twelve circles at the same place, Tracing four preprinted spirals/meanders \newline
\textbf{Other:} Drawing circles in air around virtual point, Pronation/supination (20s), Finger tapping (20s) \\
\hline
Vásquez-Correa et al. \cite{vasquez2019multimodal} & 2019 & 44 PD \newline 40 HC & 
- Wacom Cintiq 13-HD \newline
- eGaIT system (3D gyroscopes and accelerometers) & 
\textbf{Gait:} 20m walk with stop after 10m, 40m walk with stop every 10m \newline
\textbf{Handwriting:} Circle, cube, rectangles, house, diamond, Rey-Osterrieth figure, spiral with/without template \newline
\textbf{Speech:} Repetition of syllables (/pa-ta-ka/, etc.), Sentence reading, story, monologue \\
\hline
Prashanth et al. \cite{prashant_high_early_PD_detection_ppmi_dataset} & 2016 & 401 PD \newline 183 HC & - & 
\textbf{Other:} University of Pennsylvania Smell Identification Test (UPSIT), REM sleep Behavior Disorder Screening Questionnaire (RBDSQ) \\
\hline
PD-Posture-Gait Dataset \cite{physionet140} & 2024 & 14 HC & 
- Dual RGB-D cameras on robotic walker \newline
- Inertial-based motion capture system & 
\textbf{Gait:} Walking at 3 speeds in 3 scenarios/locations, 30 fps \newline
\textbf{Other:} Skeleton joint data as ground truth \newline
Supports pose estimation, tracking, forecasting, and posture analysis \\
\hline
REMAP Dataset \cite{morgan2023multimodal} & 2024 & 12 PD \newline 12 HC & 
- Full-body skeleton data \newline
- Bilateral wrist-worn accelerometers & 
\textbf{Gait:} Sit-to-Stand (STS), turning, non-turning actions \newline
\textbf{Other:} Clinical rating scores (range-based), demographic data \newline
Controlled and open-access pseudonymized datasets available \\
\hline
Multimodal-FOG Dataset \cite{li2021} & 2024 & 12 PD & 
- EEG, EMG, ECG, SC, ACC sensors \newline
- Integrated commercial and self-designed hardware & 
\textbf{Gait:} Walking tasks in hospital settings designed to induce FoG \newline
\textbf{Other:} Physiological data (EEG, EMG, ECG, SC), annotated FoG episodes by physicians \newline
3 hours 42 minutes of valid multimodal recordings \\
\hline
Multi-PD-FOG Dataset \cite{souza2022} & 2024 & 35 PD & 
- Inertial Measurement Units (128 Hz) \newline
- Commercial digital camera (Sony, 30 Hz) & 
\textbf{Gait}: Video and IMU (triaxial acceleration and angular velocity) recordings during turning-in-place tasks. Includes clinical scales (PD severity, number/duration of FOG episodes), medication state, and demographic info. \\ \hline

Multi-Symptom Dataset \cite{li2024} & 2021 & 35 PD & 
- Inertial Measurement Units (128 Hz) \newline
- Digital camera (Sony, 30 Hz) & 
Video and IMU data (triaxial acceleration and angular velocity) were recorded during turning-in-place tasks for individuals with PD in the ON medication state. This includes demographic details, clinical scales, PD severity, the number and duration of FOG episodes, and medication state. \\ \hline
Multimodal-medical Dataset \cite{xue2024} & 2024 & 51,269 & 
- MRI scanners \newline
- Clinical and neurological exam sources & 
Individual-level demographics, medical history, neuropsychological tests, physical/neurological exams, lab results, medications, and multisequence MRI scans. Aggregated from 9 cohorts (ADNI, PPMI, OASIS, etc.). Includes 19,849 NC, 9,357 MCI, and 22,063 dementia participants. \\ \hline
 \end{tabular}
\end{table*}

\begin{table*}[htp!]
\centering
\caption{Methodological review with the Multimodal Dataset}
\label{tab:EMG_data_performance}
\begin{tabular}{|p{2cm}|p{.7cm}|p{2cm}|p{3cm}|p{1cm}|p{2cm}|p{2cm}|p{1cm}|}

\hline
Author & Year & Dataset Name & Modality Names &  Sample & Feature Extraction & Classifier & Performance \\ \hline
Rao et al. \cite{rao2020parkinson_voic_spiral} & 2024 & Multimodal & Voice, spiral, Handwriting & 40 & Median  & Majority Voting & 90-95 \\ \hline
Sadhu et al. \cite{sadhu2022telehealth} & 2022 & - & Handset Signal & 4-PD, 5-HC & Speed, Ampl, Hesitation, Halts & KNN, SVM, DT & 93 \\ \hline
Gil-Mart{\'i}n et al. \cite{gilmartin2019parkinsons} & 2019 & - & Handset Signal & 62-PD, 15-HC & Coord, Pressure, and Grip Angle & CNN & 96.5 \\ \hline
Zhao et al. \cite{zhao2021multimodal} & 2022 & - & Gait Signal & 93-PD, 73-HC & Spatio-temporal  & HMM & 98.93 \\ \hline
Zhao et al. \cite{zhao2022associated} & 2022 & - & Gait Signal & 93-PD, 73-HC & Spatio-temporal  & ASTCapsNet & 97.31 \\ \hline
Mohaghegh and Gascon \cite{mohaghegh2021identifying} & 2021 & - & Handwriting and Voice & 81-PD, 85-HC & Spatial Features and MFCC & DeiT, AST & 92.37 \\ \hline
Ma et al. \cite{ma2021deep} & 2021 & - & Speech & 34-PD, 20-HC & Original and Deep Features & Deep Ensemble model & 99.67 \\ \hline
PDMultiMC \cite{tsanas2011nonlinear_data_PDMultiMC} & 2018 & PDMul \newline tiMC & Multimodal Clinical data & 32 & 32 & - & 96.87 \cite{islam2024review_sruvey_handwriting_voice_data} \\ \hline
Li et al. \cite{li2024}     & 2024  & Audio, motion and facial data & Image, signal & - & - & STN+RGA & 95.08 \\ \hline
Zhao et al. \cite{zhao2021multimodal} & 2022  & Gait & VGRF + time series &  - & - & CorrMNN & 99.31 \\ \hline
Zhao et al. \cite{zhao2022associated}    & 2022  & Gait & Images, skeleton, force signal &  - & - & ASTCapsNet & 97.31 \\ \hline
Huo et al. \cite{Huo2020}     & 2020  & Bradykinesia, rigidity and tremor & Force signal, IMU, MMU &  - & - & Voting Classifier & 85.40 \\ \hline
Wang et al. \cite{Wang2021hierarchical} & 2021  & Bradykinesia and tremor & Accelerometer and gyroscope data &  - & - & HMMs & 89.28 \\ \hline
Junaid et al. \cite{Junaid2023}  & 2023  & Motor and non-motor function data & Time-series, non time-series & - & - & LGBM, RF & 94.89 \\ \hline
Pahuja et al. \cite{Pahuja2022}  & 2022  & MRI and CSF & Neuroimaging and biological features &  - & - & CNN & 93.33 \\ \hline
Cheriet et al. \cite{cheriet2023multi} & 2023  & Gait & Videos, time series & - & - & Multi-speed transformer network & 96.90 \\ \hline
Li et al. \cite{Li2023cnn}   & 2023  & Gait & Inertial, stride, and pressure data &  - & - & CNN-BiLSTM & 98.89 \\ \hline
Faiem et al. \cite{Faiem2024}   & 2024  & Gait & GRF + time series & - &  - & Cross-attention, Transformer & 97.30 \\ \hline
Wang et al. \cite{Wang2024freezing} & 2024  & Freezing of Gait & Acceleration of ankle, thigh, and trunk &  - & - & MCT-Net & 96.21 \\ \hline
Xue et al. \cite{xue2024}     & 2024  & Multimodal-medical-data & Medical history, medication use, neuropsychological assessments, functional evaluations and multimodal neuroimaging &  - & - & 3D transformer-based Models & 96 (AUC) \\ \hline
Shanmugam et al. \cite{shanmugam2025hybrid} & 2025 & Multimodal & Voice Signals, Hand-Drawn Spiral Images &  - & Voice Features (Gaussian Filter), Spiral Image Features (Bilateral Filter, Augmentation) & ZFNet-LHO-DRN (Majority Voting) & 89.8 \\ \hline
Archila et al. \cite{archila2025riemannian} & 2025 & Gait and Eye Motion Video & Gait, Ocular Pursuit Motion & 32 & Frame-Level Convolutional Features, Riemannian Means & Geometrical Riemannian Neural Network & 96.00 \\ \hline
Yang et al. \cite{yang2025multimodal} & 2024 & FOG Detection Dataset & IMU, GSR, ECG & - &  IMU, GSR, ECG Features & Transformer-based Model, XGBoost, Temporal Convolutional Network &  77.10 \\ 
\hline
\end{tabular}
\end{table*}

\subsection{Dataset Description}
In the domain of Parkinson's Disease (PD) recognition, various datasets have been utilized, each containing multimodal data to enhance the diagnostic accuracy. Table \ref{tab:multimodal_PD_datasets} provides an overview of several datasets and \ref{tab:EMG_data_performance} shows the methodology with various dataset with lastes performance accuracy. For instance, the PDMultiMC dataset \cite{tsanas2011nonlinear_data_PDMultiMC} incorporates multimodal clinical data from 32 subjects (PD and CN), offering a variety of features for both PD detection and monitoring. Another example is the mPower database \cite{bot2016mpower}, which collects data from smartphone sensors, capturing signals related to motor and cognitive function from PD patients. The dataset also includes patient severity levels to offer a more granular understanding of disease progression. Additionally, the HandPDMultiMC dataset \cite{tsanas2011nonlinear_data_PDMultiMC} further expands on multimodal clinical data, targeting hand motor functions.
Other datasets, such as the ones used by Smith et al. \cite{Smith2015} and Cook et al. \cite{Cook2015}, include genetic and behavioural data. These datasets are unique in that they focus on both motor function and daily activities, respectively, offering broader insight into PD symptomatology beyond traditional clinical measurements.
\subsubsection{Behavioral}
Cook et al. \cite{Cook2015} explored using smart home and machine learning technologies to monitor behavioral changes in Parkinson's Disease (PD) patients. Their focus was on aiding clinical assessment and distinguishing healthy older adults (HOA) from those with cognitive or physical impairments, including mild cognitive impairment (MCI). The results showed that smart homes and wearable devices effectively monitored PD patient activity and highlighted differences between HOAs, PD, and MCI. However, limitations arose with multi-resident settings and interrupted activities.
Shamir et al. \cite{Shamir2015} introduced a Clinical Decision Support System (CDSS) to integrate patient symptoms and medications for three key functions: information retrieval, treatment visualization, and dosage recommendations. Using NB, SVM, and RF, they predicted 86\% of motor improvements one year post-surgery.

\subsection{Preprocessing}
Preprocessing plays a crucial role in multimodal PD recognition systems, as it helps clean, normalize, and structure the raw data for optimal analysis. For EEG and genetic data, preprocessing often involves artefact removal, normalization, and feature extraction. In the case of behavioural data, noise filtering, activity segmentation, and signal enhancement techniques are frequently employed.
For example, Shamir et al. \cite{Shamir2015} used a Clinical Decision Support System (CDSS) to process patient-specific symptoms and medication details. The preprocessing step in their approach incorporated data retrieval, cleaning, and structuring to create visualizations and provide treatment recommendations. Similarly, the smartphone sensor data from mPower \cite{bot2016mpower} require normalization and feature extraction from raw sensor outputs to transform signals into meaningful attributes, such as motor function metrics or cognitive markers.
In genetics-based datasets, such as the one used by Smith et al. \cite{Smith2015}, evolutionary algorithms are applied during preprocessing to identify genetic markers relevant to PD. This preprocessing step includes filtering out irrelevant genetic variations and focusing on key biomarkers that correlate strongly with the disease.

\subsection{Multimodal Dataset-Based PD Recognition Using ML Approach}
Machine learning techniques have been widely applied to multimodal datasets for Parkinson's Disease recognition. The combination of different data types—such as genetics, behaviour, and sensor data—offers a more holistic view of PD symptoms, enabling more accurate classification and detection. For example, Smith et al. \cite{Smith2015} employed evolutionary algorithms to classify genetic data collected from commercial sensors. The algorithms achieved a performance accuracy of 78\% when distinguishing between PD patients and healthy controls. Similarly, Cook et al. \cite{Cook2015} used behavioural data and manual decision methods to observe daily activity changes in PD patients, enhancing the understanding of PD progression in older adults.
Shamir et al. \cite{Shamir2015} applied machine learning algorithms, including Support Vector Machines (SVM), Naïve Bayes (NB), and RF, to multimodal clinical data in their CDSS framework. The classifiers achieved varying performance results, with SVM yielding the highest accuracy of 71\% for predicting treatment outcomes.
The PDMultiMC dataset \cite{tsanas2011nonlinear_data_PDMultiMC}, incorporating multimodal clinical data from 32 subjects, achieved an impressive accuracy of 96.87\% using traditional machine learning classifiers, demonstrating the power of multimodal data for PD diagnosis.
Rao et al. (2024) proposed a multimodal approach for Parkinson's Disease (PD) detection by combining voice, spiral, and handwriting data. The study involved 40 participants, classified as either PD or healthy controls (CN). They used a normalization technique by calculating the median of normalized values (0-1) for the input data. The classification was performed using a majority voting method, achieving an accuracy range between 90\% and 95\%. This approach highlights the effectiveness of leveraging multiple modalities to enhance PD detection accuracy.

\subsection{Multimodal Dataset-Based PD Recognition Using Deep Learning}
Deep learning techniques have also been applied to multimodal PD recognition systems, particularly when dealing with complex, high-dimensional data like EEG or sensor signals. These models can automatically learn feature representations from raw data, eliminating the need for manual feature extraction and enhancing model performance.
For example, the mPower dataset \cite{bot2016mpower}, containing smartphone sensor signals, has been a key focus for deep learning research. Deep learning architectures such as Convolutional Neural Networks (CNNs) and Long Short-Term Memory (LSTM) networks have been used to capture spatial and temporal dependencies in the sensor data. These models have proven effective in predicting PD severity, although the lack of feature extraction remains a challenge. The study by Shamir et al. \cite{Shamir2015} implemented a deep learning-based Clinical Decision Support System (CDSS), using multiple machine learning models (SVM, NB, and RF) to predict treatment outcomes for PD patients. The deep learning component allowed the system to incorporate a wide range of patient-specific symptoms and medication data, ultimately achieving high predictive performance. Multimodal clinical datasets, like PDMultiMC \cite{tsanas2011nonlinear_data_PDMultiMC}, further emphasize the importance of using deep learning models to process and classify data. These datasets, which combine genetic, behavioural, and clinical data, present opportunities for deep learning models to extract richer and more complex patterns for PD diagnosis, achieving accuracies up to 96.87\%.  Both machine learning and deep learning techniques have demonstrated significant potential in utilizing multimodal datasets for Parkinson's Disease recognition. The combination of genetic, behavioural, and sensor data provides a comprehensive understanding of the disease, paving the way for more accurate and reliable diagnostic tools.

\subsection{Challenges and Future Direction}
Multimodal PD recognition systems face significant challenges, including managing data variability across different sensors, synchronizing data streams, and addressing computational complexity. Variability in sensor data can impact the detection of subtle PD symptoms such as tremor, rigidity, bradykinesia, and postural instability. For instance, tremor-related speech fluctuations may be hard to detect without complementary motion sensors, and rigidity may not significantly affect speech but can alter voice modulation. Additionally, speech patterns caused by bradykinesia, such as slowness or monotony, must be paired with motion data (e.g., gait analysis) to provide a complete assessment.
Synchronizing these diverse sensor streams is critical to capture a holistic picture of a patient's condition. The computational complexity of processing and integrating multiple data sources can hinder real-time applications. 
To address these challenges, future research should focus on developing deep learning (DL) architectures capable of handling data variability across sensors and synchronizing streams efficiently. Advanced integration techniques are needed to enhance recognition accuracy and system robustness, particularly for multimodal systems that combine speech, motion, and other physiological signals.
Additionally, the ability to integrate speech data with motion sensor data for symptoms like tremor, rigidity, and postural stability will improve diagnostic accuracy and the detection of subtle changes in symptoms. For example, combining speech data with motion sensors can provide a more comprehensive understanding of tremor severity, while combining speech patterns with gait analysis can enhance bradykinesia detection. Future work should also focus on overcoming the limitations of small and imbalanced datasets by leveraging transfer learning and data augmentation strategies, ultimately improving model generalization and clinical applicability.
By developing more efficient algorithms and integrating multimodal data streams, future systems can provide a deeper and more accurate understanding of PD symptoms, aiding in earlier diagnosis, better disease tracking, and personalized treatment.

\section{Discussion}
In this study, we review the current advancements in PD (PD) across diverse data modalities. We present a comprehensive framework that addresses key elements of feature extraction using ML, as well as end-to-end DL modules. Each data modality is explored in detail, including primary datasets, preprocessing techniques, proposed architectures, fusion methodologies, state-of-the-art performance, and the challenges and future trends specific to each modality. For dynamic PD, we emphasize the importance of managing temporal contextual features. This section aims to guide future research by highlighting the challenges and potential directions involving 3D DL models. We evaluate the associated problems, complexities, and the need for practical solutions within the field, providing a clear pathway for future advancements.

We identify and discuss new research gaps and offer guidance to overcome challenges in next-generation PD technologies. Multimodal datasets, which integrate hand-crafted features with new CNN features combined with RGB, skeleton, and depth-based information, are examined for their efficacy in improving gesture recognition accuracy. In the skeleton modality section, we focus on GCNs and their application to large-scale, real-time PD. This has become a focal point of the research community due to its potential to solve significant challenges. We highlight the need for gesture localization within realistic, uncut, and extended videos, predicting that emerging challenges such as early recognition, multi-task learning, gesture captioning, recognition from low-resolution sequences, and life log devices will gain increased attention in the coming years. Our discussion underscores the necessity of addressing these challenges to advance PD systems, with a particular emphasis on multimodal approaches and the integration of new technologies. By providing a detailed overview of the current state and future directions, we hope this study serves as a guideline for researchers and practitioners in the field of PD.

\subsection{Diverse Modality-Based PD Recognition: Challenges and Limitations}
This paper has reviewed the state-of-the-art machine and deep learning techniques for Parkinson’s Disease (PD) diagnosis with diverse domains, including MRI, Video, Sensor, Handwriting, Voice, EEG, and Multimodality datasets, which mainly involve screening, staging, and biomarker identification. While significant progress has been made, particularly in the classification of subjects as PD-positive or healthy, several challenges and limitations persist in this field, hindering broader clinical adoption and impact.
First, the size and availability of datasets remain a critical bottleneck. Most existing datasets, including EEG signals, MRI scans, handwriting samples, and speech recordings, are limited in size, which restricts the ability of machine and deep learning models to generalize effectively. Insufficient data increases the risk of overfitting and reduces the robustness of the models. Second, many of the datasets used in current research are not commonly employed by clinicians for PD diagnosis and staging. Established clinical tools, such as Unified Parkinson’s Disease Rating Scale (UPDRS) scores. This mismatch diminishes the clinical relevance of the proposed approaches.
Another significant challenge is the lack of explainability in current deep learning models. While these methods achieve high accuracy, they often function as "black boxes," making it difficult to interpret or visualize the features contributing to their decisions. This lack of transparency reduces trust among medical professionals, who require detailed explanations for diagnoses and classifications.
Moreover, the quality of annotations provided by medical experts is crucial for supervised deep-learning models. However, obtaining accurate, high-quality annotations for large, multimodal datasets is often cost-prohibitive and time-intensive, further limiting the development of robust models.
The integration of multimodal data presents both an opportunity and a challenge. While combining data from different modalities, such as speech, motion, and EEG signals, has the potential to enhance diagnostic accuracy, achieving effective fusion of predictions from different models remains a complex task. Clinicians often rely on multiple diagnostic markers, and single-modality approaches may fail to capture the full spectrum of PD symptoms, reducing their practical utility. Finally, many available datasets are collected from patients already exhibiting clinical symptoms, limiting the ability of models to detect early-stage PD. This highlights the need for new datasets and modalities capable of identifying early biomarkers before significant neurodegeneration occurs. Collaboration between artificial intelligence researchers and medical professionals is essential to identify and validate such early indicators.
Despite these challenges, the integration of multimodal data, increased dataset availability, and advancements in explainable AI hold promise for overcoming these limitations, paving the way for more effective and reliable PD diagnosis systems.

\subsection{Diverse Modality-Based PD Recognition: Future Trends} Parkinson’s Disease (PD) recognition and assessment could greatly benefit from methods that align with how doctors make decisions, by observing tangible indicators. For instance, gait and body movement-based PD recognition, combined with scoring systems, offer impactful societal applications. Utilizing multi-camera setups instead of a single camera can address information loss during movement recording. Recent advancements like Opensim and OpenCap Explorer, commonly used in sports for camera calibration, could be adapted for multi-camera calibration in PD recognition. Additionally, integrating cameras with other sensors could replace foot sensors, which occasionally face reliability issues.
Movement-based assessments, such as finger tapping, hand movements, supination-pronation, postural stability, and actions like rising from a chair, could be captured using RGB cameras, EMG sensors, or EEG sensors. These methods could provide comprehensive data for PD diagnosis. For MRI-based assessments, current slicing methods risk losing information. Implementing 3D MRI classification, where the entire MRI scan is treated as a single sample, could retain voxel-level details, enhancing diagnostic accuracy. Furthermore, sleep disorders—a prevalent non-motor symptom of PD—present unique diagnostic opportunities. Changes in REM and NREM sleep, such as altered spectral power and baseline activity, are promising biomarkers for early cognitive impairment detection. Integrating modalities like DaTscan and sleep EEG with advanced AI techniques, including GradCAM, integrated gradients, and XRAI, could improve biomarker identification and disease progression understanding. Synthetic data generation and augmentation could overcome dataset limitations while ensuring privacy. For practical clinical adoption, AI models must be validated in real-world settings with clinician feedback. Beyond classification, AI should contribute to identifying biomarkers, progression risk factors, and treatment efficacy, offering transformative potential for personalized PD management.

\section{Conclusion} \label{sect8}
Parkinson's Disease (PD) remains a significant global health challenge, with millions affected worldwide. Early and accurate diagnosis is critical for slowing disease progression and improving the quality of life for patients and caregivers. This comprehensive review highlights the advancements in machine and deep learning techniques for PD diagnosis, staging, and biomarker identification from 2016 to 2024. The review spans various data modalities, including MRI, gait-based sensory and pose data, handwriting analysis, speech tests, EEG, and multimodal fusion approaches. The studies demonstrate significant progress in the application of artificial intelligence, particularly in the use of hand-crafted features and deep learning models, which have improved diagnostic accuracy. However, notable gaps remain, especially in the continuous and real-time recognition of PD symptoms. Multimodal approaches show great potential in addressing these limitations, combining diverse data sources to provide more comprehensive and reliable assessments. Despite these advancements, challenges such as small datasets, variability in data quality, and the complexity of multimodal data integration remain. Future research should focus on overcoming these challenges by developing robust, scalable, and generalizable models. Furthermore, the integration of real-time data collection and processing could significantly enhance the utility of PD recognition systems. This review serves as a foundation for future work in developing more effective and practical AI-driven tools for PD diagnosis and monitoring, guiding the field towards better healthcare solutions for Parkinson's patients.


\section*{ABBREVIATIONS}
\begin{table}[H]
\label{Appendix_A}
\setlength{\tabcolsep}{3pt}
\begin{tabular}{ll}
MLP& Multilayer Perceptron\\
MR& Mendelian Randomization \\
AUs& Facial Action Units\\
COA& Escape Coati Optimization Algorithm \\
OPF & Optimum Path Forest \\
ENSM & Ensemble Method\\
PPMI &Parkinson’s Progression Markers Initiative \\
SMMKL & Soft margin multiple kernel learning \\
Q-BTDNN& Q-Backpropagated time \\ delay neural network \\
CASIA &Institute of Automation, \\ Chinese Academy of Sciences \\
TD & tremor-dominant\\
PIGD& postural instability gait difficulty\\
KFD & kernel Fisher Discriminant \\
HCI & Human-Computer Interfacing \\
BCI & Brain-Computer Interface \\
EEG & Electroencephalography \\
MEG & Magnetoencephalogram \\
RQ & Research Question \\ 
PD & Parkinson Disease Recognition \\
CNN & Convolutional Neural Network\\
ADDSL & Annotated Dataset for Danish Sign Language\\
ML & Machine Learning\\
DL & Deep Learning\\
HMM & Hidden Markov Model \\
HOG & Histogram of Oriented Gradient\\
PCA & Principal Component Analysis\\
CRNN & Convolutional Recurrent Neural Network\\
LSTM & Long Short-Term Memory \\
Bi-LSTM & Bidirectional Long Short-Term Memory \\
SVM & Support Vector Machine\\
RTDPDS & Real-Time Dynamic PD System \\
SMKD &Self-Mutual Knowledge Distillation \\
CTC&connectionist temporal classification \\
MUD& Massey University Dataset \\
SSC-DNN & Spotted Hyena-based Sine Cosine \\ & Chimp Optimization Algorithm \\ & with Deep Neural Network\\
DMD & Dynamic Mode  Decomposition \\
MsMHA-VTN & Multiscaled Multi-Head \\ & Attention Video Transformer  Network \\
HSL&Hongkong Sign Language \\
FPHA & First Person Hand Action\\
STr-GCN& Spatial Graph Convolutional \\ & Network and Transformer \\ &
Graph Encoder for 3D PD\\
MF-HAN& Multimodal Fusion \\ &Hierarchical Self-Attention Network\\
SMLT& Simultaneous Multi-Loss Training \\
ResGCNeXt&Efficient Graph Convolution Network\\
HDCAM & Hierarchical Depth-wise \\ & Convolution and Attention Mechanism\\
\end{tabular}
\end{table}
\newpage
\bibliographystyle{unsrt}
\bibliography{reference}

\begin{IEEEbiography}[{\includegraphics[width=1in,height=1.25in,clip,keepaspectratio]{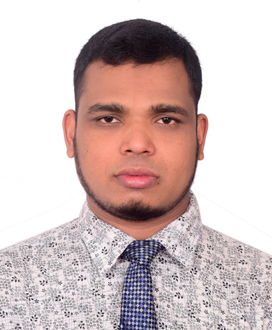}}]{Abu Saleh Musa Miah} received the B.Sc.Engg. and M.Sc.Engg. degrees in computer science and engineering from the Department of Computer Science and Engineering, University of Rajshahi, Rajshahi-6205, Bangladesh, in 2014 and 2015, respectively, achieving the first merit position. He received his Ph.D. in computer science and engineering from the University of Aizu, Japan, in 2024, under a scholarship from the Japanese government (MEXT). He assumed the positions of Lecturer and Assistant Professor at the Department of Computer Science and Engineering, Bangladesh Army University of Science and Technology (BAUST), Saidpur, Bangladesh, in 2018 and 2021, respectively. Currently, he has been working as a visiting researcher (postdoc) at the University of Aizu since April 1, 2024. His research interests include AI, ML, DL, Human Activity Recognition (HCR), Hand Gesture Recognition (HGR), Movement Disorder Detection, Parkinson's Disease (PD), HCI, BCI, and Neurological Disorder Detection. He has authored and co-authored more than 50 publications in widely cited journals and conferences.
\end{IEEEbiography}

\begin{IEEEbiography}[{\includegraphics[width=1in,height=1.25in, clip,keepaspectratio]{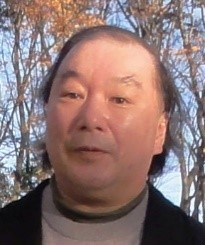}}]
{TARO SUZUKI (Member, ACM)} received the B.S. degree from the Science University of Tokyo in 1987, the M.E. degree from the University of Tsukuba in 1993, and the D.S. degree from the Tokyo University in 1998. He was an engineer at Mitsubishi Electric Toubu Computer Company during 1987-1991. He was a Research Associate at the Institute of Electronics and Information Science and TARA center, University of Tsukuba during 1995-1998, at School of Information Science, JAIST until 2000, and at Research Institute of Electric Communication, Tohoku University until 2001, respectively. He was an Assistant Professor at the School of Computer Science and Engineering, the University of Aizu until 2006, where he was an Associate Professor until 2012, where he has been a Senior Associate Professor since 2013. His research interests include language theory with automata, functional programming, formal methods, and foundation of deep learning and reinforcement learning.
\end{IEEEbiography}

\begin{IEEEbiography}[{\includegraphics[width=1in,height=1.25in, clip,keepaspectratio]{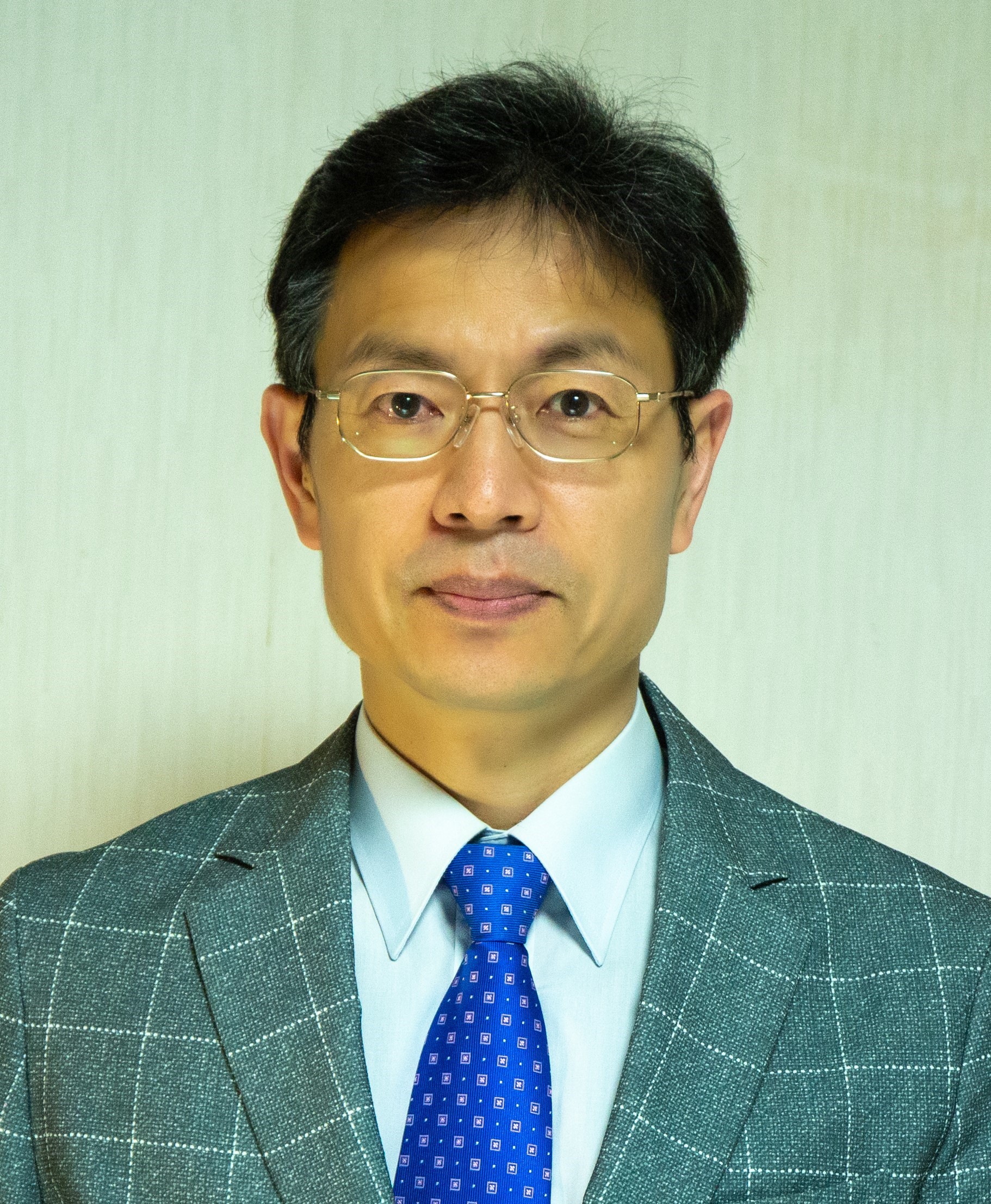}}]
{Jungpil Shin} (Senior Member, IEEE) received the B.Sc. degree in computer science and statistics and the M.Sc. degree in computer science from Pusan National University, South Korea, in 1990 and 1994, respectively, and the Ph.D. degree in computer science and communication engineering from Kyushu University, Japan, in 1999, under a scholarship from the Japanese Government (MEXT). He was an Associate Professor, a Senior Associate Professor, and a Full Professor with the School of Computer Science and Engineering, The University of Aizu, Japan, in 1999, 2004, and 2019, respectively. He has co-authored more than 420 published papers for widely cited journals and conferences. His research interests include pattern recognition, image processing, computer vision, machine learning, human–computer interaction, non-touch interfaces, human gesture recognition, automatic control, Parkinson’s disease diagnosis, ADHD diagnosis, user authentication, machine intelligence, bioinformatics, and handwriting analysis, recognition, and synthesis. He is a member of ACM, IEICE, IPSJ, KISS, and KIPS. He serves as an Editorial Board Member for Scientific Reports. He was included among the top 2\% of scientists worldwide edition of Stanford University/Elsevier, in 2024. He served as the general chair, the program chair, and a committee member for numerous international conferences. He serves as an Editor for IEEE journals, Springer, Sage, Taylor and Francis, Sensors (MDPI), Electronics (MDPI), and Tech Science. He serves as a reviewer for several major IEEE and SCI journals.
\end{IEEEbiography}

\EOD

\end{document}